\documentclass[reprint,amsmath,amssymb,aps]{revtex4-2}
\usepackage{graphicx}
\usepackage{dcolumn}
\usepackage{bm}
\usepackage{tabstackengine}
\usepackage[dvipsnames]{xcolor}
\usepackage[normalem]{ulem}

\newcommand{\etal}{\textit{et al}.}
\newcommand{\tkafm}{3\textbf{k} AFM}
\begin{document}

\title{Phonon thermal transport in UO$_2$ via self-consistent perturbation theory}
\author{Shuxiang Zhou$^1$, Enda Xiao$^2$, Hao Ma$^{3,4}$, Krzysztof Gofryk$^1$, Chao Jiang$^1$, Michael E. Manley$^3$, David H. Hurley$^1$, and Chris A. Marianetti$^5$}
\affiliation{$^1$Idaho National Laboratory, Idaho Falls, Idaho 83415, USA\\$^2$Department of Chemistry, Columbia University, New York, New York 10027, USA\\$^3$Oak Ridge National Laboratory, Oak Ridge, Tennessee 37831, USA\\$^4$Department of Thermal Science and Energy Engineering, University of Science and Technology of China, Hefei, Anhui 230026, China\\$^5$Department of Applied Physics and Applied Mathematics, Columbia University, New York, New York 10027, USA}

\begin{abstract}

Computing thermal transport from first-principles in UO$_2$ is complicated due to the challenges associated 
with Mott
physics. Here we use irreducible derivative approaches to compute the cubic and quartic
phonon interactions in UO$_2$ from first-principles, and we perform enhanced thermal transport
computations by evaluating the
phonon Green's function via self-consistent diagrammatic perturbation theory. 
Our predicted phonon lifetimes at $T=600$ K agree well with our inelastic neutron scattering measurements 
across the entire Brillouin zone, 
and our thermal
conductivity predictions agree well with previous measurements.
Both the changes due to thermal expansion and self-consistent contributions are nontrivial
at high temperatures, though the effects tend to cancel, and interband transitions yield a substantial 
contribution.

\end{abstract}

\maketitle


Uranium dioxide (UO$_2$) has attracted a great deal of research interest since the 1950s, both as a standard nuclear fuel and as a fundamental system of rich physics induced by the partially filled \emph{f} shell \cite{landerFiftyYearsIt2020, hurleyThermalEnergyTransport2022}. As thermal transport is critical in
nuclear fuels, phonon thermal transport in UO$_2$ has been extensively studied both by experiments \cite{gofrykAnisotropicThermalConductivity2014a,
finkThermophysicalPropertiesUranium2000, batesVisibleInfraredAbsorption1965,
godfreyThermalConductivityUranium1965, ronchiEffectBurnupThermal2004} and from first-principles \cite{yinOriginLowThermal2008a,kaur_thermal_2013,meiFirstprinciplesStudyThermophysical2014,wangThermalConductivityUO22015, torresThermalConductivityDiffusion2019, torres_comparative_2020, yangReducedAnharmonicPhonon2022}. However, wide-ranging results were obtained from first-principles computations, and a robust consensus has not yet merged (see Sec. II in Supplemental Material (SM) \cite{Supplemental_Material} for a detailed discussion of the different approaches).
While low-temperature thermal conductivity is substantially complicated by magnons, defects, and boundary effects, room temperature and
beyond should be dominated by phonon thermal transport. However, accurately computing phonon interactions in
UO$_2$ from first-principles is complicated due to the complex interplay of Mott physics, magnetic order, and 
spin-orbit coupling (SOC). 
Here we circumvent these technical challenges by employing $f$-orbital occupation matrix control (OMC) \cite{dorado_textdfttextu_2009,amadon_ensuremathgamma_2008,jomard_structural_2008,zhou_crystal_2011} and the 3\textbf{k} antiferromagnetic (AFM) ground state obtained by our previous study \cite{zhouCapturingGroundState2022a}, which provides a 
robust description of the ground state and phonons as compared to experiment. 

Phonon thermal conductivity has been reliably computed in band insulators by solving the linearized phonon Peierls-Boltzmann transport equation (BTE) from first-principles using scattering rates computed within leading order perturbation theory \cite{broidoLatticeThermalConductivity2005,broidoIntrinsicLatticeThermal2007, broidoThermalConductivityDiamond2012, chaputDirectSolutionLinearized2013}. This \emph{de facto} standard approach for computing phonon thermal conductivity, as implemented by multiple publicly available software packages \cite{liShengBTESolverBoltzmann2014a, tadanoAnharmonicForceConstants2014, togoDistributionsPhononLifetimes2015a, chernatynskiyPhononTransportSimulator2015}, solves the BTE using cubic phonon interactions and the imaginary part of the bare bubble diagram. Naturally, non-trivial inaccuracy will occur under extreme conditions (e.g., high temperatures) where perturbation theory is inadequate. More recently, quartic phonon interactions have been incorporated using the imaginary part of the sunset diagram \cite{fengQuantumMechanicalPrediction2016, fengFourphononScatteringSignificantly2017}, and the contribution of interband phonon transitions has been addressed by a generalization of the BTE, known as the Wigner transport equation (WTE) \cite{simoncelliUnifiedTheoryThermal2019, simoncelliWignerFormulationThermal2022}. 
Here, we go beyond the current state of the art for computing thermal conductivity, which only uses
the imaginary parts of the bubble and sunset diagrams,
by using self-consistent diagrammatic perturbation theory to compute the single phonon Green’s
function \cite{xiaoAnharmonicPhononBehavior2023}.

In the present work, we use density functional theory plus $U$ (DFT+$U$) \cite{anisimovFirstprinciplesCalculationsElectronic1997} to compute the
cubic and quartic phonon interactions, which are then used to compute the inelastic neutron scattering (INS)
function and thermal conductivity. 
Both the scattering function and the thermal conductivity are computed using
increasingly sophisticated levels of theory, including bare perturbation theory
and self-consistent perturbation theory \cite{xiaoAnharmonicPhononBehavior2023}. For the latter, two different levels
of self-consistency are employed: the Hartree-Fock (HF) approximation for
phonons and quasiparticle perturbation (QP) theory. The former is the
traditional variational approach of Hooton \cite{hootonLINewTreatment1955}, where the four phonon loop diagram
is evaluated self-consistently, and the latter self-consistently evaluates both
the four phonon loop diagram and the
real part of the three phonon bubble diagram  \cite{xiaoAnharmonicPhononBehavior2023}. Following Ref.~\cite{xiaoAnharmonicPhononBehavior2023}, the self-consistency scheme and the subsequent diagrams evaluated to construct the phonon self-energy are indicated by the notation $\mathcal{S}^A_{ijk...}$, where $A\in\{o, HF, QP\}$ labels the self-consistency scheme and $i, j, k, . . .$
indicate all diagrams evaluated post self-consistency (see Figure 1 and S1 of Ref \cite{xiaoAnharmonicPhononBehavior2023} for schematics of diagrams). The colloquial diagram names
bubble, loop, and sunset are abbreviated as $b$, $l$, and $s$, respectively, while 
the self-consistency schemes
$o$, $HF$, and $QP$
correspond to the bare, Hartree-Fock, and quasiparticle
Green’s function, respectively.
For example, the imaginary part of the phonon self-energy used in the standard thermal conductivity approach \cite{broidoIntrinsicLatticeThermal2007} is 
obtained from  $\mathcal{S}^o_{b}$; and the approach in Ref.~\cite{fengFourphononScatteringSignificantly2017}, which employs
quartic phonon interactions using the imaginary part of the sunset diagram, is obtained from $\mathcal{S}^o_{bs}$. 
For each scheme we employ, both BTE and WTE are applied within the relaxation time
approximation (RTA). 
For $\mathcal{S}^o_{b}$, the full solution to the BTE 
is also obtained, yielding results very close to the RTA (see Sec. III of SM \cite{Supplemental_Material}), as is
consistent with previous results for ThO$_2$ \cite{xiaoValidatingFirstprinciplesPhonon2022} and CaF$_2$ \cite{qiLatticeDynamicsThermal2016, xiaoValidatingFirstprinciplesPhonon2022} .
To include the effects of the thermal expansion, the phonons and phonon interactions are computed at three expanded volumes, according to the experimental thermal expansion coefficients at $T= 360$, $600$, and $1000$ K \cite{mominHighTemperatureXray1991}. These computed results are linearly interpolated or extrapolated to temperatures from $0$ to $1400$ K.

\begin{figure}[t]
\includegraphics[width=0.91\columnwidth]{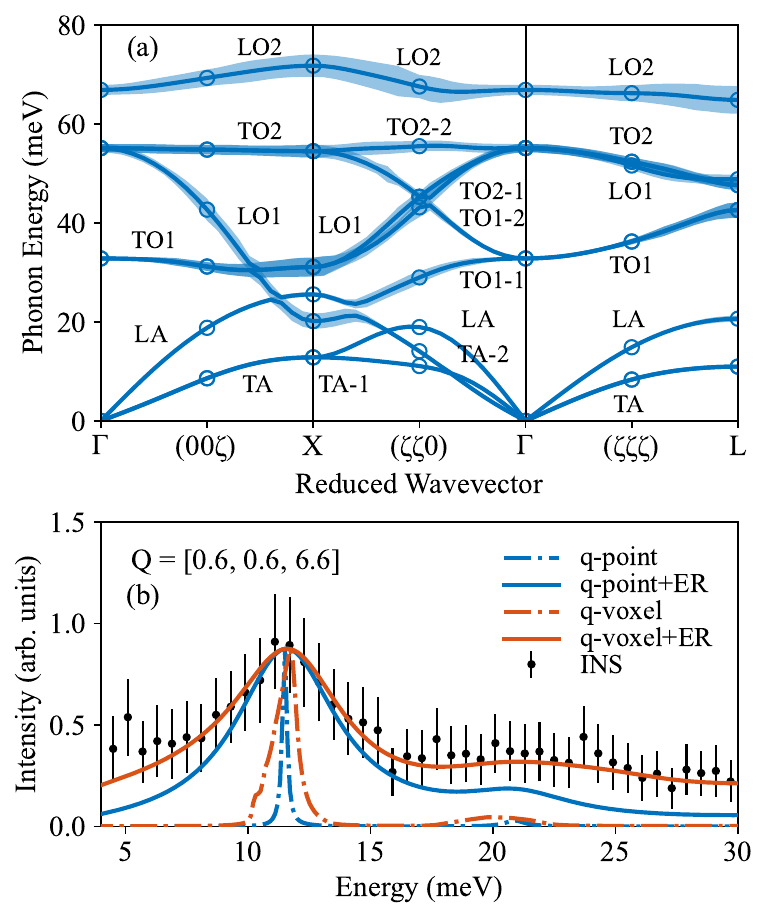}
\caption{\label{fig:cutplot} 
(a) The unfolded phonon dispersion at $T=0$ K, computed using  GGA+$U$+SOC ($U=4$ eV) in the \tkafm{} state. The hollow points were directly computed using DFT, while the 
corresponding curves are Fourier interpolations. The width of the line shading represents the FWHM computed at $T=600$ K
using $\mathcal{S}_{b}^{o}$. (b) $S(\mathbf{Q},\omega)$ at $\mathbf{Q} = [0.6, 0.6,
6.6]$ and $T=600$ K, computed using $\mathcal{S}_{b}^{o}$.  For
direct comparison, the INS instrumental energy resolution (ER) is accounted for in
the solid curves \cite{abernathy2012design}.  The $q$-voxel dimensions used in both the
measurement and computation are 0.075, 0.2, and 0.2 reciprocal lattice units
(r.l.u.) along the [L, L, L],  [0, 0, L], and [-H, H, 0] directions,
respectively. }
\end{figure}


Our DFT+$U$ calculations were carried out using the projector augmented-wave (PAW) method   \cite{blochl_projector_1994, kresse_ultrasoft_1999}, as implemented in the Vienna ab initio Simulation Package (VASP) code   \cite{kresse_ab_1993, kresse_efficient_1996}. The exchange correlation functional employed in our DFT+$U$ calculations was the generalized gradient approximation (GGA) as formulated by Perdew, Burke, and Ernzerhof (PBE) \cite{perdew_generalized_1996}, due to its overall better accuracy for phonons in UO$_2$ (see Sec. IX of SM \cite{Supplemental_Material}).  
We used the rotationally invariant DFT+$U$ approach of Dudarev \etal{} \cite{dudarev_electron-energy-loss_1998}, which only employs a single effective interaction, and $U=4$ eV was used throughout. SOC was included in all calculations. We customized the VASP code to initialize and monitor the occupation matrices during the calculations \cite{zhouCapturingGroundState2022a}, and the initial values of the occupation matrices were taken from our previous work (i.e.,  $\mathbb{S}_0$) \cite{zhouCapturingGroundState2022a}. The cubic and quartic phonon interactions were calculated via the bundled
irreducible derivative (BID) approach  \cite{fu_group_2019}. More information on the phonon interaction calculations, including supercell size, \emph{k}-point mesh, and Born effective charges, is provided in Sec. III of SM \cite{Supplemental_Material}. Details of the thermal conductivity calculations are also
provided in Sec. III of  SM \cite{Supplemental_Material}.

The scattering function $S(\mathbf{Q}, \omega)$ at $T = 600$ K was measured using  the Angular Range Chopper Spectrometer (ARCS) with an incident neutron energy $E_i=120$ meV and ARCS-100-1.5-AST Fermi chopper \cite{lin2019energy}. Further details of the UO$_2$ crystal and ARCS measurements have been reported previously \cite{bryan2020nonlinear}. The ARCS energy resolution functions \cite{lin2019energy} were used in fitting the phonon peaks, and the reported widths are the intrinsic full-width half-maximum (FWHM) values that have been corrected for the instrument contribution. The ARCS instrument measures a large volume in Q and E, which contains many Brillouin zones, and the data analysis allows for an adjustable size of $q$-voxel, a finite volume in reciprocal space
associated with some $q$-point, 
in all crystallographic directions. The $q$-voxel size is normally chosen to be as small as
possible, with the minimum being dictated by having sufficient counting statistics, and the resulting
scattering function is normally inherently broadened due to this issue
\cite{xiaoValidatingFirstprinciplesPhonon2022}. To make a meaningful comparison against
experiment, 
the usage of the experimental $q$-voxel must be accounted for within theory.  Details of the computation are reported in Ref.
\cite{xiaoValidatingFirstprinciplesPhonon2022}, and the $q$-voxel information is
included in Sec. V of SM \cite{Supplemental_Material}.


\begin{figure}[t]
\includegraphics[width=1.00\columnwidth]{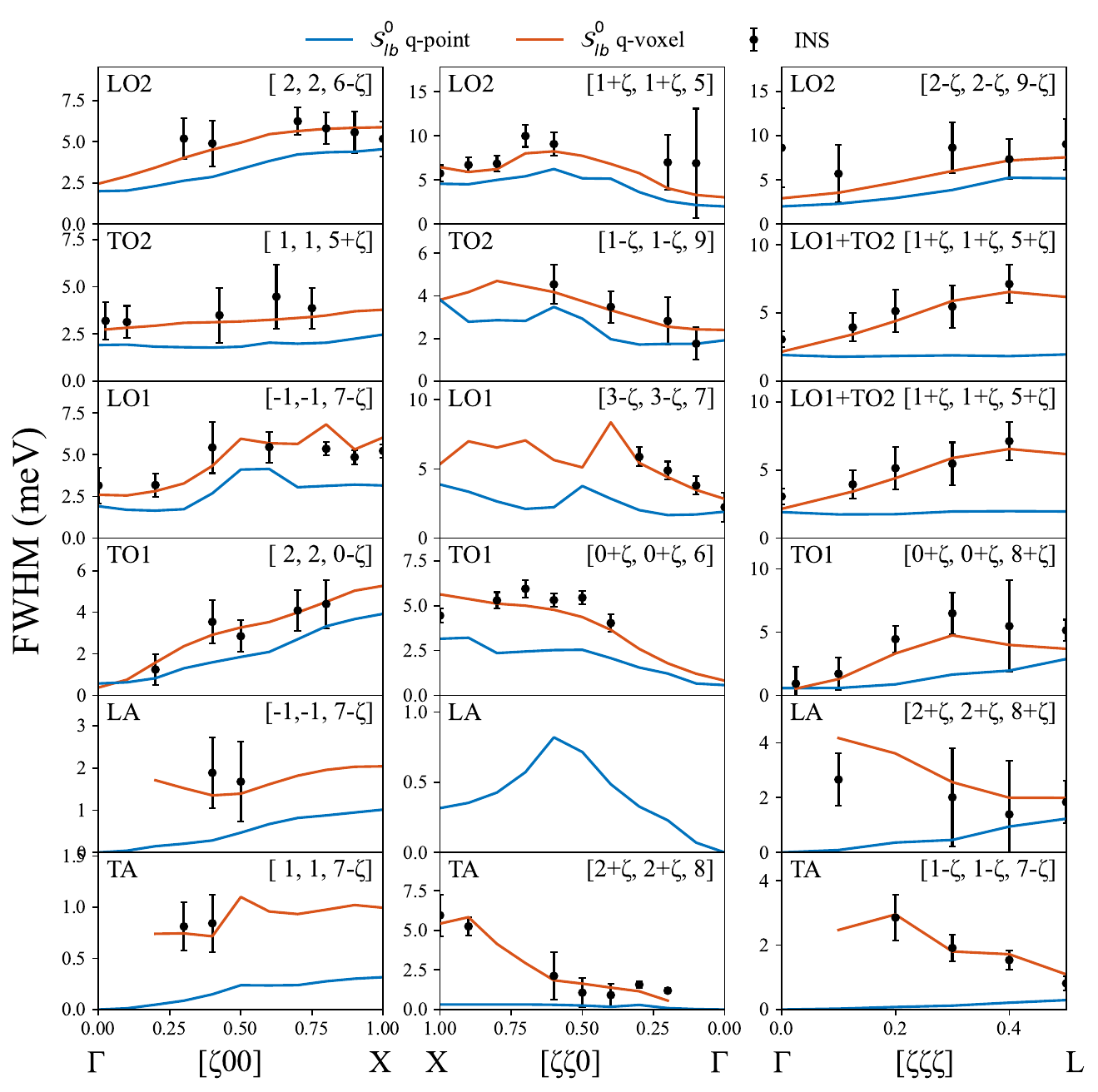}
\caption{\label{fig:linewidth} 
FWHMs of the $S(\mathbf{Q},\omega)$ peaks as a function of $\bm q$ in various zones for UO$_2$ at
$T=600$ K. The  $\mathcal{S}_{lb}^{o}$  $q$-point and $q$-voxel results are shown as blue and red curves, respectively. INS results are shown as black points. }
\end{figure}

\begin{figure*}[t]
\includegraphics[width=0.91\columnwidth]{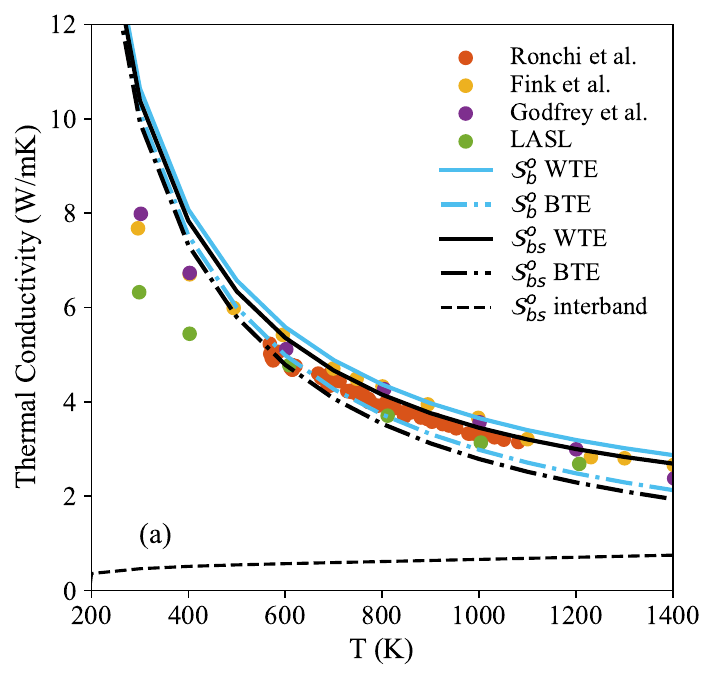}
\includegraphics[width=0.91\columnwidth]{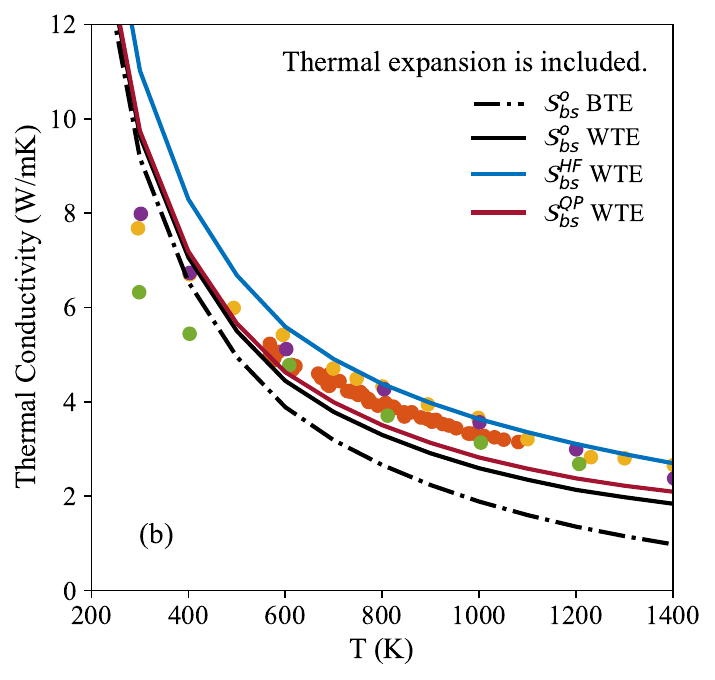}
\caption{\label{fig:conductivity} Thermal conductivity computed using GGA+$U$+SOC ($U=4$ eV) and comparing with experiments \cite{finkThermophysicalPropertiesUranium2000, batesVisibleInfraredAbsorption1965, godfreyThermalConductivityUranium1965, ronchiEffectBurnupThermal2004}. 
Panel (a) presents the bare perturbation theory results without including thermal expansion. Panel (b) presents the results with thermal expansion and self-consistent perturbation theory. }
\end{figure*}

We begin by considering the INS scattering function, which was measured at
$T=600$ K. The scattering function can be decomposed into components from
$n$-phonon contributions, and the dominant peaks in the spectra arise from 
the one phonon
contributions. The one phonon scattering function can be obtained from the phonon self-energy,
which can be computed using standard tools from many-body physics \cite{xiaoValidatingFirstprinciplesPhonon2022}. 
Given that $T=600$ K is still a modest temperature, we will demonstrate that the 
bare bubble and loop diagrams are still sufficient to reasonably capture the INS scattering function peak width.

We begin by plotting the phonon
linewidths computed using $\mathcal{S}_b^o$, which are overlaid on the phonon
dispersion (see Fig. \ref{fig:cutplot} (a)). The branch naming convention follows
Ref.  \cite{pang_phonon_2013}. Generally, the acoustic branches
have a much smaller FWHM than the optical branches, as is expected. For the INS
scattering function, we first consider the particular example of
$\mathbf{Q}=[0.6, 0.6, 6.6]$, where we individually illustrate the effects
of the $q$-voxel and energy resolution (see Fig. \ref{fig:cutplot} (b)).
The peak at approximately 12 meV corresponds to the TA mode, whereas the LA mode at
approximately 20 meV is barely observable due to the weighting factors in the
scattering function.  Clearly, both the $q$-voxel and energy resolution must be
considered to make a meaningful comparison with INS measurements. The
favorable agreement suggests that the level of theory we are using is robust,
but a detailed comparison across the entire Brillouin zone is still needed.

We now proceed to comprehensively compare the computational and experimental
results for the FWHMs of the scattering function peaks across the Brillouin zone
(see Fig. \ref{fig:linewidth}).  Following standard INS conventions, the energy
resolution is removed from the peak width, and the theoretical results are
presented for both the $q$-point and the $q$-voxel that was used in INS.
Overall, there is favorable agreement across all modes and $q$-paths,
indicating that the cubic phonon interactions computed using DFT+$U$ and the
bubble diagram used to evaluate the self-energy are sufficient to describe
experiment at $T=600$ K. We reevaluated Fig. \ref{fig:linewidth} using $\mathcal{S}_{lb}^{HF}$ and 
 $\mathcal{S}_{lb}^{QP}$, while accounting for thermal expansion, and the net changes
 are found to be modest at $T=600$ K (see Sec. VIII in SM \cite{Supplemental_Material}).

The thermal conductivity will first be explored at up to $T=1400$ K using bare perturbation theory (see Fig. \ref{fig:conductivity}(a)),
where the imaginary parts of the bubble and sunset diagrams will be considered.
Recall that we are not accounting for magnons, defects, or boundary effects, and thus our results will not
describe experiment below $T\approx400$ K. 
The $\mathcal{S}_b^o$ BTE results using the imaginary part of the bubble diagram 
are reasonable  above $T=400$ K
as compared to experiment \cite{finkThermophysicalPropertiesUranium2000, batesVisibleInfraredAbsorption1965, godfreyThermalConductivityUranium1965, ronchiEffectBurnupThermal2004}. 
This favorable agreement is anticipated from our preceding 
favorable comparison with INS. 
However, the result systematically underpredicts experiment at high temperatures,
and therefore it is compelling to include the quartic phonon interactions.
We begin by using the imaginary parts of both
the bare bubble and sunset diagrams ($\mathcal{S}_{bs}^o$), where the latter purely uses quartic interactions,
and the result is pushed further away from experiment by a small amount. This same effect was previously observed
in Ref. \cite{yangReducedAnharmonicPhonon2022} (see SM \cite{Supplemental_Material}, Fig. S1), though the 
value of their results are strongly underpredicted relative to our own (see Sec. II in SM \cite{Supplemental_Material}). 
Interestingly,
above room temperature, roughly 30{\%} of the thermal conductivity
arises from phonon modes with energies above 26 meV, which are predominantly optical modes, and therefore the
optical modes do play a nontrivial role in thermal transport (see Sec. VI in SM \cite{Supplemental_Material}).
We proceed by including interband phonon contributions via the WTE, which
increases the thermal conductivity monotonically with temperature, yielding
good agreement with experiment. At $T = 1400$ K, 
the interband portion contributes about 30\% of the total thermal
conductivity, which is non-trivial. 

While the current approximation yields reasonable agreement with experiment, it is important
to use a logically consistent approach where the real portion of the self-energy is not discarded
and the effects of thermal expansion are included (see Fig. \ref{fig:conductivity}(b)). 
When thermal expansion is included in the thermal conductivity calculation, a non-trivial decrease in thermal conductivity is observed: at $T=1400$ K, the decrease of the $\mathcal{S}_{bs}^o$ BTE result due to thermal expansion is nearly 50\%. This strong decrease may explain the relatively low predicted thermal conductivity in
Ref. \cite{pang_phonon_2013}, which used $\mathcal{S}_{b}^o$ BTE and included thermal expansion. 
While the interband contribution contained in the $\mathcal{S}_{bs}^o$ WTE result increases the predicted value, the thermal conductivity is still underpredicted as compared to experiment. However, it is still necessary
to account for the real part of the phonon self-energy, and explore
the possibility of accounting for higher order diagrams via self-consistent perturbation theory \cite{xiaoAnharmonicPhononBehavior2023}.

Nominally, we would expect the QP result to be better than
HF as it sums additional diagrams which are known to be relevant \cite{xiaoAnharmonicPhononBehavior2023}. 
We first consider $\mathcal{S}_{bs}^{HF}$ WTE, which notably increases
the thermal conductivity at high temperatures. This result is anticipated, given that the HF
approximation nominally increases the effective harmonic frequencies, decreasing the phonon
lifetime contribution from the bubble. The $\mathcal{S}_{bs}^{QP}$ WTE result is shifted downards from the HF result, slightly
above the bare perturbation theory result using the bubble and sunset diagrams. 
The underestimation
of the experimental thermal conductivity by $\mathcal{S}_{bs}^{QP}$ might be accounted for by including more diagrams,
possibly requiring diagrams with phonon interactions beyond fourth order.
Ideally, one would 
sum all possible diagrams and obtain the exact phonon self-energy, which can be done in the classical
limit using molecular dynamics \cite{xiaoAnharmonicPhononBehavior2023}. 
Another possibility for the discrepancy is that the
PBE exchange correlation functional is not sufficiently describing the phonon interactions.


In summary, we have computed the scattering function and thermal
conductivity of UO$_2$ from first-principles using various levels of
self-consistent perturbation theory and compared to our own INS experiments and
existing thermal conductivity experiments. The relevant contributions of this
work include accurately describing the phonon interactions in UO$_2$ from first
principles and illustrating the effects of improving the quality of the single
particle phonon Green's function on the thermal conductivity.  Favorable
agreement between our theory and INS experiment is obtained for the FWHM of
the scattering function across the Brillouin zone. 
In terms of quantitatively computing the thermal conductivity at high temperatures,
we find that thermal expansion decreases the thermal conductivity while
interband transitions increases the thermal conductivity, and these effects are
of similar magnitude. Including quartic phonon interactions at the level of the
bare sunset diagram causes a small decrease in thermal conductivity, while
the self-consistent perturbation theory yielded moderate and appreciable increases
for the quasiparticle and Hartree-Fock procedures, respectively.
Aside from low temperatures where magnons and defects play important roles, phonon
thermal transport in UO$_2$ is now well characterized from first-principles.


This work is supported by the Center for Thermal Energy Transport under Irradiation, an Energy Frontier Research Center funded by the U.S. Department of Energy (DOE) Office of Basic Energy Sciences. This research used resources at the Spallation Neutron Source, a DOE Office of Science User Facility operated by the ORNL. This research made use of Idaho National Laboratory computing resources, which are supported by the DOE Office of Nuclear Energy and the Nuclear Science User Facilities under contract no. DE-AC07-05ID14517.
This research also used resources of the National Energy Research Scientific Computing Center, a DOE Office of Science User Facility supported by the Office of Science of the U.S. Department of Energy under Contract No. DE-AC02-05CH11231.
The unfolding of phonons and phonon interactions was supported by grant DE-SC0016507 funded by the U.S. Department of Energy, Office of Science.

\bibliographystyle{apsrev4-1} 
\bibliography{uo2} 

\begin{thebibliography}{59}%
\makeatletter
\providecommand \@ifxundefined [1]{%
 \@ifx{#1\undefined}
}%
\providecommand \@ifnum [1]{%
 \ifnum #1\expandafter \@firstoftwo
 \else \expandafter \@secondoftwo
 \fi
}%
\providecommand \@ifx [1]{%
 \ifx #1\expandafter \@firstoftwo
 \else \expandafter \@secondoftwo
 \fi
}%
\providecommand \natexlab [1]{#1}%
\providecommand \enquote  [1]{``#1''}%
\providecommand \bibnamefont  [1]{#1}%
\providecommand \bibfnamefont [1]{#1}%
\providecommand \citenamefont [1]{#1}%
\providecommand \href@noop [0]{\@secondoftwo}%
\providecommand \href [0]{\begingroup \@sanitize@url \@href}%
\providecommand \@href[1]{\@@startlink{#1}\@@href}%
\providecommand \@@href[1]{\endgroup#1\@@endlink}%
\providecommand \@sanitize@url [0]{\catcode `\\12\catcode `\$12\catcode
  `\&12\catcode `\#12\catcode `\^12\catcode `\_12\catcode `\%12\relax}%
\providecommand \@@startlink[1]{}%
\providecommand \@@endlink[0]{}%
\providecommand \url  [0]{\begingroup\@sanitize@url \@url }%
\providecommand \@url [1]{\endgroup\@href {#1}{\urlprefix }}%
\providecommand \urlprefix  [0]{URL }%
\providecommand \Eprint [0]{\href }%
\providecommand \doibase [0]{http://dx.doi.org/}%
\providecommand \selectlanguage [0]{\@gobble}%
\providecommand \bibinfo  [0]{\@secondoftwo}%
\providecommand \bibfield  [0]{\@secondoftwo}%
\providecommand \translation [1]{[#1]}%
\providecommand \BibitemOpen [0]{}%
\providecommand \bibitemStop [0]{}%
\providecommand \bibitemNoStop [0]{.\EOS\space}%
\providecommand \EOS [0]{\spacefactor3000\relax}%
\providecommand \BibitemShut  [1]{\csname bibitem#1\endcsname}%
\let\auto@bib@innerbib\@empty
\bibitem [{\citenamefont {Lander}\ and\ \citenamefont
  {Caciuffo}(2020)}]{landerFiftyYearsIt2020}%
  \BibitemOpen
  \bibfield  {author} {\bibinfo {author} {\bibfnamefont {G.~H.}\ \bibnamefont
  {Lander}}\ and\ \bibinfo {author} {\bibfnamefont {R.}~\bibnamefont
  {Caciuffo}},\ }\href {\doibase 10.1088/1361-648X/ab1dc5} {\bibfield
  {journal} {\bibinfo  {journal} {Journal of Physics: Condensed Matter}\
  }\textbf {\bibinfo {volume} {32}},\ \bibinfo {pages} {374001} (\bibinfo
  {year} {2020})}\BibitemShut {NoStop}%
\bibitem [{\citenamefont {Hurley}\ \emph {et~al.}(2022)\citenamefont {Hurley},
  \citenamefont {{El-Azab}}, \citenamefont {Bryan}, \citenamefont {Cooper},
  \citenamefont {Dennett}, \citenamefont {Gofryk}, \citenamefont {He},
  \citenamefont {Khafizov}, \citenamefont {Lander}, \citenamefont {Manley},
  \citenamefont {Mann}, \citenamefont {Marianetti}, \citenamefont {Rickert},
  \citenamefont {Selim}, \citenamefont {Tonks},\ and\ \citenamefont
  {Wharry}}]{hurleyThermalEnergyTransport2022}%
  \BibitemOpen
  \bibfield  {author} {\bibinfo {author} {\bibfnamefont {D.~H.}\ \bibnamefont
  {Hurley}}, \bibinfo {author} {\bibfnamefont {A.}~\bibnamefont {{El-Azab}}},
  \bibinfo {author} {\bibfnamefont {M.~S.}\ \bibnamefont {Bryan}}, \bibinfo
  {author} {\bibfnamefont {M.~W.~D.}\ \bibnamefont {Cooper}}, \bibinfo {author}
  {\bibfnamefont {C.~A.}\ \bibnamefont {Dennett}}, \bibinfo {author}
  {\bibfnamefont {K.}~\bibnamefont {Gofryk}}, \bibinfo {author} {\bibfnamefont
  {L.}~\bibnamefont {He}}, \bibinfo {author} {\bibfnamefont {M.}~\bibnamefont
  {Khafizov}}, \bibinfo {author} {\bibfnamefont {G.~H.}\ \bibnamefont
  {Lander}}, \bibinfo {author} {\bibfnamefont {M.~E.}\ \bibnamefont {Manley}},
  \bibinfo {author} {\bibfnamefont {J.~M.}\ \bibnamefont {Mann}}, \bibinfo
  {author} {\bibfnamefont {C.~A.}\ \bibnamefont {Marianetti}}, \bibinfo
  {author} {\bibfnamefont {K.}~\bibnamefont {Rickert}}, \bibinfo {author}
  {\bibfnamefont {F.~A.}\ \bibnamefont {Selim}}, \bibinfo {author}
  {\bibfnamefont {M.~R.}\ \bibnamefont {Tonks}}, \ and\ \bibinfo {author}
  {\bibfnamefont {J.~P.}\ \bibnamefont {Wharry}},\ }\href {\doibase
  10.1021/acs.chemrev.1c00262} {\bibfield  {journal} {\bibinfo  {journal}
  {Chemical Reviews}\ }\textbf {\bibinfo {volume} {122}},\ \bibinfo {pages}
  {3711} (\bibinfo {year} {2022})}\BibitemShut {NoStop}%
\bibitem [{\citenamefont {Gofryk}\ \emph {et~al.}(2014)\citenamefont {Gofryk},
  \citenamefont {Du}, \citenamefont {Stanek}, \citenamefont {Lashley},
  \citenamefont {Liu}, \citenamefont {Schulze}, \citenamefont {Smith},
  \citenamefont {Safarik}, \citenamefont {Byler}, \citenamefont {McClellan},
  \citenamefont {Uberuaga}, \citenamefont {Scott},\ and\ \citenamefont
  {Andersson}}]{gofrykAnisotropicThermalConductivity2014a}%
  \BibitemOpen
  \bibfield  {author} {\bibinfo {author} {\bibfnamefont {K.}~\bibnamefont
  {Gofryk}}, \bibinfo {author} {\bibfnamefont {S.}~\bibnamefont {Du}}, \bibinfo
  {author} {\bibfnamefont {C.~R.}\ \bibnamefont {Stanek}}, \bibinfo {author}
  {\bibfnamefont {J.~C.}\ \bibnamefont {Lashley}}, \bibinfo {author}
  {\bibfnamefont {X.-Y.}\ \bibnamefont {Liu}}, \bibinfo {author} {\bibfnamefont
  {R.~K.}\ \bibnamefont {Schulze}}, \bibinfo {author} {\bibfnamefont {J.~L.}\
  \bibnamefont {Smith}}, \bibinfo {author} {\bibfnamefont {D.~J.}\ \bibnamefont
  {Safarik}}, \bibinfo {author} {\bibfnamefont {D.~D.}\ \bibnamefont {Byler}},
  \bibinfo {author} {\bibfnamefont {K.~J.}\ \bibnamefont {McClellan}}, \bibinfo
  {author} {\bibfnamefont {B.~P.}\ \bibnamefont {Uberuaga}}, \bibinfo {author}
  {\bibfnamefont {B.~L.}\ \bibnamefont {Scott}}, \ and\ \bibinfo {author}
  {\bibfnamefont {D.~A.}\ \bibnamefont {Andersson}},\ }\href {\doibase
  10.1038/ncomms5551} {\bibfield  {journal} {\bibinfo  {journal} {Nature
  Communications}\ }\textbf {\bibinfo {volume} {5}},\ \bibinfo {pages} {4551}
  (\bibinfo {year} {2014})}\BibitemShut {NoStop}%
\bibitem [{\citenamefont
  {Fink}(2000)}]{finkThermophysicalPropertiesUranium2000}%
  \BibitemOpen
  \bibfield  {author} {\bibinfo {author} {\bibfnamefont {J.~K.}\ \bibnamefont
  {Fink}},\ }\href {\doibase 10.1016/S0022-3115(99)00273-1} {\bibfield
  {journal} {\bibinfo  {journal} {Journal of Nuclear Materials}\ }\textbf
  {\bibinfo {volume} {279}},\ \bibinfo {pages} {1} (\bibinfo {year}
  {2000})}\BibitemShut {NoStop}%
\bibitem [{\citenamefont {Bates}(1965)}]{batesVisibleInfraredAbsorption1965}%
  \BibitemOpen
  \bibfield  {author} {\bibinfo {author} {\bibfnamefont {J.~L.}\ \bibnamefont
  {Bates}},\ }\href {\doibase 10.13182/NSE65-A21011} {\bibfield  {journal}
  {\bibinfo  {journal} {Nuclear Science and Engineering}\ }\textbf {\bibinfo
  {volume} {21}},\ \bibinfo {pages} {26} (\bibinfo {year} {1965})}\BibitemShut
  {NoStop}%
\bibitem [{\citenamefont {Godfrey}\ \emph {et~al.}(1965)\citenamefont
  {Godfrey}, \citenamefont {Fulkerson}, \citenamefont {Kollie}, \citenamefont
  {Moore},\ and\ \citenamefont
  {McELROY}}]{godfreyThermalConductivityUranium1965}%
  \BibitemOpen
  \bibfield  {author} {\bibinfo {author} {\bibfnamefont {T.~G.}\ \bibnamefont
  {Godfrey}}, \bibinfo {author} {\bibfnamefont {W.}~\bibnamefont {Fulkerson}},
  \bibinfo {author} {\bibfnamefont {T.~G.}\ \bibnamefont {Kollie}}, \bibinfo
  {author} {\bibfnamefont {J.~P.}\ \bibnamefont {Moore}}, \ and\ \bibinfo
  {author} {\bibfnamefont {D.~L.}\ \bibnamefont {McELROY}},\ }\href {\doibase
  10.1111/j.1151-2916.1965.tb14745.x} {\bibfield  {journal} {\bibinfo
  {journal} {Journal of the American Ceramic Society}\ }\textbf {\bibinfo
  {volume} {48}},\ \bibinfo {pages} {297} (\bibinfo {year} {1965})}\BibitemShut
  {NoStop}%
\bibitem [{\citenamefont {Ronchi}\ \emph {et~al.}(2004)\citenamefont {Ronchi},
  \citenamefont {Sheindlin}, \citenamefont {Staicu},\ and\ \citenamefont
  {Kinoshita}}]{ronchiEffectBurnupThermal2004}%
  \BibitemOpen
  \bibfield  {author} {\bibinfo {author} {\bibfnamefont {C.}~\bibnamefont
  {Ronchi}}, \bibinfo {author} {\bibfnamefont {M.}~\bibnamefont {Sheindlin}},
  \bibinfo {author} {\bibfnamefont {D.}~\bibnamefont {Staicu}}, \ and\ \bibinfo
  {author} {\bibfnamefont {M.}~\bibnamefont {Kinoshita}},\ }\href {\doibase
  10.1016/j.jnucmat.2004.01.018} {\bibfield  {journal} {\bibinfo  {journal}
  {Journal of Nuclear Materials}\ }\textbf {\bibinfo {volume} {327}},\ \bibinfo
  {pages} {58} (\bibinfo {year} {2004})}\BibitemShut {NoStop}%
\bibitem [{\citenamefont {Yin}\ and\ \citenamefont
  {Savrasov}(2008)}]{yinOriginLowThermal2008a}%
  \BibitemOpen
  \bibfield  {author} {\bibinfo {author} {\bibfnamefont {Q.}~\bibnamefont
  {Yin}}\ and\ \bibinfo {author} {\bibfnamefont {S.~Y.}\ \bibnamefont
  {Savrasov}},\ }\href {\doibase 10.1103/PhysRevLett.100.225504} {\bibfield
  {journal} {\bibinfo  {journal} {Physical Review Letters}\ }\textbf {\bibinfo
  {volume} {100}},\ \bibinfo {pages} {225504} (\bibinfo {year}
  {2008})}\BibitemShut {NoStop}%
\bibitem [{\citenamefont {Kaur}\ \emph {et~al.}(2013)\citenamefont {Kaur},
  \citenamefont {Panigrahi},\ and\ \citenamefont
  {Valsakumar}}]{kaur_thermal_2013}%
  \BibitemOpen
  \bibfield  {author} {\bibinfo {author} {\bibfnamefont {G.}~\bibnamefont
  {Kaur}}, \bibinfo {author} {\bibfnamefont {P.}~\bibnamefont {Panigrahi}}, \
  and\ \bibinfo {author} {\bibfnamefont {M.~C.}\ \bibnamefont {Valsakumar}},\
  }\href {\doibase 10.1088/0965-0393/21/6/065014} {\bibfield  {journal}
  {\bibinfo  {journal} {Modelling and Simulation in Materials Science and
  Engineering}\ }\textbf {\bibinfo {volume} {21}},\ \bibinfo {pages} {065014}
  (\bibinfo {year} {2013})}\BibitemShut {NoStop}%
\bibitem [{\citenamefont {Mei}\ \emph {et~al.}(2014)\citenamefont {Mei},
  \citenamefont {Stan},\ and\ \citenamefont
  {Yang}}]{meiFirstprinciplesStudyThermophysical2014}%
  \BibitemOpen
  \bibfield  {author} {\bibinfo {author} {\bibfnamefont {Z.-G.}\ \bibnamefont
  {Mei}}, \bibinfo {author} {\bibfnamefont {M.}~\bibnamefont {Stan}}, \ and\
  \bibinfo {author} {\bibfnamefont {J.}~\bibnamefont {Yang}},\ }\href {\doibase
  10.1016/j.jallcom.2014.03.091} {\bibfield  {journal} {\bibinfo  {journal}
  {Journal of Alloys and Compounds}\ }\textbf {\bibinfo {volume} {603}},\
  \bibinfo {pages} {282} (\bibinfo {year} {2014})}\BibitemShut {NoStop}%
\bibitem [{\citenamefont {Wang}\ \emph {et~al.}(2015)\citenamefont {Wang},
  \citenamefont {Zheng}, \citenamefont {Qu}, \citenamefont {Li},\ and\
  \citenamefont {Zhang}}]{wangThermalConductivityUO22015}%
  \BibitemOpen
  \bibfield  {author} {\bibinfo {author} {\bibfnamefont {B.-T.}\ \bibnamefont
  {Wang}}, \bibinfo {author} {\bibfnamefont {J.-J.}\ \bibnamefont {Zheng}},
  \bibinfo {author} {\bibfnamefont {X.}~\bibnamefont {Qu}}, \bibinfo {author}
  {\bibfnamefont {W.-D.}\ \bibnamefont {Li}}, \ and\ \bibinfo {author}
  {\bibfnamefont {P.}~\bibnamefont {Zhang}},\ }\href {\doibase
  10.1016/j.jallcom.2014.12.204} {\bibfield  {journal} {\bibinfo  {journal}
  {Journal of Alloys and Compounds}\ }\textbf {\bibinfo {volume} {628}},\
  \bibinfo {pages} {267} (\bibinfo {year} {2015})}\BibitemShut {NoStop}%
\bibitem [{\citenamefont {Torres}\ and\ \citenamefont
  {Kaloni}(2019)}]{torresThermalConductivityDiffusion2019}%
  \BibitemOpen
  \bibfield  {author} {\bibinfo {author} {\bibfnamefont {E.}~\bibnamefont
  {Torres}}\ and\ \bibinfo {author} {\bibfnamefont {T.~P.}\ \bibnamefont
  {Kaloni}},\ }\href {\doibase 10.1016/j.jnucmat.2019.04.040} {\bibfield
  {journal} {\bibinfo  {journal} {Journal of Nuclear Materials}\ }\textbf
  {\bibinfo {volume} {521}},\ \bibinfo {pages} {137} (\bibinfo {year}
  {2019})}\BibitemShut {NoStop}%
\bibitem [{\citenamefont {Torres}\ \emph {et~al.}(2020)\citenamefont {Torres},
  \citenamefont {CheikNjifon}, \citenamefont {Kaloni},\ and\ \citenamefont
  {Pencer}}]{torres_comparative_2020}%
  \BibitemOpen
  \bibfield  {author} {\bibinfo {author} {\bibfnamefont {E.}~\bibnamefont
  {Torres}}, \bibinfo {author} {\bibfnamefont {I.}~\bibnamefont {CheikNjifon}},
  \bibinfo {author} {\bibfnamefont {T.~P.}\ \bibnamefont {Kaloni}}, \ and\
  \bibinfo {author} {\bibfnamefont {J.}~\bibnamefont {Pencer}},\ }\href
  {\doibase 10.1016/j.commatsci.2020.109594} {\bibfield  {journal} {\bibinfo
  {journal} {Computational Materials Science}\ }\textbf {\bibinfo {volume}
  {177}},\ \bibinfo {pages} {109594} (\bibinfo {year} {2020})}\BibitemShut
  {NoStop}%
\bibitem [{\citenamefont {Yang}\ \emph {et~al.}(2022)\citenamefont {Yang},
  \citenamefont {Tiwari},\ and\ \citenamefont
  {Feng}}]{yangReducedAnharmonicPhonon2022}%
  \BibitemOpen
  \bibfield  {author} {\bibinfo {author} {\bibfnamefont {X.}~\bibnamefont
  {Yang}}, \bibinfo {author} {\bibfnamefont {J.}~\bibnamefont {Tiwari}}, \ and\
  \bibinfo {author} {\bibfnamefont {T.}~\bibnamefont {Feng}},\ }\href {\doibase
  10.1016/j.mtphys.2022.100689} {\bibfield  {journal} {\bibinfo  {journal}
  {Materials Today Physics}\ }\textbf {\bibinfo {volume} {24}},\ \bibinfo
  {pages} {100689} (\bibinfo {year} {2022})}\BibitemShut {NoStop}%
\bibitem [{Sup()}]{Supplemental_Material}%
  \BibitemOpen
  \href@noop {} {}\bibinfo {note} {See Supplemental Materials at [link] for
  information about previous DFT+$U$ studies on thermal conductivity of UO$_2$,
  the computational details and results of thermal conductivity, and the values
  of computed irreducible derivatives in this work. See also Refs.
  \cite{wang_phonon_2013, gonzeDynamicalMatricesBorn1997a,
  mathisGeneralizedQuasiharmonicApproximation2022, idiri_behavior_2004,
  santini_multipolar_2009, bryan_impact_2019, international1965thermodynamic,
  ronchiThermalConductivityUranium1999a, kimLatticeThermalConductivity2014,
  pavlovMeasurementInterpretationThermophysical2017}}\BibitemShut {NoStop}%
\bibitem [{\citenamefont {Dorado}\ \emph {et~al.}(2009)\citenamefont {Dorado},
  \citenamefont {Amadon}, \citenamefont {Freyss},\ and\ \citenamefont
  {Bertolus}}]{dorado_textdfttextu_2009}%
  \BibitemOpen
  \bibfield  {author} {\bibinfo {author} {\bibfnamefont {B.}~\bibnamefont
  {Dorado}}, \bibinfo {author} {\bibfnamefont {B.}~\bibnamefont {Amadon}},
  \bibinfo {author} {\bibfnamefont {M.}~\bibnamefont {Freyss}}, \ and\ \bibinfo
  {author} {\bibfnamefont {M.}~\bibnamefont {Bertolus}},\ }\href {\doibase
  10.1103/PhysRevB.79.235125} {\bibfield  {journal} {\bibinfo  {journal}
  {Physical Review B}\ }\textbf {\bibinfo {volume} {79}},\ \bibinfo {pages}
  {235125} (\bibinfo {year} {2009})}\BibitemShut {NoStop}%
\bibitem [{\citenamefont {Amadon}\ \emph {et~al.}(2008)\citenamefont {Amadon},
  \citenamefont {Jollet},\ and\ \citenamefont
  {Torrent}}]{amadon_ensuremathgamma_2008}%
  \BibitemOpen
  \bibfield  {author} {\bibinfo {author} {\bibfnamefont {B.}~\bibnamefont
  {Amadon}}, \bibinfo {author} {\bibfnamefont {F.}~\bibnamefont {Jollet}}, \
  and\ \bibinfo {author} {\bibfnamefont {M.}~\bibnamefont {Torrent}},\ }\href
  {\doibase 10.1103/PhysRevB.77.155104} {\bibfield  {journal} {\bibinfo
  {journal} {Physical Review B}\ }\textbf {\bibinfo {volume} {77}},\ \bibinfo
  {pages} {155104} (\bibinfo {year} {2008})}\BibitemShut {NoStop}%
\bibitem [{\citenamefont {Jomard}\ \emph {et~al.}(2008)\citenamefont {Jomard},
  \citenamefont {Amadon}, \citenamefont {Bottin},\ and\ \citenamefont
  {Torrent}}]{jomard_structural_2008}%
  \BibitemOpen
  \bibfield  {author} {\bibinfo {author} {\bibfnamefont {G.}~\bibnamefont
  {Jomard}}, \bibinfo {author} {\bibfnamefont {B.}~\bibnamefont {Amadon}},
  \bibinfo {author} {\bibfnamefont {F.}~\bibnamefont {Bottin}}, \ and\ \bibinfo
  {author} {\bibfnamefont {M.}~\bibnamefont {Torrent}},\ }\href {\doibase
  10.1103/PhysRevB.78.075125} {\bibfield  {journal} {\bibinfo  {journal}
  {Physical Review B}\ }\textbf {\bibinfo {volume} {78}},\ \bibinfo {pages}
  {075125} (\bibinfo {year} {2008})}\BibitemShut {NoStop}%
\bibitem [{\citenamefont {Zhou}\ and\ \citenamefont
  {Ozoliņš}(2011)}]{zhou_crystal_2011}%
  \BibitemOpen
  \bibfield  {author} {\bibinfo {author} {\bibfnamefont {F.}~\bibnamefont
  {Zhou}}\ and\ \bibinfo {author} {\bibfnamefont {V.}~\bibnamefont
  {Ozoliņš}},\ }\href {\doibase 10.1103/PhysRevB.83.085106} {\bibfield
  {journal} {\bibinfo  {journal} {Physical Review B}\ }\textbf {\bibinfo
  {volume} {83}},\ \bibinfo {pages} {085106} (\bibinfo {year}
  {2011})}\BibitemShut {NoStop}%
\bibitem [{\citenamefont {Zhou}\ \emph {et~al.}(2022)\citenamefont {Zhou},
  \citenamefont {Ma}, \citenamefont {Xiao}, \citenamefont {Gofryk},
  \citenamefont {Jiang}, \citenamefont {Manley}, \citenamefont {Hurley},\ and\
  \citenamefont {Marianetti}}]{zhouCapturingGroundState2022a}%
  \BibitemOpen
  \bibfield  {author} {\bibinfo {author} {\bibfnamefont {S.}~\bibnamefont
  {Zhou}}, \bibinfo {author} {\bibfnamefont {H.}~\bibnamefont {Ma}}, \bibinfo
  {author} {\bibfnamefont {E.}~\bibnamefont {Xiao}}, \bibinfo {author}
  {\bibfnamefont {K.}~\bibnamefont {Gofryk}}, \bibinfo {author} {\bibfnamefont
  {C.}~\bibnamefont {Jiang}}, \bibinfo {author} {\bibfnamefont {M.~E.}\
  \bibnamefont {Manley}}, \bibinfo {author} {\bibfnamefont {D.~H.}\
  \bibnamefont {Hurley}}, \ and\ \bibinfo {author} {\bibfnamefont {C.~A.}\
  \bibnamefont {Marianetti}},\ }\href {\doibase 10.1103/PhysRevB.106.125134}
  {\bibfield  {journal} {\bibinfo  {journal} {Physical Review B}\ }\textbf
  {\bibinfo {volume} {106}},\ \bibinfo {pages} {125134} (\bibinfo {year}
  {2022})}\BibitemShut {NoStop}%
\bibitem [{\citenamefont {Broido}\ \emph {et~al.}(2005)\citenamefont {Broido},
  \citenamefont {Ward},\ and\ \citenamefont
  {Mingo}}]{broidoLatticeThermalConductivity2005}%
  \BibitemOpen
  \bibfield  {author} {\bibinfo {author} {\bibfnamefont {D.~A.}\ \bibnamefont
  {Broido}}, \bibinfo {author} {\bibfnamefont {A.}~\bibnamefont {Ward}}, \ and\
  \bibinfo {author} {\bibfnamefont {N.}~\bibnamefont {Mingo}},\ }\href
  {\doibase 10.1103/PhysRevB.72.014308} {\bibfield  {journal} {\bibinfo
  {journal} {Physical Review B}\ }\textbf {\bibinfo {volume} {72}},\ \bibinfo
  {pages} {014308} (\bibinfo {year} {2005})}\BibitemShut {NoStop}%
\bibitem [{\citenamefont {Broido}\ \emph {et~al.}(2007)\citenamefont {Broido},
  \citenamefont {Malorny}, \citenamefont {Birner}, \citenamefont {Mingo},\ and\
  \citenamefont {Stewart}}]{broidoIntrinsicLatticeThermal2007}%
  \BibitemOpen
  \bibfield  {author} {\bibinfo {author} {\bibfnamefont {D.~A.}\ \bibnamefont
  {Broido}}, \bibinfo {author} {\bibfnamefont {M.}~\bibnamefont {Malorny}},
  \bibinfo {author} {\bibfnamefont {G.}~\bibnamefont {Birner}}, \bibinfo
  {author} {\bibfnamefont {N.}~\bibnamefont {Mingo}}, \ and\ \bibinfo {author}
  {\bibfnamefont {D.~A.}\ \bibnamefont {Stewart}},\ }\href {\doibase
  10.1063/1.2822891} {\bibfield  {journal} {\bibinfo  {journal} {Applied
  Physics Letters}\ }\textbf {\bibinfo {volume} {91}},\ \bibinfo {pages}
  {231922} (\bibinfo {year} {2007})}\BibitemShut {NoStop}%
\bibitem [{\citenamefont {Broido}\ \emph {et~al.}(2012)\citenamefont {Broido},
  \citenamefont {Lindsay},\ and\ \citenamefont
  {Ward}}]{broidoThermalConductivityDiamond2012}%
  \BibitemOpen
  \bibfield  {author} {\bibinfo {author} {\bibfnamefont {D.~A.}\ \bibnamefont
  {Broido}}, \bibinfo {author} {\bibfnamefont {L.}~\bibnamefont {Lindsay}}, \
  and\ \bibinfo {author} {\bibfnamefont {A.}~\bibnamefont {Ward}},\ }\href
  {\doibase 10.1103/PhysRevB.86.115203} {\bibfield  {journal} {\bibinfo
  {journal} {Physical Review B}\ }\textbf {\bibinfo {volume} {86}},\ \bibinfo
  {pages} {115203} (\bibinfo {year} {2012})}\BibitemShut {NoStop}%
\bibitem [{\citenamefont {Chaput}(2013)}]{chaputDirectSolutionLinearized2013}%
  \BibitemOpen
  \bibfield  {author} {\bibinfo {author} {\bibfnamefont {L.}~\bibnamefont
  {Chaput}},\ }\href {\doibase 10.1103/PhysRevLett.110.265506} {\bibfield
  {journal} {\bibinfo  {journal} {Physical Review Letters}\ }\textbf {\bibinfo
  {volume} {110}},\ \bibinfo {pages} {265506} (\bibinfo {year}
  {2013})}\BibitemShut {NoStop}%
\bibitem [{\citenamefont {Li}\ \emph {et~al.}(2014)\citenamefont {Li},
  \citenamefont {Carrete}, \citenamefont {A.~Katcho},\ and\ \citenamefont
  {Mingo}}]{liShengBTESolverBoltzmann2014a}%
  \BibitemOpen
  \bibfield  {author} {\bibinfo {author} {\bibfnamefont {W.}~\bibnamefont
  {Li}}, \bibinfo {author} {\bibfnamefont {J.}~\bibnamefont {Carrete}},
  \bibinfo {author} {\bibfnamefont {N.}~\bibnamefont {A.~Katcho}}, \ and\
  \bibinfo {author} {\bibfnamefont {N.}~\bibnamefont {Mingo}},\ }\href
  {\doibase 10.1016/j.cpc.2014.02.015} {\bibfield  {journal} {\bibinfo
  {journal} {Computer Physics Communications}\ }\textbf {\bibinfo {volume}
  {185}},\ \bibinfo {pages} {1747} (\bibinfo {year} {2014})}\BibitemShut
  {NoStop}%
\bibitem [{\citenamefont {Tadano}\ \emph {et~al.}(2014)\citenamefont {Tadano},
  \citenamefont {Gohda},\ and\ \citenamefont
  {Tsuneyuki}}]{tadanoAnharmonicForceConstants2014}%
  \BibitemOpen
  \bibfield  {author} {\bibinfo {author} {\bibfnamefont {T.}~\bibnamefont
  {Tadano}}, \bibinfo {author} {\bibfnamefont {Y.}~\bibnamefont {Gohda}}, \
  and\ \bibinfo {author} {\bibfnamefont {S.}~\bibnamefont {Tsuneyuki}},\ }\href
  {\doibase 10.1088/0953-8984/26/22/225402} {\bibfield  {journal} {\bibinfo
  {journal} {Journal of Physics: Condensed Matter}\ }\textbf {\bibinfo {volume}
  {26}},\ \bibinfo {pages} {225402} (\bibinfo {year} {2014})}\BibitemShut
  {NoStop}%
\bibitem [{\citenamefont {Togo}\ \emph {et~al.}(2015)\citenamefont {Togo},
  \citenamefont {Chaput},\ and\ \citenamefont
  {Tanaka}}]{togoDistributionsPhononLifetimes2015a}%
  \BibitemOpen
  \bibfield  {author} {\bibinfo {author} {\bibfnamefont {A.}~\bibnamefont
  {Togo}}, \bibinfo {author} {\bibfnamefont {L.}~\bibnamefont {Chaput}}, \ and\
  \bibinfo {author} {\bibfnamefont {I.}~\bibnamefont {Tanaka}},\ }\href
  {\doibase 10.1103/PhysRevB.91.094306} {\bibfield  {journal} {\bibinfo
  {journal} {Physical Review B}\ }\textbf {\bibinfo {volume} {91}},\ \bibinfo
  {pages} {094306} (\bibinfo {year} {2015})}\BibitemShut {NoStop}%
\bibitem [{\citenamefont {Chernatynskiy}\ and\ \citenamefont
  {Phillpot}(2015)}]{chernatynskiyPhononTransportSimulator2015}%
  \BibitemOpen
  \bibfield  {author} {\bibinfo {author} {\bibfnamefont {A.}~\bibnamefont
  {Chernatynskiy}}\ and\ \bibinfo {author} {\bibfnamefont {S.~R.}\ \bibnamefont
  {Phillpot}},\ }\href {\doibase 10.1016/j.cpc.2015.01.008} {\bibfield
  {journal} {\bibinfo  {journal} {Computer Physics Communications}\ }\textbf
  {\bibinfo {volume} {192}},\ \bibinfo {pages} {196} (\bibinfo {year}
  {2015})}\BibitemShut {NoStop}%
\bibitem [{\citenamefont {Feng}\ and\ \citenamefont
  {Ruan}(2016)}]{fengQuantumMechanicalPrediction2016}%
  \BibitemOpen
  \bibfield  {author} {\bibinfo {author} {\bibfnamefont {T.}~\bibnamefont
  {Feng}}\ and\ \bibinfo {author} {\bibfnamefont {X.}~\bibnamefont {Ruan}},\
  }\href {\doibase 10.1103/PhysRevB.93.045202} {\bibfield  {journal} {\bibinfo
  {journal} {Physical Review B}\ }\textbf {\bibinfo {volume} {93}},\ \bibinfo
  {pages} {045202} (\bibinfo {year} {2016})}\BibitemShut {NoStop}%
\bibitem [{\citenamefont {Feng}\ \emph {et~al.}(2017)\citenamefont {Feng},
  \citenamefont {Lindsay},\ and\ \citenamefont
  {Ruan}}]{fengFourphononScatteringSignificantly2017}%
  \BibitemOpen
  \bibfield  {author} {\bibinfo {author} {\bibfnamefont {T.}~\bibnamefont
  {Feng}}, \bibinfo {author} {\bibfnamefont {L.}~\bibnamefont {Lindsay}}, \
  and\ \bibinfo {author} {\bibfnamefont {X.}~\bibnamefont {Ruan}},\ }\href
  {\doibase 10.1103/PhysRevB.96.161201} {\bibfield  {journal} {\bibinfo
  {journal} {Physical Review B}\ }\textbf {\bibinfo {volume} {96}},\ \bibinfo
  {pages} {161201} (\bibinfo {year} {2017})}\BibitemShut {NoStop}%
\bibitem [{\citenamefont {Simoncelli}\ \emph {et~al.}(2019)\citenamefont
  {Simoncelli}, \citenamefont {Marzari},\ and\ \citenamefont
  {Mauri}}]{simoncelliUnifiedTheoryThermal2019}%
  \BibitemOpen
  \bibfield  {author} {\bibinfo {author} {\bibfnamefont {M.}~\bibnamefont
  {Simoncelli}}, \bibinfo {author} {\bibfnamefont {N.}~\bibnamefont {Marzari}},
  \ and\ \bibinfo {author} {\bibfnamefont {F.}~\bibnamefont {Mauri}},\ }\href
  {\doibase 10.1038/s41567-019-0520-x} {\bibfield  {journal} {\bibinfo
  {journal} {Nature Physics}\ }\textbf {\bibinfo {volume} {15}},\ \bibinfo
  {pages} {809} (\bibinfo {year} {2019})}\BibitemShut {NoStop}%
\bibitem [{\citenamefont {Simoncelli}\ \emph {et~al.}(2022)\citenamefont
  {Simoncelli}, \citenamefont {Marzari},\ and\ \citenamefont
  {Mauri}}]{simoncelliWignerFormulationThermal2022}%
  \BibitemOpen
  \bibfield  {author} {\bibinfo {author} {\bibfnamefont {M.}~\bibnamefont
  {Simoncelli}}, \bibinfo {author} {\bibfnamefont {N.}~\bibnamefont {Marzari}},
  \ and\ \bibinfo {author} {\bibfnamefont {F.}~\bibnamefont {Mauri}},\ }\href
  {\doibase 10.1103/PhysRevX.12.041011} {\bibfield  {journal} {\bibinfo
  {journal} {Physical Review X}\ }\textbf {\bibinfo {volume} {12}},\ \bibinfo
  {pages} {041011} (\bibinfo {year} {2022})}\BibitemShut {NoStop}%
\bibitem [{\citenamefont {Xiao}\ and\ \citenamefont
  {Marianetti}(2023)}]{xiaoAnharmonicPhononBehavior2023}%
  \BibitemOpen
  \bibfield  {author} {\bibinfo {author} {\bibfnamefont {E.}~\bibnamefont
  {Xiao}}\ and\ \bibinfo {author} {\bibfnamefont {C.~A.}\ \bibnamefont
  {Marianetti}},\ }\href {\doibase 10.1103/PhysRevB.107.094303} {\bibfield
  {journal} {\bibinfo  {journal} {Physical Review B}\ }\textbf {\bibinfo
  {volume} {107}},\ \bibinfo {pages} {094303} (\bibinfo {year}
  {2023})}\BibitemShut {NoStop}%
\bibitem [{\citenamefont {Anisimov}\ \emph {et~al.}(1997)\citenamefont
  {Anisimov}, \citenamefont {Aryasetiawan},\ and\ \citenamefont
  {Lichtenstein}}]{anisimovFirstprinciplesCalculationsElectronic1997}%
  \BibitemOpen
  \bibfield  {author} {\bibinfo {author} {\bibfnamefont {V.~I.}\ \bibnamefont
  {Anisimov}}, \bibinfo {author} {\bibfnamefont {F.}~\bibnamefont
  {Aryasetiawan}}, \ and\ \bibinfo {author} {\bibfnamefont {A.~I.}\
  \bibnamefont {Lichtenstein}},\ }\href {\doibase 10.1088/0953-8984/9/4/002}
  {\bibfield  {journal} {\bibinfo  {journal} {Journal of Physics: Condensed
  Matter}\ }\textbf {\bibinfo {volume} {9}},\ \bibinfo {pages} {767} (\bibinfo
  {year} {1997})}\BibitemShut {NoStop}%
\bibitem [{\citenamefont {Hooton}(1955)}]{hootonLINewTreatment1955}%
  \BibitemOpen
  \bibfield  {author} {\bibinfo {author} {\bibfnamefont {D.}~\bibnamefont
  {Hooton}},\ }\href {\doibase 10.1080/14786440408520575} {\bibfield  {journal}
  {\bibinfo  {journal} {The London, Edinburgh, and Dublin Philosophical
  Magazine and Journal of Science}\ }\textbf {\bibinfo {volume} {46}},\
  \bibinfo {pages} {422} (\bibinfo {year} {1955})}\BibitemShut {NoStop}%
\bibitem [{\citenamefont {Xiao}\ \emph {et~al.}(2022)\citenamefont {Xiao},
  \citenamefont {Ma}, \citenamefont {Bryan}, \citenamefont {Fu}, \citenamefont
  {Mann}, \citenamefont {Winn}, \citenamefont {Abernathy}, \citenamefont
  {Hermann}, \citenamefont {Khanolkar}, \citenamefont {Dennett}, \citenamefont
  {Hurley}, \citenamefont {Manley},\ and\ \citenamefont
  {Marianetti}}]{xiaoValidatingFirstprinciplesPhonon2022}%
  \BibitemOpen
  \bibfield  {author} {\bibinfo {author} {\bibfnamefont {E.}~\bibnamefont
  {Xiao}}, \bibinfo {author} {\bibfnamefont {H.}~\bibnamefont {Ma}}, \bibinfo
  {author} {\bibfnamefont {M.~S.}\ \bibnamefont {Bryan}}, \bibinfo {author}
  {\bibfnamefont {L.}~\bibnamefont {Fu}}, \bibinfo {author} {\bibfnamefont
  {J.~M.}\ \bibnamefont {Mann}}, \bibinfo {author} {\bibfnamefont
  {B.}~\bibnamefont {Winn}}, \bibinfo {author} {\bibfnamefont {D.~L.}\
  \bibnamefont {Abernathy}}, \bibinfo {author} {\bibfnamefont {R.~P.}\
  \bibnamefont {Hermann}}, \bibinfo {author} {\bibfnamefont {A.~R.}\
  \bibnamefont {Khanolkar}}, \bibinfo {author} {\bibfnamefont {C.~A.}\
  \bibnamefont {Dennett}}, \bibinfo {author} {\bibfnamefont {D.~H.}\
  \bibnamefont {Hurley}}, \bibinfo {author} {\bibfnamefont {M.~E.}\
  \bibnamefont {Manley}}, \ and\ \bibinfo {author} {\bibfnamefont {C.~A.}\
  \bibnamefont {Marianetti}},\ }\href {\doibase 10.1103/PhysRevB.106.144310}
  {\bibfield  {journal} {\bibinfo  {journal} {Physical Review B}\ }\textbf
  {\bibinfo {volume} {106}},\ \bibinfo {pages} {144310} (\bibinfo {year}
  {2022})}\BibitemShut {NoStop}%
\bibitem [{\citenamefont {Qi}\ \emph {et~al.}(2016)\citenamefont {Qi},
  \citenamefont {Zhang}, \citenamefont {Cheng}, \citenamefont {Chen},
  \citenamefont {Wei},\ and\ \citenamefont
  {Cai}}]{qiLatticeDynamicsThermal2016}%
  \BibitemOpen
  \bibfield  {author} {\bibinfo {author} {\bibfnamefont {Y.-Y.}\ \bibnamefont
  {Qi}}, \bibinfo {author} {\bibfnamefont {T.}~\bibnamefont {Zhang}}, \bibinfo
  {author} {\bibfnamefont {Y.}~\bibnamefont {Cheng}}, \bibinfo {author}
  {\bibfnamefont {X.-R.}\ \bibnamefont {Chen}}, \bibinfo {author}
  {\bibfnamefont {D.-Q.}\ \bibnamefont {Wei}}, \ and\ \bibinfo {author}
  {\bibfnamefont {L.-C.}\ \bibnamefont {Cai}},\ }\href {\doibase
  10.1063/1.4942841} {\bibfield  {journal} {\bibinfo  {journal} {Journal of
  Applied Physics}\ }\textbf {\bibinfo {volume} {119}},\ \bibinfo {pages}
  {095103} (\bibinfo {year} {2016})}\BibitemShut {NoStop}%
\bibitem [{\citenamefont {Momin}\ \emph {et~al.}(1991)\citenamefont {Momin},
  \citenamefont {Mirza},\ and\ \citenamefont
  {Mathews}}]{mominHighTemperatureXray1991}%
  \BibitemOpen
  \bibfield  {author} {\bibinfo {author} {\bibfnamefont {A.~C.}\ \bibnamefont
  {Momin}}, \bibinfo {author} {\bibfnamefont {E.~B.}\ \bibnamefont {Mirza}}, \
  and\ \bibinfo {author} {\bibfnamefont {M.~D.}\ \bibnamefont {Mathews}},\
  }\href {\doibase 10.1016/0022-3115(91)90521-8} {\bibfield  {journal}
  {\bibinfo  {journal} {Journal of Nuclear Materials}\ }\textbf {\bibinfo
  {volume} {185}},\ \bibinfo {pages} {308} (\bibinfo {year}
  {1991})}\BibitemShut {NoStop}%
\bibitem [{\citenamefont {Abernathy}\ \emph {et~al.}(2012)\citenamefont
  {Abernathy}, \citenamefont {Stone}, \citenamefont {Loguillo}, \citenamefont
  {Lucas}, \citenamefont {Delaire}, \citenamefont {Tang}, \citenamefont {Lin},\
  and\ \citenamefont {Fultz}}]{abernathy2012design}%
  \BibitemOpen
  \bibfield  {author} {\bibinfo {author} {\bibfnamefont {D.~L.}\ \bibnamefont
  {Abernathy}}, \bibinfo {author} {\bibfnamefont {M.~B.}\ \bibnamefont
  {Stone}}, \bibinfo {author} {\bibfnamefont {M.}~\bibnamefont {Loguillo}},
  \bibinfo {author} {\bibfnamefont {M.}~\bibnamefont {Lucas}}, \bibinfo
  {author} {\bibfnamefont {O.}~\bibnamefont {Delaire}}, \bibinfo {author}
  {\bibfnamefont {X.}~\bibnamefont {Tang}}, \bibinfo {author} {\bibfnamefont
  {J.}~\bibnamefont {Lin}}, \ and\ \bibinfo {author} {\bibfnamefont
  {B.}~\bibnamefont {Fultz}},\ }\href@noop {} {\bibfield  {journal} {\bibinfo
  {journal} {Review of Scientific Instruments}\ }\textbf {\bibinfo {volume}
  {83}},\ \bibinfo {pages} {015114} (\bibinfo {year} {2012})}\BibitemShut
  {NoStop}%
\bibitem [{\citenamefont {Blöchl}(1994)}]{blochl_projector_1994}%
  \BibitemOpen
  \bibfield  {author} {\bibinfo {author} {\bibfnamefont {P.~E.}\ \bibnamefont
  {Blöchl}},\ }\href {\doibase 10.1103/PhysRevB.50.17953} {\bibfield
  {journal} {\bibinfo  {journal} {Physical Review B}\ }\textbf {\bibinfo
  {volume} {50}},\ \bibinfo {pages} {17953} (\bibinfo {year}
  {1994})}\BibitemShut {NoStop}%
\bibitem [{\citenamefont {Kresse}\ and\ \citenamefont
  {Joubert}(1999)}]{kresse_ultrasoft_1999}%
  \BibitemOpen
  \bibfield  {author} {\bibinfo {author} {\bibfnamefont {G.}~\bibnamefont
  {Kresse}}\ and\ \bibinfo {author} {\bibfnamefont {D.}~\bibnamefont
  {Joubert}},\ }\href {\doibase 10.1103/PhysRevB.59.1758} {\bibfield  {journal}
  {\bibinfo  {journal} {Physical Review B}\ }\textbf {\bibinfo {volume} {59}},\
  \bibinfo {pages} {1758} (\bibinfo {year} {1999})}\BibitemShut {NoStop}%
\bibitem [{\citenamefont {Kresse}\ and\ \citenamefont
  {Hafner}(1993)}]{kresse_ab_1993}%
  \BibitemOpen
  \bibfield  {author} {\bibinfo {author} {\bibfnamefont {G.}~\bibnamefont
  {Kresse}}\ and\ \bibinfo {author} {\bibfnamefont {J.}~\bibnamefont
  {Hafner}},\ }\href {http://link.aps.org/doi/10.1103/PhysRevB.47.558}
  {\bibfield  {journal} {\bibinfo  {journal} {Physical Review B}\ }\textbf
  {\bibinfo {volume} {47}},\ \bibinfo {pages} {558} (\bibinfo {year}
  {1993})}\BibitemShut {NoStop}%
\bibitem [{\citenamefont {Kresse}\ and\ \citenamefont
  {Furthmüller}(1996)}]{kresse_efficient_1996}%
  \BibitemOpen
  \bibfield  {author} {\bibinfo {author} {\bibfnamefont {G.}~\bibnamefont
  {Kresse}}\ and\ \bibinfo {author} {\bibfnamefont {J.}~\bibnamefont
  {Furthmüller}},\ }\href {\doibase 10.1103/PhysRevB.54.11169} {\bibfield
  {journal} {\bibinfo  {journal} {Physical Review B}\ }\textbf {\bibinfo
  {volume} {54}},\ \bibinfo {pages} {11169} (\bibinfo {year}
  {1996})}\BibitemShut {NoStop}%
\bibitem [{\citenamefont {Perdew}\ \emph {et~al.}(1996)\citenamefont {Perdew},
  \citenamefont {Burke},\ and\ \citenamefont
  {Ernzerhof}}]{perdew_generalized_1996}%
  \BibitemOpen
  \bibfield  {author} {\bibinfo {author} {\bibfnamefont {J.~P.}\ \bibnamefont
  {Perdew}}, \bibinfo {author} {\bibfnamefont {K.}~\bibnamefont {Burke}}, \
  and\ \bibinfo {author} {\bibfnamefont {M.}~\bibnamefont {Ernzerhof}},\ }\href
  {\doibase 10.1103/PhysRevLett.77.3865} {\bibfield  {journal} {\bibinfo
  {journal} {Physical Review Letters}\ }\textbf {\bibinfo {volume} {77}},\
  \bibinfo {pages} {3865} (\bibinfo {year} {1996})}\BibitemShut {NoStop}%
\bibitem [{\citenamefont {Dudarev}\ \emph {et~al.}(1998)\citenamefont
  {Dudarev}, \citenamefont {Botton}, \citenamefont {Savrasov}, \citenamefont
  {Humphreys},\ and\ \citenamefont
  {Sutton}}]{dudarev_electron-energy-loss_1998}%
  \BibitemOpen
  \bibfield  {author} {\bibinfo {author} {\bibfnamefont {S.~L.}\ \bibnamefont
  {Dudarev}}, \bibinfo {author} {\bibfnamefont {G.~A.}\ \bibnamefont {Botton}},
  \bibinfo {author} {\bibfnamefont {S.~Y.}\ \bibnamefont {Savrasov}}, \bibinfo
  {author} {\bibfnamefont {C.~J.}\ \bibnamefont {Humphreys}}, \ and\ \bibinfo
  {author} {\bibfnamefont {A.~P.}\ \bibnamefont {Sutton}},\ }\href {\doibase
  10.1103/PhysRevB.57.1505} {\bibfield  {journal} {\bibinfo  {journal}
  {Physical Review B}\ }\textbf {\bibinfo {volume} {57}},\ \bibinfo {pages}
  {1505} (\bibinfo {year} {1998})}\BibitemShut {NoStop}%
\bibitem [{\citenamefont {Fu}\ \emph {et~al.}(2019)\citenamefont {Fu},
  \citenamefont {Kornbluth}, \citenamefont {Cheng},\ and\ \citenamefont
  {Marianetti}}]{fu_group_2019}%
  \BibitemOpen
  \bibfield  {author} {\bibinfo {author} {\bibfnamefont {L.}~\bibnamefont
  {Fu}}, \bibinfo {author} {\bibfnamefont {M.}~\bibnamefont {Kornbluth}},
  \bibinfo {author} {\bibfnamefont {Z.}~\bibnamefont {Cheng}}, \ and\ \bibinfo
  {author} {\bibfnamefont {C.~A.}\ \bibnamefont {Marianetti}},\ }\href
  {\doibase 10.1103/PhysRevB.100.014303} {\bibfield  {journal} {\bibinfo
  {journal} {Physical Review B}\ }\textbf {\bibinfo {volume} {100}},\ \bibinfo
  {pages} {014303} (\bibinfo {year} {2019})}\BibitemShut {NoStop}%
\bibitem [{\citenamefont {Lin}\ \emph {et~al.}(2019)\citenamefont {Lin},
  \citenamefont {Banerjee}, \citenamefont {Islam}, \citenamefont {Le},\ and\
  \citenamefont {Abernathy}}]{lin2019energy}%
  \BibitemOpen
  \bibfield  {author} {\bibinfo {author} {\bibfnamefont {J.~Y.}\ \bibnamefont
  {Lin}}, \bibinfo {author} {\bibfnamefont {A.}~\bibnamefont {Banerjee}},
  \bibinfo {author} {\bibfnamefont {F.}~\bibnamefont {Islam}}, \bibinfo
  {author} {\bibfnamefont {M.~D.}\ \bibnamefont {Le}}, \ and\ \bibinfo {author}
  {\bibfnamefont {D.~L.}\ \bibnamefont {Abernathy}},\ }\href@noop {} {\bibfield
   {journal} {\bibinfo  {journal} {Physica B: Condensed Matter}\ }\textbf
  {\bibinfo {volume} {562}},\ \bibinfo {pages} {26} (\bibinfo {year}
  {2019})}\BibitemShut {NoStop}%
\bibitem [{\citenamefont {Bryan}\ \emph {et~al.}(2020)\citenamefont {Bryan},
  \citenamefont {Fu}, \citenamefont {Rickert}, \citenamefont {Turner},
  \citenamefont {Prusnick}, \citenamefont {Mann}, \citenamefont {Abernathy},
  \citenamefont {Marianetti},\ and\ \citenamefont
  {Manley}}]{bryan2020nonlinear}%
  \BibitemOpen
  \bibfield  {author} {\bibinfo {author} {\bibfnamefont {M.~S.}\ \bibnamefont
  {Bryan}}, \bibinfo {author} {\bibfnamefont {L.}~\bibnamefont {Fu}}, \bibinfo
  {author} {\bibfnamefont {K.}~\bibnamefont {Rickert}}, \bibinfo {author}
  {\bibfnamefont {D.}~\bibnamefont {Turner}}, \bibinfo {author} {\bibfnamefont
  {T.~A.}\ \bibnamefont {Prusnick}}, \bibinfo {author} {\bibfnamefont {J.~M.}\
  \bibnamefont {Mann}}, \bibinfo {author} {\bibfnamefont {D.~L.}\ \bibnamefont
  {Abernathy}}, \bibinfo {author} {\bibfnamefont {C.~A.}\ \bibnamefont
  {Marianetti}}, \ and\ \bibinfo {author} {\bibfnamefont {M.~E.}\ \bibnamefont
  {Manley}},\ }\href@noop {} {\bibfield  {journal} {\bibinfo  {journal}
  {Communications Physics}\ }\textbf {\bibinfo {volume} {3}},\ \bibinfo {pages}
  {1} (\bibinfo {year} {2020})}\BibitemShut {NoStop}%
\bibitem [{\citenamefont {Pang}\ \emph {et~al.}(2013)\citenamefont {Pang},
  \citenamefont {Buyers}, \citenamefont {Chernatynskiy}, \citenamefont
  {Lumsden}, \citenamefont {Larson},\ and\ \citenamefont
  {Phillpot}}]{pang_phonon_2013}%
  \BibitemOpen
  \bibfield  {author} {\bibinfo {author} {\bibfnamefont {J.~W.~L.}\
  \bibnamefont {Pang}}, \bibinfo {author} {\bibfnamefont {W.~J.~L.}\
  \bibnamefont {Buyers}}, \bibinfo {author} {\bibfnamefont {A.}~\bibnamefont
  {Chernatynskiy}}, \bibinfo {author} {\bibfnamefont {M.~D.}\ \bibnamefont
  {Lumsden}}, \bibinfo {author} {\bibfnamefont {B.~C.}\ \bibnamefont {Larson}},
  \ and\ \bibinfo {author} {\bibfnamefont {S.~R.}\ \bibnamefont {Phillpot}},\
  }\href {\doibase 10.1103/PhysRevLett.110.157401} {\bibfield  {journal}
  {\bibinfo  {journal} {Physical Review Letters}\ }\textbf {\bibinfo {volume}
  {110}},\ \bibinfo {pages} {157401} (\bibinfo {year} {2013})}\BibitemShut
  {NoStop}%
\bibitem [{\citenamefont {Wang}\ \emph {et~al.}(2013)\citenamefont {Wang},
  \citenamefont {Zhang}, \citenamefont {Lizárraga}, \citenamefont {Di~Marco},\
  and\ \citenamefont {Eriksson}}]{wang_phonon_2013}%
  \BibitemOpen
  \bibfield  {author} {\bibinfo {author} {\bibfnamefont {B.-T.}\ \bibnamefont
  {Wang}}, \bibinfo {author} {\bibfnamefont {P.}~\bibnamefont {Zhang}},
  \bibinfo {author} {\bibfnamefont {R.}~\bibnamefont {Lizárraga}}, \bibinfo
  {author} {\bibfnamefont {I.}~\bibnamefont {Di~Marco}}, \ and\ \bibinfo
  {author} {\bibfnamefont {O.}~\bibnamefont {Eriksson}},\ }\href {\doibase
  10.1103/PhysRevB.88.104107} {\bibfield  {journal} {\bibinfo  {journal}
  {Physical Review B}\ }\textbf {\bibinfo {volume} {88}},\ \bibinfo {pages}
  {104107} (\bibinfo {year} {2013})}\BibitemShut {NoStop}%
\bibitem [{\citenamefont {Gonze}\ and\ \citenamefont
  {Lee}(1997)}]{gonzeDynamicalMatricesBorn1997a}%
  \BibitemOpen
  \bibfield  {author} {\bibinfo {author} {\bibfnamefont {X.}~\bibnamefont
  {Gonze}}\ and\ \bibinfo {author} {\bibfnamefont {C.}~\bibnamefont {Lee}},\
  }\href {\doibase 10.1103/PhysRevB.55.10355} {\bibfield  {journal} {\bibinfo
  {journal} {Physical Review B}\ }\textbf {\bibinfo {volume} {55}},\ \bibinfo
  {pages} {10355} (\bibinfo {year} {1997})}\BibitemShut {NoStop}%
\bibitem [{\citenamefont {Mathis}\ \emph {et~al.}(2022)\citenamefont {Mathis},
  \citenamefont {Khanolkar}, \citenamefont {Fu}, \citenamefont {Bryan},
  \citenamefont {Dennett}, \citenamefont {Rickert}, \citenamefont {Mann},
  \citenamefont {Winn}, \citenamefont {Abernathy}, \citenamefont {Manley},
  \citenamefont {Hurley},\ and\ \citenamefont
  {Marianetti}}]{mathisGeneralizedQuasiharmonicApproximation2022}%
  \BibitemOpen
  \bibfield  {author} {\bibinfo {author} {\bibfnamefont {M.~A.}\ \bibnamefont
  {Mathis}}, \bibinfo {author} {\bibfnamefont {A.}~\bibnamefont {Khanolkar}},
  \bibinfo {author} {\bibfnamefont {L.}~\bibnamefont {Fu}}, \bibinfo {author}
  {\bibfnamefont {M.~S.}\ \bibnamefont {Bryan}}, \bibinfo {author}
  {\bibfnamefont {C.~A.}\ \bibnamefont {Dennett}}, \bibinfo {author}
  {\bibfnamefont {K.}~\bibnamefont {Rickert}}, \bibinfo {author} {\bibfnamefont
  {J.~M.}\ \bibnamefont {Mann}}, \bibinfo {author} {\bibfnamefont
  {B.}~\bibnamefont {Winn}}, \bibinfo {author} {\bibfnamefont {D.~L.}\
  \bibnamefont {Abernathy}}, \bibinfo {author} {\bibfnamefont {M.~E.}\
  \bibnamefont {Manley}}, \bibinfo {author} {\bibfnamefont {D.~H.}\
  \bibnamefont {Hurley}}, \ and\ \bibinfo {author} {\bibfnamefont {C.~A.}\
  \bibnamefont {Marianetti}},\ }\href {\doibase 10.1103/PhysRevB.106.014314}
  {\bibfield  {journal} {\bibinfo  {journal} {Physical Review B}\ }\textbf
  {\bibinfo {volume} {106}},\ \bibinfo {pages} {014314} (\bibinfo {year}
  {2022})}\BibitemShut {NoStop}%
\bibitem [{\citenamefont {Idiri}\ \emph {et~al.}(2004)\citenamefont {Idiri},
  \citenamefont {Le~Bihan}, \citenamefont {Heathman},\ and\ \citenamefont
  {Rebizant}}]{idiri_behavior_2004}%
  \BibitemOpen
  \bibfield  {author} {\bibinfo {author} {\bibfnamefont {M.}~\bibnamefont
  {Idiri}}, \bibinfo {author} {\bibfnamefont {T.}~\bibnamefont {Le~Bihan}},
  \bibinfo {author} {\bibfnamefont {S.}~\bibnamefont {Heathman}}, \ and\
  \bibinfo {author} {\bibfnamefont {J.}~\bibnamefont {Rebizant}},\ }\href
  {\doibase 10.1103/PhysRevB.70.014113} {\bibfield  {journal} {\bibinfo
  {journal} {Physical Review B}\ }\textbf {\bibinfo {volume} {70}},\ \bibinfo
  {pages} {014113} (\bibinfo {year} {2004})}\BibitemShut {NoStop}%
\bibitem [{\citenamefont {Santini}\ \emph {et~al.}(2009)\citenamefont
  {Santini}, \citenamefont {Carretta}, \citenamefont {Amoretti}, \citenamefont
  {Caciuffo}, \citenamefont {Magnani},\ and\ \citenamefont
  {Lander}}]{santini_multipolar_2009}%
  \BibitemOpen
  \bibfield  {author} {\bibinfo {author} {\bibfnamefont {P.}~\bibnamefont
  {Santini}}, \bibinfo {author} {\bibfnamefont {S.}~\bibnamefont {Carretta}},
  \bibinfo {author} {\bibfnamefont {G.}~\bibnamefont {Amoretti}}, \bibinfo
  {author} {\bibfnamefont {R.}~\bibnamefont {Caciuffo}}, \bibinfo {author}
  {\bibfnamefont {N.}~\bibnamefont {Magnani}}, \ and\ \bibinfo {author}
  {\bibfnamefont {G.~H.}\ \bibnamefont {Lander}},\ }\href {\doibase
  10.1103/RevModPhys.81.807} {\bibfield  {journal} {\bibinfo  {journal}
  {Reviews of Modern Physics}\ }\textbf {\bibinfo {volume} {81}},\ \bibinfo
  {pages} {807} (\bibinfo {year} {2009})}\BibitemShut {NoStop}%
\bibitem [{\citenamefont {Bryan}\ \emph {et~al.}(2019)\citenamefont {Bryan},
  \citenamefont {Pang}, \citenamefont {Larson}, \citenamefont {Chernatynskiy},
  \citenamefont {Abernathy}, \citenamefont {Gofryk},\ and\ \citenamefont
  {Manley}}]{bryan_impact_2019}%
  \BibitemOpen
  \bibfield  {author} {\bibinfo {author} {\bibfnamefont {M.~S.}\ \bibnamefont
  {Bryan}}, \bibinfo {author} {\bibfnamefont {J.~W.~L.}\ \bibnamefont {Pang}},
  \bibinfo {author} {\bibfnamefont {B.~C.}\ \bibnamefont {Larson}}, \bibinfo
  {author} {\bibfnamefont {A.}~\bibnamefont {Chernatynskiy}}, \bibinfo {author}
  {\bibfnamefont {D.~L.}\ \bibnamefont {Abernathy}}, \bibinfo {author}
  {\bibfnamefont {K.}~\bibnamefont {Gofryk}}, \ and\ \bibinfo {author}
  {\bibfnamefont {M.~E.}\ \bibnamefont {Manley}},\ }\href {\doibase
  10.1103/PhysRevMaterials.3.065405} {\bibfield  {journal} {\bibinfo  {journal}
  {Physical Review Materials}\ }\textbf {\bibinfo {volume} {3}},\ \bibinfo
  {pages} {065405} (\bibinfo {year} {2019})}\BibitemShut {NoStop}%
\bibitem [{\citenamefont {Agency}(1965)}]{international1965thermodynamic}%
  \BibitemOpen
  \bibfield  {author} {\bibinfo {author} {\bibfnamefont {I.~A.~E.}\
  \bibnamefont {Agency}},\ }\href@noop {} {\emph {\bibinfo {title}
  {Thermodynamic and transport properties of uranium dioxide and related
  phases}}}\ (\bibinfo {year} {1965})\BibitemShut {NoStop}%
\bibitem [{\citenamefont {Ronchi}\ \emph {et~al.}(1999)\citenamefont {Ronchi},
  \citenamefont {Sheindlin}, \citenamefont {Musella},\ and\ \citenamefont
  {Hyland}}]{ronchiThermalConductivityUranium1999a}%
  \BibitemOpen
  \bibfield  {author} {\bibinfo {author} {\bibfnamefont {C.}~\bibnamefont
  {Ronchi}}, \bibinfo {author} {\bibfnamefont {M.}~\bibnamefont {Sheindlin}},
  \bibinfo {author} {\bibfnamefont {M.}~\bibnamefont {Musella}}, \ and\
  \bibinfo {author} {\bibfnamefont {G.~J.}\ \bibnamefont {Hyland}},\ }\href
  {\doibase 10.1063/1.369159} {\bibfield  {journal} {\bibinfo  {journal}
  {Journal of Applied Physics}\ }\textbf {\bibinfo {volume} {85}},\ \bibinfo
  {pages} {776} (\bibinfo {year} {1999})}\BibitemShut {NoStop}%
\bibitem [{\citenamefont {Kim}\ \emph {et~al.}(2014)\citenamefont {Kim},
  \citenamefont {Kim},\ and\ \citenamefont
  {Kaviany}}]{kimLatticeThermalConductivity2014}%
  \BibitemOpen
  \bibfield  {author} {\bibinfo {author} {\bibfnamefont {H.}~\bibnamefont
  {Kim}}, \bibinfo {author} {\bibfnamefont {M.~H.}\ \bibnamefont {Kim}}, \ and\
  \bibinfo {author} {\bibfnamefont {M.}~\bibnamefont {Kaviany}},\ }\href
  {\doibase 10.1063/1.4869669} {\bibfield  {journal} {\bibinfo  {journal}
  {Journal of Applied Physics}\ }\textbf {\bibinfo {volume} {115}},\ \bibinfo
  {pages} {123510} (\bibinfo {year} {2014})}\BibitemShut {NoStop}%
\bibitem [{\citenamefont {Pavlov}\ \emph {et~al.}(2017)\citenamefont {Pavlov},
  \citenamefont {Wenman}, \citenamefont {Vlahovic}, \citenamefont {Robba},
  \citenamefont {Konings}, \citenamefont {Van~Uffelen},\ and\ \citenamefont
  {Grimes}}]{pavlovMeasurementInterpretationThermophysical2017}%
  \BibitemOpen
  \bibfield  {author} {\bibinfo {author} {\bibfnamefont {T.~R.}\ \bibnamefont
  {Pavlov}}, \bibinfo {author} {\bibfnamefont {M.~R.}\ \bibnamefont {Wenman}},
  \bibinfo {author} {\bibfnamefont {L.}~\bibnamefont {Vlahovic}}, \bibinfo
  {author} {\bibfnamefont {D.}~\bibnamefont {Robba}}, \bibinfo {author}
  {\bibfnamefont {R.~J.~M.}\ \bibnamefont {Konings}}, \bibinfo {author}
  {\bibfnamefont {P.}~\bibnamefont {Van~Uffelen}}, \ and\ \bibinfo {author}
  {\bibfnamefont {R.~W.}\ \bibnamefont {Grimes}},\ }\href {\doibase
  10.1016/j.actamat.2017.07.060} {\bibfield  {journal} {\bibinfo  {journal}
  {Acta Materialia}\ }\textbf {\bibinfo {volume} {139}},\ \bibinfo {pages}
  {138} (\bibinfo {year} {2017})}\BibitemShut {NoStop}%
\end{thebibliography}%


\begin{thebibliography}{24}%
\makeatletter
\providecommand \@ifxundefined [1]{%
 \@ifx{#1\undefined}
}%
\providecommand \@ifnum [1]{%
 \ifnum #1\expandafter \@firstoftwo
 \else \expandafter \@secondoftwo
 \fi
}%
\providecommand \@ifx [1]{%
 \ifx #1\expandafter \@firstoftwo
 \else \expandafter \@secondoftwo
 \fi
}%
\providecommand \natexlab [1]{#1}%
\providecommand \enquote  [1]{``#1''}%
\providecommand \bibnamefont  [1]{#1}%
\providecommand \bibfnamefont [1]{#1}%
\providecommand \citenamefont [1]{#1}%
\providecommand \href@noop [0]{\@secondoftwo}%
\providecommand \href [0]{\begingroup \@sanitize@url \@href}%
\providecommand \@href[1]{\@@startlink{#1}\@@href}%
\providecommand \@@href[1]{\endgroup#1\@@endlink}%
\providecommand \@sanitize@url [0]{\catcode `\\12\catcode `\$12\catcode
  `\&12\catcode `\#12\catcode `\^12\catcode `\_12\catcode `\%12\relax}%
\providecommand \@@startlink[1]{}%
\providecommand \@@endlink[0]{}%
\providecommand \url  [0]{\begingroup\@sanitize@url \@url }%
\providecommand \@url [1]{\endgroup\@href {#1}{\urlprefix }}%
\providecommand \urlprefix  [0]{URL }%
\providecommand \Eprint [0]{\href }%
\providecommand \doibase [0]{http://dx.doi.org/}%
\providecommand \selectlanguage [0]{\@gobble}%
\providecommand \bibinfo  [0]{\@secondoftwo}%
\providecommand \bibfield  [0]{\@secondoftwo}%
\providecommand \translation [1]{[#1]}%
\providecommand \BibitemOpen [0]{}%
\providecommand \bibitemStop [0]{}%
\providecommand \bibitemNoStop [0]{.\EOS\space}%
\providecommand \EOS [0]{\spacefactor3000\relax}%
\providecommand \BibitemShut  [1]{\csname bibitem#1\endcsname}%
\let\auto@bib@innerbib\@empty
\bibitem [{\citenamefont {Mei}\ \emph {et~al.}(2014)\citenamefont {Mei},
  \citenamefont {Stan},\ and\ \citenamefont
  {Yang}}]{meiFirstprinciplesStudyThermophysical2014}%
  \BibitemOpen
  \bibfield  {author} {\bibinfo {author} {\bibfnamefont {Z.-G.}\ \bibnamefont
  {Mei}}, \bibinfo {author} {\bibfnamefont {M.}~\bibnamefont {Stan}}, \ and\
  \bibinfo {author} {\bibfnamefont {J.}~\bibnamefont {Yang}},\ }\href {\doibase
  10.1016/j.jallcom.2014.03.091} {\bibfield  {journal} {\bibinfo  {journal}
  {Journal of Alloys and Compounds}\ }\textbf {\bibinfo {volume} {603}},\
  \bibinfo {pages} {282} (\bibinfo {year} {2014})}\BibitemShut {NoStop}%
\bibitem [{\citenamefont {Wang}\ \emph {et~al.}(2015)\citenamefont {Wang},
  \citenamefont {Zheng}, \citenamefont {Qu}, \citenamefont {Li},\ and\
  \citenamefont {Zhang}}]{wangThermalConductivityUO22015}%
  \BibitemOpen
  \bibfield  {author} {\bibinfo {author} {\bibfnamefont {B.-T.}\ \bibnamefont
  {Wang}}, \bibinfo {author} {\bibfnamefont {J.-J.}\ \bibnamefont {Zheng}},
  \bibinfo {author} {\bibfnamefont {X.}~\bibnamefont {Qu}}, \bibinfo {author}
  {\bibfnamefont {W.-D.}\ \bibnamefont {Li}}, \ and\ \bibinfo {author}
  {\bibfnamefont {P.}~\bibnamefont {Zhang}},\ }\href {\doibase
  10.1016/j.jallcom.2014.12.204} {\bibfield  {journal} {\bibinfo  {journal}
  {Journal of Alloys and Compounds}\ }\textbf {\bibinfo {volume} {628}},\
  \bibinfo {pages} {267} (\bibinfo {year} {2015})}\BibitemShut {NoStop}%
\bibitem [{\citenamefont {Pang}\ \emph {et~al.}(2013)\citenamefont {Pang},
  \citenamefont {Buyers}, \citenamefont {Chernatynskiy}, \citenamefont
  {Lumsden}, \citenamefont {Larson},\ and\ \citenamefont
  {Phillpot}}]{pang_phonon_2013}%
  \BibitemOpen
  \bibfield  {author} {\bibinfo {author} {\bibfnamefont {J.~W.~L.}\
  \bibnamefont {Pang}}, \bibinfo {author} {\bibfnamefont {W.~J.~L.}\
  \bibnamefont {Buyers}}, \bibinfo {author} {\bibfnamefont {A.}~\bibnamefont
  {Chernatynskiy}}, \bibinfo {author} {\bibfnamefont {M.~D.}\ \bibnamefont
  {Lumsden}}, \bibinfo {author} {\bibfnamefont {B.~C.}\ \bibnamefont {Larson}},
  \ and\ \bibinfo {author} {\bibfnamefont {S.~R.}\ \bibnamefont {Phillpot}},\
  }\href {\doibase 10.1103/PhysRevLett.110.157401} {\bibfield  {journal}
  {\bibinfo  {journal} {Physical Review Letters}\ }\textbf {\bibinfo {volume}
  {110}},\ \bibinfo {pages} {157401} (\bibinfo {year} {2013})}\BibitemShut
  {NoStop}%
\bibitem [{\citenamefont {Kaur}\ \emph {et~al.}(2013)\citenamefont {Kaur},
  \citenamefont {Panigrahi},\ and\ \citenamefont
  {Valsakumar}}]{kaur_thermal_2013}%
  \BibitemOpen
  \bibfield  {author} {\bibinfo {author} {\bibfnamefont {G.}~\bibnamefont
  {Kaur}}, \bibinfo {author} {\bibfnamefont {P.}~\bibnamefont {Panigrahi}}, \
  and\ \bibinfo {author} {\bibfnamefont {M.~C.}\ \bibnamefont {Valsakumar}},\
  }\href {\doibase 10.1088/0965-0393/21/6/065014} {\bibfield  {journal}
  {\bibinfo  {journal} {Modelling and Simulation in Materials Science and
  Engineering}\ }\textbf {\bibinfo {volume} {21}},\ \bibinfo {pages} {065014}
  (\bibinfo {year} {2013})}\BibitemShut {NoStop}%
\bibitem [{\citenamefont {Yang}\ \emph {et~al.}(2022)\citenamefont {Yang},
  \citenamefont {Tiwari},\ and\ \citenamefont
  {Feng}}]{yangReducedAnharmonicPhonon2022}%
  \BibitemOpen
  \bibfield  {author} {\bibinfo {author} {\bibfnamefont {X.}~\bibnamefont
  {Yang}}, \bibinfo {author} {\bibfnamefont {J.}~\bibnamefont {Tiwari}}, \ and\
  \bibinfo {author} {\bibfnamefont {T.}~\bibnamefont {Feng}},\ }\href {\doibase
  10.1016/j.mtphys.2022.100689} {\bibfield  {journal} {\bibinfo  {journal}
  {Materials Today Physics}\ }\textbf {\bibinfo {volume} {24}},\ \bibinfo
  {pages} {100689} (\bibinfo {year} {2022})}\BibitemShut {NoStop}%
\bibitem [{\citenamefont {Torres}\ and\ \citenamefont
  {Kaloni}(2019)}]{torresThermalConductivityDiffusion2019}%
  \BibitemOpen
  \bibfield  {author} {\bibinfo {author} {\bibfnamefont {E.}~\bibnamefont
  {Torres}}\ and\ \bibinfo {author} {\bibfnamefont {T.~P.}\ \bibnamefont
  {Kaloni}},\ }\href {\doibase 10.1016/j.jnucmat.2019.04.040} {\bibfield
  {journal} {\bibinfo  {journal} {Journal of Nuclear Materials}\ }\textbf
  {\bibinfo {volume} {521}},\ \bibinfo {pages} {137} (\bibinfo {year}
  {2019})}\BibitemShut {NoStop}%
\bibitem [{\citenamefont {Torres}\ \emph {et~al.}(2020)\citenamefont {Torres},
  \citenamefont {CheikNjifon}, \citenamefont {Kaloni},\ and\ \citenamefont
  {Pencer}}]{torres_comparative_2020}%
  \BibitemOpen
  \bibfield  {author} {\bibinfo {author} {\bibfnamefont {E.}~\bibnamefont
  {Torres}}, \bibinfo {author} {\bibfnamefont {I.}~\bibnamefont {CheikNjifon}},
  \bibinfo {author} {\bibfnamefont {T.~P.}\ \bibnamefont {Kaloni}}, \ and\
  \bibinfo {author} {\bibfnamefont {J.}~\bibnamefont {Pencer}},\ }\href
  {\doibase 10.1016/j.commatsci.2020.109594} {\bibfield  {journal} {\bibinfo
  {journal} {Computational Materials Science}\ }\textbf {\bibinfo {volume}
  {177}},\ \bibinfo {pages} {109594} (\bibinfo {year} {2020})}\BibitemShut
  {NoStop}%
\bibitem [{\citenamefont
  {Fink}(2000)}]{finkThermophysicalPropertiesUranium2000}%
  \BibitemOpen
  \bibfield  {author} {\bibinfo {author} {\bibfnamefont {J.~K.}\ \bibnamefont
  {Fink}},\ }\href {\doibase 10.1016/S0022-3115(99)00273-1} {\bibfield
  {journal} {\bibinfo  {journal} {Journal of Nuclear Materials}\ }\textbf
  {\bibinfo {volume} {279}},\ \bibinfo {pages} {1} (\bibinfo {year}
  {2000})}\BibitemShut {NoStop}%
\bibitem [{\citenamefont {Bates}(1965)}]{batesVisibleInfraredAbsorption1965}%
  \BibitemOpen
  \bibfield  {author} {\bibinfo {author} {\bibfnamefont {J.~L.}\ \bibnamefont
  {Bates}},\ }\href {\doibase 10.13182/NSE65-A21011} {\bibfield  {journal}
  {\bibinfo  {journal} {Nuclear Science and Engineering}\ }\textbf {\bibinfo
  {volume} {21}},\ \bibinfo {pages} {26} (\bibinfo {year} {1965})}\BibitemShut
  {NoStop}%
\bibitem [{\citenamefont {Godfrey}\ \emph {et~al.}(1965)\citenamefont
  {Godfrey}, \citenamefont {Fulkerson}, \citenamefont {Kollie}, \citenamefont
  {Moore},\ and\ \citenamefont
  {McELROY}}]{godfreyThermalConductivityUranium1965}%
  \BibitemOpen
  \bibfield  {author} {\bibinfo {author} {\bibfnamefont {T.~G.}\ \bibnamefont
  {Godfrey}}, \bibinfo {author} {\bibfnamefont {W.}~\bibnamefont {Fulkerson}},
  \bibinfo {author} {\bibfnamefont {T.~G.}\ \bibnamefont {Kollie}}, \bibinfo
  {author} {\bibfnamefont {J.~P.}\ \bibnamefont {Moore}}, \ and\ \bibinfo
  {author} {\bibfnamefont {D.~L.}\ \bibnamefont {McELROY}},\ }\href {\doibase
  10.1111/j.1151-2916.1965.tb14745.x} {\bibfield  {journal} {\bibinfo
  {journal} {Journal of the American Ceramic Society}\ }\textbf {\bibinfo
  {volume} {48}},\ \bibinfo {pages} {297} (\bibinfo {year} {1965})}\BibitemShut
  {NoStop}%
\bibitem [{\citenamefont {Ronchi}\ \emph {et~al.}(2004)\citenamefont {Ronchi},
  \citenamefont {Sheindlin}, \citenamefont {Staicu},\ and\ \citenamefont
  {Kinoshita}}]{ronchiEffectBurnupThermal2004}%
  \BibitemOpen
  \bibfield  {author} {\bibinfo {author} {\bibfnamefont {C.}~\bibnamefont
  {Ronchi}}, \bibinfo {author} {\bibfnamefont {M.}~\bibnamefont {Sheindlin}},
  \bibinfo {author} {\bibfnamefont {D.}~\bibnamefont {Staicu}}, \ and\ \bibinfo
  {author} {\bibfnamefont {M.}~\bibnamefont {Kinoshita}},\ }\href {\doibase
  10.1016/j.jnucmat.2004.01.018} {\bibfield  {journal} {\bibinfo  {journal}
  {Journal of Nuclear Materials}\ }\textbf {\bibinfo {volume} {327}},\ \bibinfo
  {pages} {58} (\bibinfo {year} {2004})}\BibitemShut {NoStop}%
\bibitem [{\citenamefont {Wang}\ \emph {et~al.}(2013)\citenamefont {Wang},
  \citenamefont {Zhang}, \citenamefont {Lizárraga}, \citenamefont {Di~Marco},\
  and\ \citenamefont {Eriksson}}]{wang_phonon_2013}%
  \BibitemOpen
  \bibfield  {author} {\bibinfo {author} {\bibfnamefont {B.-T.}\ \bibnamefont
  {Wang}}, \bibinfo {author} {\bibfnamefont {P.}~\bibnamefont {Zhang}},
  \bibinfo {author} {\bibfnamefont {R.}~\bibnamefont {Lizárraga}}, \bibinfo
  {author} {\bibfnamefont {I.}~\bibnamefont {Di~Marco}}, \ and\ \bibinfo
  {author} {\bibfnamefont {O.}~\bibnamefont {Eriksson}},\ }\href {\doibase
  10.1103/PhysRevB.88.104107} {\bibfield  {journal} {\bibinfo  {journal}
  {Physical Review B}\ }\textbf {\bibinfo {volume} {88}},\ \bibinfo {pages}
  {104107} (\bibinfo {year} {2013})}\BibitemShut {NoStop}%
\bibitem [{\citenamefont {Zhou}\ \emph {et~al.}(2022)\citenamefont {Zhou},
  \citenamefont {Ma}, \citenamefont {Xiao}, \citenamefont {Gofryk},
  \citenamefont {Jiang}, \citenamefont {Manley}, \citenamefont {Hurley},\ and\
  \citenamefont {Marianetti}}]{zhouCapturingGroundState2022a}%
  \BibitemOpen
  \bibfield  {author} {\bibinfo {author} {\bibfnamefont {S.}~\bibnamefont
  {Zhou}}, \bibinfo {author} {\bibfnamefont {H.}~\bibnamefont {Ma}}, \bibinfo
  {author} {\bibfnamefont {E.}~\bibnamefont {Xiao}}, \bibinfo {author}
  {\bibfnamefont {K.}~\bibnamefont {Gofryk}}, \bibinfo {author} {\bibfnamefont
  {C.}~\bibnamefont {Jiang}}, \bibinfo {author} {\bibfnamefont {M.~E.}\
  \bibnamefont {Manley}}, \bibinfo {author} {\bibfnamefont {D.~H.}\
  \bibnamefont {Hurley}}, \ and\ \bibinfo {author} {\bibfnamefont {C.~A.}\
  \bibnamefont {Marianetti}},\ }\href {\doibase 10.1103/PhysRevB.106.125134}
  {\bibfield  {journal} {\bibinfo  {journal} {Physical Review B}\ }\textbf
  {\bibinfo {volume} {106}},\ \bibinfo {pages} {125134} (\bibinfo {year}
  {2022})}\BibitemShut {NoStop}%
\bibitem [{\citenamefont {Fu}\ \emph {et~al.}(2019)\citenamefont {Fu},
  \citenamefont {Kornbluth}, \citenamefont {Cheng},\ and\ \citenamefont
  {Marianetti}}]{fu_group_2019}%
  \BibitemOpen
  \bibfield  {author} {\bibinfo {author} {\bibfnamefont {L.}~\bibnamefont
  {Fu}}, \bibinfo {author} {\bibfnamefont {M.}~\bibnamefont {Kornbluth}},
  \bibinfo {author} {\bibfnamefont {Z.}~\bibnamefont {Cheng}}, \ and\ \bibinfo
  {author} {\bibfnamefont {C.~A.}\ \bibnamefont {Marianetti}},\ }\href
  {\doibase 10.1103/PhysRevB.100.014303} {\bibfield  {journal} {\bibinfo
  {journal} {Physical Review B}\ }\textbf {\bibinfo {volume} {100}},\ \bibinfo
  {pages} {014303} (\bibinfo {year} {2019})}\BibitemShut {NoStop}%
\bibitem [{\citenamefont {Gonze}\ and\ \citenamefont
  {Lee}(1997)}]{gonzeDynamicalMatricesBorn1997a}%
  \BibitemOpen
  \bibfield  {author} {\bibinfo {author} {\bibfnamefont {X.}~\bibnamefont
  {Gonze}}\ and\ \bibinfo {author} {\bibfnamefont {C.}~\bibnamefont {Lee}},\
  }\href {\doibase 10.1103/PhysRevB.55.10355} {\bibfield  {journal} {\bibinfo
  {journal} {Physical Review B}\ }\textbf {\bibinfo {volume} {55}},\ \bibinfo
  {pages} {10355} (\bibinfo {year} {1997})}\BibitemShut {NoStop}%
\bibitem [{\citenamefont {Mathis}\ \emph {et~al.}(2022)\citenamefont {Mathis},
  \citenamefont {Khanolkar}, \citenamefont {Fu}, \citenamefont {Bryan},
  \citenamefont {Dennett}, \citenamefont {Rickert}, \citenamefont {Mann},
  \citenamefont {Winn}, \citenamefont {Abernathy}, \citenamefont {Manley},
  \citenamefont {Hurley},\ and\ \citenamefont
  {Marianetti}}]{mathisGeneralizedQuasiharmonicApproximation2022}%
  \BibitemOpen
  \bibfield  {author} {\bibinfo {author} {\bibfnamefont {M.~A.}\ \bibnamefont
  {Mathis}}, \bibinfo {author} {\bibfnamefont {A.}~\bibnamefont {Khanolkar}},
  \bibinfo {author} {\bibfnamefont {L.}~\bibnamefont {Fu}}, \bibinfo {author}
  {\bibfnamefont {M.~S.}\ \bibnamefont {Bryan}}, \bibinfo {author}
  {\bibfnamefont {C.~A.}\ \bibnamefont {Dennett}}, \bibinfo {author}
  {\bibfnamefont {K.}~\bibnamefont {Rickert}}, \bibinfo {author} {\bibfnamefont
  {J.~M.}\ \bibnamefont {Mann}}, \bibinfo {author} {\bibfnamefont
  {B.}~\bibnamefont {Winn}}, \bibinfo {author} {\bibfnamefont {D.~L.}\
  \bibnamefont {Abernathy}}, \bibinfo {author} {\bibfnamefont {M.~E.}\
  \bibnamefont {Manley}}, \bibinfo {author} {\bibfnamefont {D.~H.}\
  \bibnamefont {Hurley}}, \ and\ \bibinfo {author} {\bibfnamefont {C.~A.}\
  \bibnamefont {Marianetti}},\ }\href {\doibase 10.1103/PhysRevB.106.014314}
  {\bibfield  {journal} {\bibinfo  {journal} {Physical Review B}\ }\textbf
  {\bibinfo {volume} {106}},\ \bibinfo {pages} {014314} (\bibinfo {year}
  {2022})}\BibitemShut {NoStop}%
\bibitem [{\citenamefont {Momin}\ \emph {et~al.}(1991)\citenamefont {Momin},
  \citenamefont {Mirza},\ and\ \citenamefont
  {Mathews}}]{mominHighTemperatureXray1991}%
  \BibitemOpen
  \bibfield  {author} {\bibinfo {author} {\bibfnamefont {A.~C.}\ \bibnamefont
  {Momin}}, \bibinfo {author} {\bibfnamefont {E.~B.}\ \bibnamefont {Mirza}}, \
  and\ \bibinfo {author} {\bibfnamefont {M.~D.}\ \bibnamefont {Mathews}},\
  }\href {\doibase 10.1016/0022-3115(91)90521-8} {\bibfield  {journal}
  {\bibinfo  {journal} {Journal of Nuclear Materials}\ }\textbf {\bibinfo
  {volume} {185}},\ \bibinfo {pages} {308} (\bibinfo {year}
  {1991})}\BibitemShut {NoStop}%
\bibitem [{\citenamefont {Idiri}\ \emph {et~al.}(2004)\citenamefont {Idiri},
  \citenamefont {Le~Bihan}, \citenamefont {Heathman},\ and\ \citenamefont
  {Rebizant}}]{idiri_behavior_2004}%
  \BibitemOpen
  \bibfield  {author} {\bibinfo {author} {\bibfnamefont {M.}~\bibnamefont
  {Idiri}}, \bibinfo {author} {\bibfnamefont {T.}~\bibnamefont {Le~Bihan}},
  \bibinfo {author} {\bibfnamefont {S.}~\bibnamefont {Heathman}}, \ and\
  \bibinfo {author} {\bibfnamefont {J.}~\bibnamefont {Rebizant}},\ }\href
  {\doibase 10.1103/PhysRevB.70.014113} {\bibfield  {journal} {\bibinfo
  {journal} {Physical Review B}\ }\textbf {\bibinfo {volume} {70}},\ \bibinfo
  {pages} {014113} (\bibinfo {year} {2004})}\BibitemShut {NoStop}%
\bibitem [{\citenamefont {Santini}\ \emph {et~al.}(2009)\citenamefont
  {Santini}, \citenamefont {Carretta}, \citenamefont {Amoretti}, \citenamefont
  {Caciuffo}, \citenamefont {Magnani},\ and\ \citenamefont
  {Lander}}]{santini_multipolar_2009}%
  \BibitemOpen
  \bibfield  {author} {\bibinfo {author} {\bibfnamefont {P.}~\bibnamefont
  {Santini}}, \bibinfo {author} {\bibfnamefont {S.}~\bibnamefont {Carretta}},
  \bibinfo {author} {\bibfnamefont {G.}~\bibnamefont {Amoretti}}, \bibinfo
  {author} {\bibfnamefont {R.}~\bibnamefont {Caciuffo}}, \bibinfo {author}
  {\bibfnamefont {N.}~\bibnamefont {Magnani}}, \ and\ \bibinfo {author}
  {\bibfnamefont {G.~H.}\ \bibnamefont {Lander}},\ }\href {\doibase
  10.1103/RevModPhys.81.807} {\bibfield  {journal} {\bibinfo  {journal}
  {Reviews of Modern Physics}\ }\textbf {\bibinfo {volume} {81}},\ \bibinfo
  {pages} {807} (\bibinfo {year} {2009})}\BibitemShut {NoStop}%
\bibitem [{\citenamefont {Bryan}\ \emph {et~al.}(2019)\citenamefont {Bryan},
  \citenamefont {Pang}, \citenamefont {Larson}, \citenamefont {Chernatynskiy},
  \citenamefont {Abernathy}, \citenamefont {Gofryk},\ and\ \citenamefont
  {Manley}}]{bryan_impact_2019}%
  \BibitemOpen
  \bibfield  {author} {\bibinfo {author} {\bibfnamefont {M.~S.}\ \bibnamefont
  {Bryan}}, \bibinfo {author} {\bibfnamefont {J.~W.~L.}\ \bibnamefont {Pang}},
  \bibinfo {author} {\bibfnamefont {B.~C.}\ \bibnamefont {Larson}}, \bibinfo
  {author} {\bibfnamefont {A.}~\bibnamefont {Chernatynskiy}}, \bibinfo {author}
  {\bibfnamefont {D.~L.}\ \bibnamefont {Abernathy}}, \bibinfo {author}
  {\bibfnamefont {K.}~\bibnamefont {Gofryk}}, \ and\ \bibinfo {author}
  {\bibfnamefont {M.~E.}\ \bibnamefont {Manley}},\ }\href {\doibase
  10.1103/PhysRevMaterials.3.065405} {\bibfield  {journal} {\bibinfo  {journal}
  {Physical Review Materials}\ }\textbf {\bibinfo {volume} {3}},\ \bibinfo
  {pages} {065405} (\bibinfo {year} {2019})}\BibitemShut {NoStop}%
\bibitem [{\citenamefont {Agency}(1965)}]{international1965thermodynamic}%
  \BibitemOpen
  \bibfield  {author} {\bibinfo {author} {\bibfnamefont {I.~A.~E.}\
  \bibnamefont {Agency}},\ }\href@noop {} {\emph {\bibinfo {title}
  {Thermodynamic and transport properties of uranium dioxide and related
  phases}}}\ (\bibinfo {year} {1965})\BibitemShut {NoStop}%
\bibitem [{\citenamefont {Ronchi}\ \emph {et~al.}(1999)\citenamefont {Ronchi},
  \citenamefont {Sheindlin}, \citenamefont {Musella},\ and\ \citenamefont
  {Hyland}}]{ronchiThermalConductivityUranium1999a}%
  \BibitemOpen
  \bibfield  {author} {\bibinfo {author} {\bibfnamefont {C.}~\bibnamefont
  {Ronchi}}, \bibinfo {author} {\bibfnamefont {M.}~\bibnamefont {Sheindlin}},
  \bibinfo {author} {\bibfnamefont {M.}~\bibnamefont {Musella}}, \ and\
  \bibinfo {author} {\bibfnamefont {G.~J.}\ \bibnamefont {Hyland}},\ }\href
  {\doibase 10.1063/1.369159} {\bibfield  {journal} {\bibinfo  {journal}
  {Journal of Applied Physics}\ }\textbf {\bibinfo {volume} {85}},\ \bibinfo
  {pages} {776} (\bibinfo {year} {1999})}\BibitemShut {NoStop}%
\bibitem [{\citenamefont {Kim}\ \emph {et~al.}(2014)\citenamefont {Kim},
  \citenamefont {Kim},\ and\ \citenamefont
  {Kaviany}}]{kimLatticeThermalConductivity2014}%
  \BibitemOpen
  \bibfield  {author} {\bibinfo {author} {\bibfnamefont {H.}~\bibnamefont
  {Kim}}, \bibinfo {author} {\bibfnamefont {M.~H.}\ \bibnamefont {Kim}}, \ and\
  \bibinfo {author} {\bibfnamefont {M.}~\bibnamefont {Kaviany}},\ }\href
  {\doibase 10.1063/1.4869669} {\bibfield  {journal} {\bibinfo  {journal}
  {Journal of Applied Physics}\ }\textbf {\bibinfo {volume} {115}},\ \bibinfo
  {pages} {123510} (\bibinfo {year} {2014})}\BibitemShut {NoStop}%
\bibitem [{\citenamefont {Pavlov}\ \emph {et~al.}(2017)\citenamefont {Pavlov},
  \citenamefont {Wenman}, \citenamefont {Vlahovic}, \citenamefont {Robba},
  \citenamefont {Konings}, \citenamefont {Van~Uffelen},\ and\ \citenamefont
  {Grimes}}]{pavlovMeasurementInterpretationThermophysical2017}%
  \BibitemOpen
  \bibfield  {author} {\bibinfo {author} {\bibfnamefont {T.~R.}\ \bibnamefont
  {Pavlov}}, \bibinfo {author} {\bibfnamefont {M.~R.}\ \bibnamefont {Wenman}},
  \bibinfo {author} {\bibfnamefont {L.}~\bibnamefont {Vlahovic}}, \bibinfo
  {author} {\bibfnamefont {D.}~\bibnamefont {Robba}}, \bibinfo {author}
  {\bibfnamefont {R.~J.~M.}\ \bibnamefont {Konings}}, \bibinfo {author}
  {\bibfnamefont {P.}~\bibnamefont {Van~Uffelen}}, \ and\ \bibinfo {author}
  {\bibfnamefont {R.~W.}\ \bibnamefont {Grimes}},\ }\href {\doibase
  10.1016/j.actamat.2017.07.060} {\bibfield  {journal} {\bibinfo  {journal}
  {Acta Materialia}\ }\textbf {\bibinfo {volume} {139}},\ \bibinfo {pages}
  {138} (\bibinfo {year} {2017})}\BibitemShut {NoStop}%
\end{thebibliography}%

\end{document}


\renewcommand{\thefigure}{S\arabic{figure}}
\renewcommand{\thetable}{S\arabic{table}}
\raggedright{\textbf{\Large Supplemental materials to}}
\title{Phonon thermal transport in UO$_2$ via self-consistent perturbation theory}
\author{Shuxiang Zhou$^1$, Enda Xiao$^2$, Hao Ma$^{3,4}$, Krzysztof Gofryk$^1$, Chao Jiang$^1$, Michael E. Manley$^3$, David H. Hurley$^1$, and Chris A. Marianetti$^5$}
\affiliation{$^1$Idaho National Laboratory, Idaho Falls, Idaho 83415, USA\\$^2$Department of Chemistry, Columbia University, New York, New York 10027, USA\\$^3$Oak Ridge National Laboratory, Oak Ridge, Tennessee 37831, USA\\$^4$Department of Thermal Science and Energy Engineering, University of Science and Technology of China, Hefei, Anhui 230026, China\\$^5$Department of Applied Physics and Applied Mathematics, Columbia University, New York, New York 10027, USA}

\maketitle

\section{\label{sec:smlabel}Lists of abbreviations and labels}
\vspace{5mm}

\justifying
\begin{longtable*}{ll}
\caption{List of abbreviations in this work.} 
\\
\hline\hline\\[-0.8em]
\endhead
Abbreviation & Definition \\
\hline\\[-0.8em]
AFM & antiferromagnetic \\
ARCS & the Angular Range Chopper Spectrometer\\
BID & the bundled irreducible derivative approach\\
BTE & the linearized phonon Peierls-Boltzmann transport equation \\
DFT & density functional theory \\
ER & energy resolution \\
FWHM & full-width half-maximum\\
GGA & the generalized gradient approximation\\
HF & the Hartree-Fock approximation \\
INS & the inelastic neutron scattering \\
OMC & occupation matrix control  \\
PAW & the projector augmented-wave method \\
PBE & GGA as formulated by Perdew, Burke, and Ernzerhof\\
QP & quasiparticle perturbation theory\\
RTA & the relaxation time approximation\\
SM & supplemental material \\
SOC & spin-orbit coupling \\
VASP &  the Vienna ab initio Simulation Package code \\
WTE & the Wigner transport equation \\

\hline\hline\\[-0.8em]
\end{longtable*}

\begin{longtable}{ll}
\caption{List of labels in this work.} 
\\
\hline\hline\\[-0.8em]
\endhead
Labels & Definition \\
\hline\\[-0.8em]
$E$ & energy\\
$\mathcal{S}^A_{ijk...}$ & the self-consistent perturbation theory scheme,   where $A\in\{o, HF, QP\}$ labels the self-consistency scheme and $i, j, k, . . .$ \\
  & indicate all diagrams evaluated post self-consistency.  The colloquial diagram names bubble, loop, and sunset are  \\
  & abbreviated as $b$, $l$, and $s$, respectively, while the self-consistency schemes $o$, $HF$, and $QP$ correspond to the bare,  \\
  & Hartree-Fock, and quasiparticle Green’s function, respectively. \\
$S(\mathbf{Q}, \omega)$ & the scattering function, where $\mathbf{Q}$ is the reciprocal lattice vector, and $\omega$ is the angular frequency.\\
$\hat{\mathbf{S}}_{BZ}$ & the invertible matrix of integers which yields the supercell\\
$T$ & temperature\\
$U$ & the $U$ value in DFT+$U$ approximation \\

\hline\hline\\[-0.8em]
\end{longtable}

\clearpage

\section{\label{sec:smlitrev}Previous DFT+U studies on thermal conductivity of UO$_2$}
\vspace{5mm}

\justifying

\begin{figure}[h]
\includegraphics[width=0.6\columnwidth]{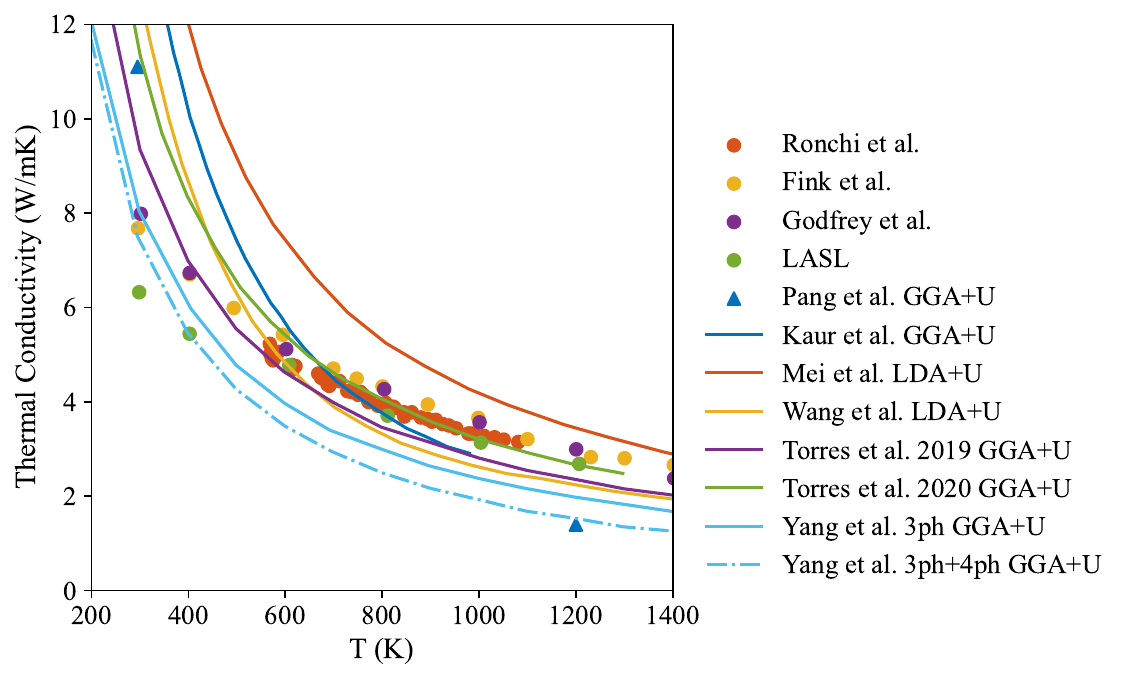}
\caption{\label{fig:smconductivity} Thermal conductivity of UO$_2$ in previous DFT+$U$ studies \cite{pang_phonon_2013,kaur_thermal_2013,meiFirstprinciplesStudyThermophysical2014,wangThermalConductivityUO22015, torresThermalConductivityDiffusion2019, torres_comparative_2020, yangReducedAnharmonicPhonon2022}, comparing with experiments \cite{finkThermophysicalPropertiesUranium2000, batesVisibleInfraredAbsorption1965, godfreyThermalConductivityUranium1965, ronchiEffectBurnupThermal2004}.}
\end{figure}

Extensive DFT+$U$ studies have computed the thermal
conductivity of UO$_2$, covering various exchange correlation functionals, including LDA+$U$ \cite{meiFirstprinciplesStudyThermophysical2014, wangThermalConductivityUO22015}, 
PBE+$U$ \cite{pang_phonon_2013, kaur_thermal_2013, yangReducedAnharmonicPhonon2022}, 
and PBEsol+$U$ \cite{torresThermalConductivityDiffusion2019,torres_comparative_2020}. The phonon interactions are computed at different orders as well:
Ref.~\cite{meiFirstprinciplesStudyThermophysical2014, wangThermalConductivityUO22015, kaur_thermal_2013} only computed the Gruneisen parameters, 
while all other studies computed the cubic phonon interactions. Additionally, Ref.~\cite{pang_phonon_2013} includes the thermal expansion using the quansiharmonic method, and
Ref.~\cite{yangReducedAnharmonicPhonon2022} computed the quartic
phonon interactions. 
From all the aforementioned studies, the computed thermal conductivity results are collected in Fig. \ref{fig:smconductivity}. Note that we only present results that are directly comparable to our own (e.g. results with boundary effects are exlcuded, etc).
Clearly, there are wide ranging results among the existing data in the literature.

\begin{figure}[t]
\includegraphics[width=0.4\columnwidth]{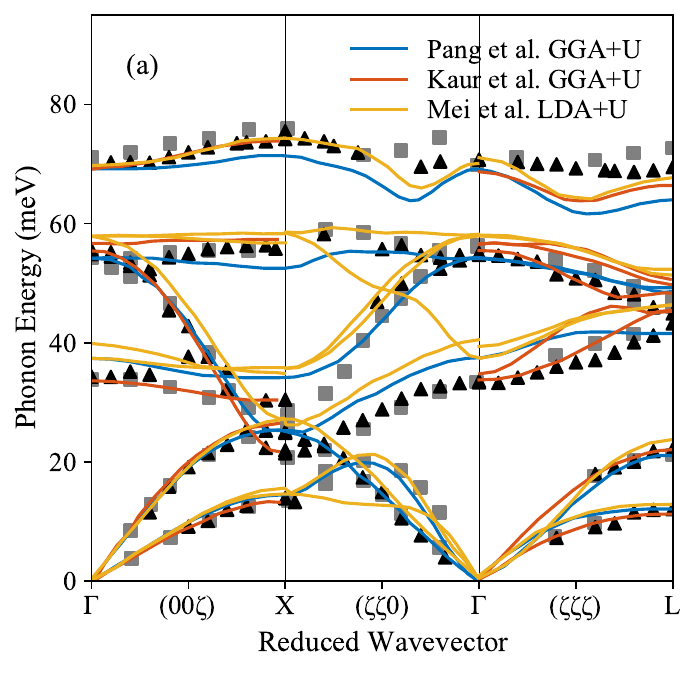}
\includegraphics[width=0.4\columnwidth]{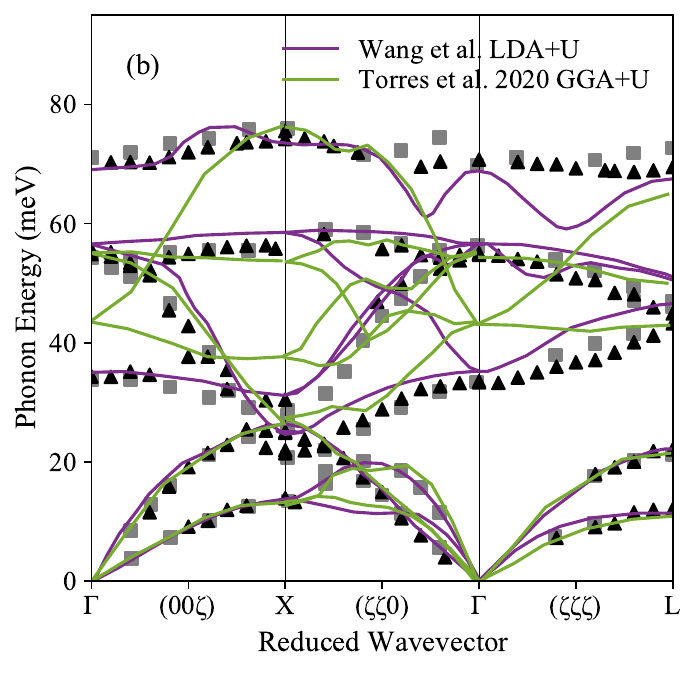}
\caption{\label{fig:smphonon} Phonon dispersion of UO$_2$ in previous DFT+$U$ studies \cite{pang_phonon_2013,kaur_thermal_2013,meiFirstprinciplesStudyThermophysical2014, wang_phonon_2013, torres_comparative_2020}, comparing with experiments \cite{pang_phonon_2013,zhouCapturingGroundState2022a}.}
\end{figure}

Given that phonon interactions are not normally tabulated in publications, and are not provided
in the aforementioned publications, it is not possible to directly scrutinize them.
However, we can assess the phonon frequencies, which are far more straightforward to compute. 
The phonons computed by the aforementioned studies are collected and compared with experiment (see Fig. \ref{fig:smphonon}) (note that Ref.~\cite{wangThermalConductivityUO22015}'s phonons are provided by Ref.~\cite{wang_phonon_2013}), with the exception of Ref.~ \cite{torresThermalConductivityDiffusion2019, yangReducedAnharmonicPhonon2022}. Ref.~\cite{torresThermalConductivityDiffusion2019} only provides folded phonon dispersion in a supercell, and Ref.~\cite{yangReducedAnharmonicPhonon2022} does not provide phonon dispersion calculated by DFT+$U$. All other studies provide phonon dispersion calculated by DFT+$U$ at zero temperature, except Ref.~\cite{pang_phonon_2013} which computes the phonons at the volume corresponding to $T = 295$ K. Multiple nontrivial anomalies can be identified, such as the inversion of the acoustic modes between $\Gamma$ and $L$ \cite{pang_phonon_2013,meiFirstprinciplesStudyThermophysical2014}, the overestimation of the LO1 and TO1 branches  \cite{pang_phonon_2013,meiFirstprinciplesStudyThermophysical2014,torres_comparative_2020}, the breaking of cubic symmetry \cite{kaur_thermal_2013,meiFirstprinciplesStudyThermophysical2014}, and large oscillations in the highest optical branch \cite{pang_phonon_2013,wang_phonon_2013}. Given these nontrivial discrepancies in the phonon frequencies, it would not be unexpected to find corresponding discrepancies in the phonon interactions.

\clearpage

\section{\label{sec:smconvergence}Computational details and convergence of Thermal conductivity}

In DFT calculations, a plane-wave cutoff energy of 550 eV was used, and the energy convergence criterion was $10^{-6}$ eV. The phonons were calculated 
in our previous work \cite{zhouCapturingGroundState2022a}, while in this work we increased the supercell size (see Sec. III of the Supplemental Materials).
The cubic and quartic irreducible derivatives were computed at six finite displacement amplitudes  
which were then used to construct quadratic error tails, ensuring that
the discretization error was properly extrapolated to zero.
Above the Néel temperature $T_N$, the space group of UO$_2$ is $Fm\bar{3}m$ and the point group symmetry of the uranium site is $O_h$. The face-centered cubic lattice vectors are encoded in a $3\times3$ row stacked
matrix $\hat{\mathbf{a}}=\frac{a_o}{2}(\hat{\mathbf{J}}-\hat{\mathbf{1}})$,
where $\hat{\mathbf{1}}$ is the identity matrix and
$\hat{\mathbf{J}}$ is a matrix in which each element is 1.
The Brillouin zone is discretized using a real space supercell
$\hat{\mathbf{S}}_{BZ}\hat{\mathbf{a}}$, where $\hat{\mathbf{S}}_{BZ}$ is an
invertible matrix of integers which produces superlattice vectors that satisfy
the point group \cite{fu_group_2019}. Below $T_N$, the \tkafm{} ordering breaks the symmetry of the $Fm\bar{3}m$ space group, and has a primitive cell of $\hat{\mathbf{S}}_{C}\hat{\mathbf{a}}$, where $\hat{\mathbf{S}}_{C}$ generates the conventional cubic cell (multiplicity of 4)
and is defined as $\hat{\mathbf{S}}_{C}=\hat{\mathbf{J}}-2\hat{\mathbf{1}}$. When computing third order irreducible derivatives (ID), we use the supercell $\hat{\mathbf{S}}_{O}\hat{\mathbf{a}}$ and extract all third order IDs consistent with the supercell, where $\hat{\mathbf{S}}_{O}=\hat{\mathbf{S}}_{C}\hat{\mathbf{S}}_{C}=4\hat{\mathbf{1}}-\hat{\mathbf{J}}$. The supercell $\hat{\mathbf{S}}_{O}\hat{\mathbf{a}}$ has a multiplicity of 16 (48 atoms).
When computing fourth order IDs, we use the primitive cell $\hat{\mathbf{S}}_{C}\hat{\mathbf{a}}$ and extract all fourth order IDs consistent with the primitive cell.
For a \tkafm{} calculation with a primitive cell $\hat{\mathbf{S}}_{C}\hat{\mathbf{a}}$, a $12\hat{\mathbf{1}}$ $\Gamma$-centered \emph{k}-point mesh was applied; for larger supercell calculations, the \emph{k}-point density was approximately held constant.
In order to compare with the INS scattering function, we unfold the \tkafm{} IDs back to the 
original primitive unit cell $\hat{\mathbf{a}}$, averaging any translational symmetry breaking due
to the antiferromagnetism. 
The Born effective charges ($Z_U^\star=5.54$ and $Z_O^\star=-2.77$) of the U and O ions
and the dielectric constant ($\epsilon=5.69$) 
were used to account for LO-TO splitting   \cite{gonzeDynamicalMatricesBorn1997a, mathisGeneralizedQuasiharmonicApproximation2022}.

\begin{figure}[h]
\includegraphics[width=0.56\columnwidth]{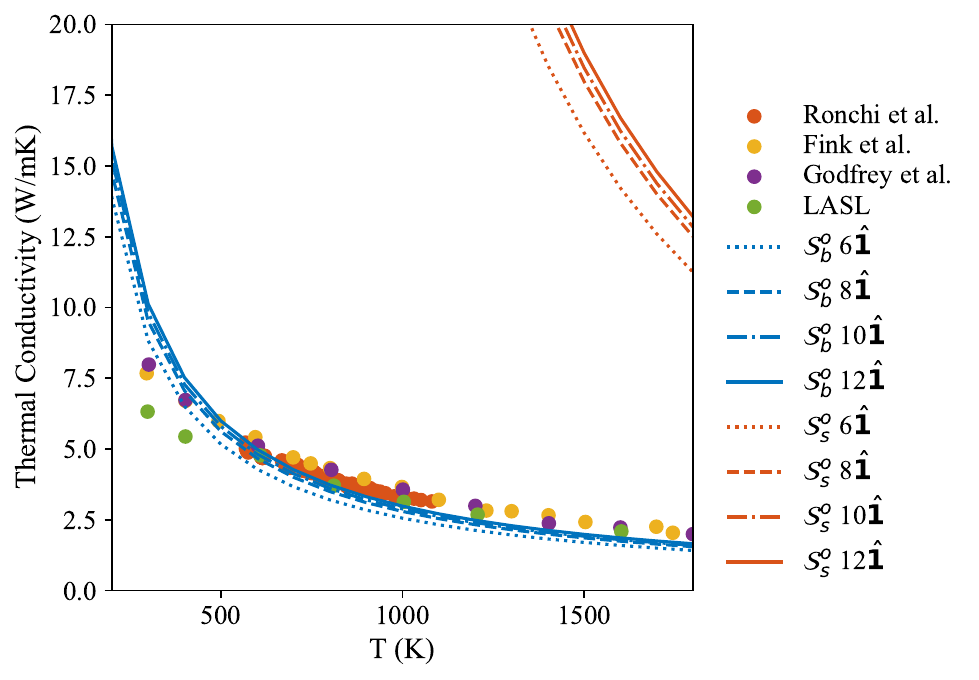}
\caption{\label{fig:smconvergence} Calculated thermal conductivity using GGA+$U$+SOC, $\mathcal{S}_b^o$ or $\mathcal{S}_s^o$ with BTE (RTA), and different interpolation grids, comparing with experiments \cite{finkThermophysicalPropertiesUranium2000, batesVisibleInfraredAbsorption1965, godfreyThermalConductivityUranium1965, ronchiEffectBurnupThermal2004}.}
\end{figure}

\begin{figure}[h]
\includegraphics[width=0.56\columnwidth]{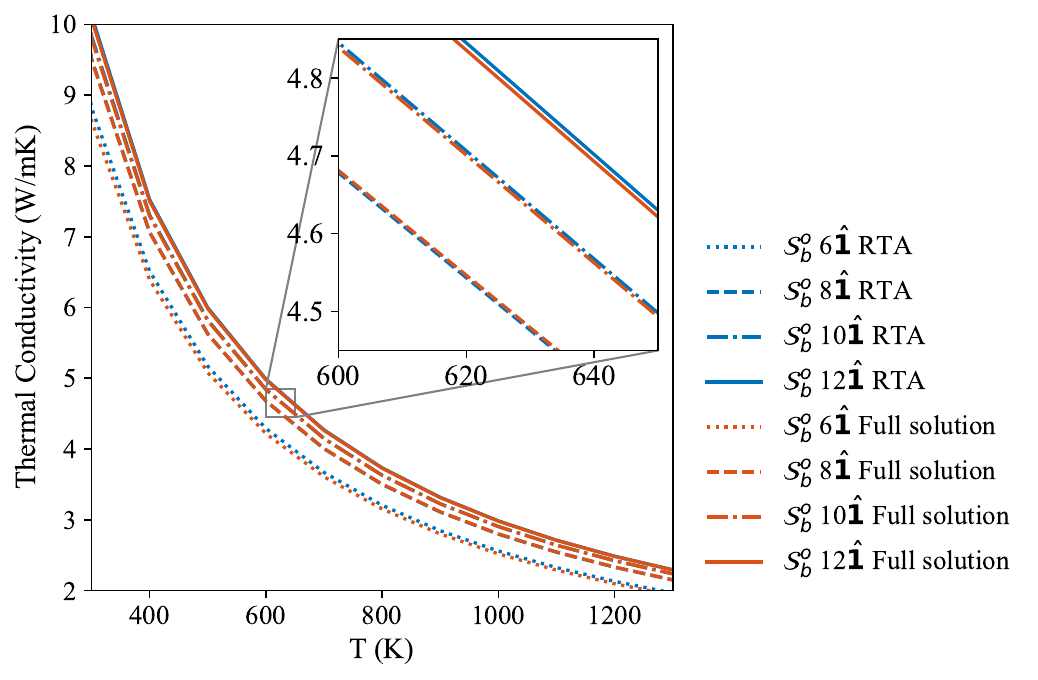}
\caption{\label{fig:smlbte} Calculated thermal conductivity using GGA+$U$+SOC, $\mathcal{S}_b^o$, and different interpolation grids. Blue and red curves present BTE (RTA) and BTE (full solution) results, respectively. }
\end{figure}

We test the convergence of thermal conductivity calculations by varying the interpolation grid from 6$\hat{\mathbf{1}}$ to 12$\hat{\mathbf{1}}$ in BTE calculations (see Fig. ~\ref{fig:smconvergence}). Both $\mathcal{S}_b^o$ and $\mathcal{S}_s^o$ results are approximately converged using 12$\hat{\mathbf{1}}$ interpolation grid, which is used in all calculations in the main manuscript. 
The linearized
Boltzmann transport equation (BTE) is also solved for $\mathcal{S}_b^o$, and yields approximately identical results to the RTA with different interpolation grids (see Fig. ~\ref{fig:smlbte}). 

\clearpage

\section{\label{sec:smhightemp}Phonon dispersion calculations}

\begin{figure}[h]
\includegraphics[width=0.37\columnwidth]{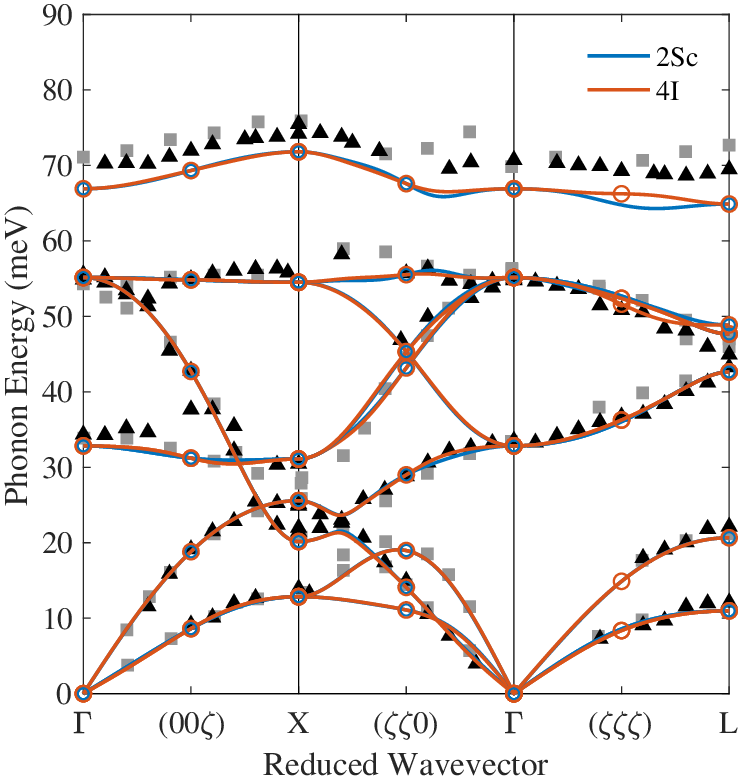}
\caption{\label{fig:phonon_4ivs2sc} Unfolded phonons of  3\textbf{k} AFM UO$_2$ computed by GGA+$U$+SOC ($U=4$ eV) (lines), compared with inelastic neutron
scattering data from Ref. \cite{pang_phonon_2013} at $T = 300$ K (grey squares) and Ref. \cite{zhouCapturingGroundState2022a} at $T = 600$ K (black triangles). Lines with different colors represent interpolations from different supercells $\hat{\mathbf{S}}_{BZ}$. The hollow points were directly computed using DFT+$U$, while the 
corresponding lines are Fourier interpolations.}
\end{figure}

The computational details of phonon dispersion calculations were already reported in our previous work \cite{zhouCapturingGroundState2022a}, and in this work we increased the supercell from $\hat{\mathbf{S}}_{BZ}=2\hat{\mathbf{S}}_{C}$ to $\hat{\mathbf{S}}_{BZ}=4\hat{\mathbf{1}}$ (see Fig. ~\ref{fig:phonon_4ivs2sc} for a comparison). 
When $\hat{\mathbf{S}}_{BZ}=4\hat{\mathbf{1}}$, the $q$-point $(0.25, 0.25, 0.25)$ is directly computed. The good agreement between $\hat{\mathbf{S}}_{BZ}=2\hat{\mathbf{S}}_{C}$ and $\hat{\mathbf{S}}_{BZ}=4\hat{\mathbf{1}}$ means the supercell size is sufficient to converge the phonon dispersion. 

\clearpage

\section{\label{sec:qvoxel}$q$-voxel dimension information }
The $q$-voxel dimensions used in Fig. 3 in main text are listed in
Tables \ref{table:uo2_001}, \ref{table:uo2_011}, and \ref{table:uo2_111}.
In all of the aforementioned tables, the first column specifies the phonon branch; 
the second column specifies a line segment in reciprocal space; 
the third, fourth, and fifth columns specify the three voxel dimensions in reciprocal lattice units.

\begin{table}[hp]
\begin{ruledtabular}
\begin{tabular}{ccccc}
(001) direction & path      & $\Delta_{[0,0,L] }$ & $\Delta_{[H,H,0]}$ & $\Delta_{[\Bar{H},H,0]}$      \\ \hline
TA              & {[ }1, 1, 7{]}$\rightarrow${[} 1, 1, 6{]}      & 0.075           & 0.2                & 0.2                \\ 
LA              & {[}-1,-1, 7{]}$\rightarrow${[}-1,-1, 6{]}  & 0.075           & 0.2                & 0.2                \\ 
TO1             & {[} 2, 2, 0{]}$\rightarrow${[} 2, 2,-1{]}     & 0.075           & 0.2                & 0.2                \\ 
LO1             & {[}-1,-1, 7{]}$\rightarrow${[}-1,-1, 6{]}  & 0.075           & 0.2                & 0.2                \\ 
TO2             & {[} 1, 1, 5{]}$\rightarrow${[} 1, 1, 6{]}      & 0.125           & 0.2                & 0.2                \\ 
LO2             & {[} 2, 2, 6{]}$\rightarrow${[} 2, 2, 5{]}      & 0.075           & 0.2                & 0.2                \\ 
\end{tabular}
\caption{Voxel information for $[\zeta00]$ direction in UO$_2$. \label{table:uo2_001} }
\end{ruledtabular}
\vspace{0.5cm}

\begin{ruledtabular}
\begin{tabular}{ccccc}
(110) direction & path      & $\Delta_{[H,H,0]}$ & $\Delta_{[0,0,L]}$ & $\Delta_{[\Bar{H},H,0]}$     \\ \hline
TA              & {[} 2, 2, 8{]}$\rightarrow${[} 3, 3, 8{]}         & 0.075     & 0.2         & 0.3      \\ 
TO1             & {[} 0, 0, 6{]}$\rightarrow${[} 1, 1, 6{]}         & 0.075     & 0.2         & 0.3      \\ 
LO1             & {[} 3, 3, 7{]}$\rightarrow${[} 2, 2, 7{]}         & 0.075     & 0.2         & 0.3      \\ 
TO2             & {[} 1, 1, 9{]}$\rightarrow${[} 0, 0, 9{]}         & 0.075     & 0.2         & 0.3      \\ 
LO2             & {[} 1, 1, 5{]}$\rightarrow${[} 2, 2, 5{]}         & 0.075     & 0.2         & 0.3      \\ 
\end{tabular}
\caption{Voxel information for $[\zeta\zeta0]$ direction in UO$_2$. \label{table:uo2_011} }
\end{ruledtabular}
\vspace{0.5cm}

\begin{ruledtabular}
\begin{tabular}{ccccc}
(111) direction & path       & $\Delta_{[H,H,H]}$ & $\Delta_{[0,0,L]}$ & $\Delta_{[\Bar{H},H,0]}$          \\ \hline
TA              & {[} 1, 1, 7{]}$\rightarrow${[}0.5,0.5,6.5{]}        & 0.075           & 0.2       & 0.2     \\ 
LA              & {[} 2, 2, 8{]}$\rightarrow${[}2.5,2.5,8.5{]}        & 0.075           & 0.2       & 0.2     \\ 
TO1             & {[} 0, 0, 8{]}$\rightarrow${[}0.5,0.5,8.5{]}        & 0.075           & 0.2       & 0.2     \\ 
LO1+TO2         & {[} 1, 1, 5{]}$\rightarrow${[}1.5,1.5,5.5{]}        & 0.075           & 0.2       & 0.2     \\ 
LO2             & {[} 2, 2, 9{]}$\rightarrow${[}1.5,1.5,8.5{]}        & 0.075           & 0.2       & 0.2     \\ 
\end{tabular}
\caption{Voxel information for $[\zeta\zeta\zeta]$ direction in UO$_2$. \label{table:uo2_111} }
\end{ruledtabular}
\end{table}

\clearpage

\section{\label{sec:smspectraltc}The spectral and cumulative thermal conductivity in BTE calculations}
\begin{figure}[h]
\includegraphics[width=0.44\columnwidth]{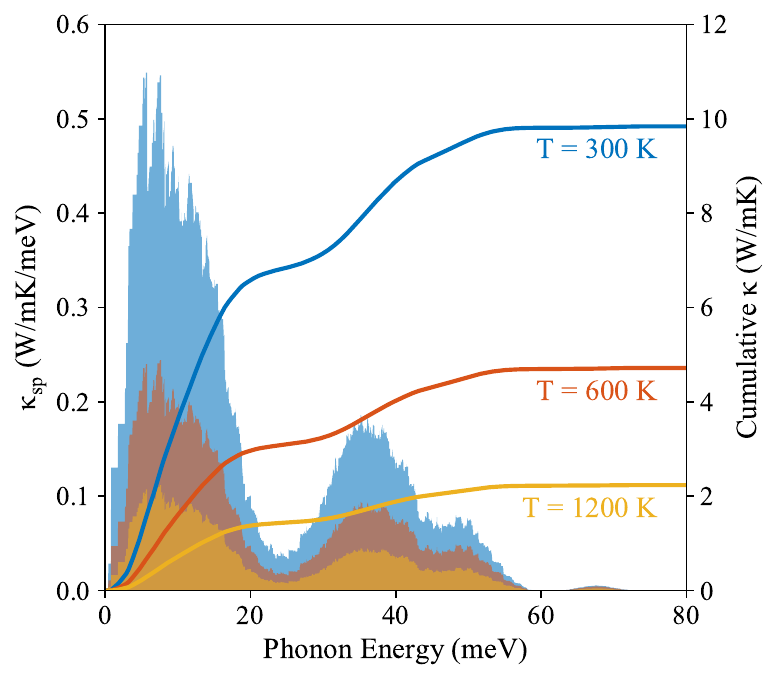}
\caption{\label{fig:mode_cond} Spectral and cumulative thermal conductivity of UO$_2$ as a function of phonon energy, computed using GGA+$U$+SOC ($U=4$ eV) at $T= 300$, $600$, and $1200$ K in $\mathcal{S}_{bs}^o$ BTE calculations.}
\end{figure}

In this section, we report the calculated spectral and cumulative thermal conductivity as
functions of phonon energy at $T=300$ K, $600$ K, and $1200$ K in $\mathcal{S}_{bs}^o$ BTE calculations, shown in Fig.
~\ref{fig:mode_cond}. At all three temperatures, roughly $70\%$ of thermal conductivity 
arises from phonon modes with energies below 24
meV, and therefore the optical modes
do play a nontrivial role in thermal transport.
Our 
GGA+$U$+SOC phonon dispersion has very good agreement with experiment \cite{zhouCapturingGroundState2022a}, except 
for a small offset in the highest
optical phonon branch, but Fig. \ref{fig:mode_cond} indicates that this
branch has a negligible influence on the thermal conductivity.

\clearpage

\section{\label{sec:smhfqp}Comparison between bare and self-consistent perturbation theory}

Here the calculated thermal conductivity using self-consistent perturbation theory, HF and QP, is analyzed separately for the bubble diagram and the bubble plus the sunset diagram. 
Note that for $\mathcal{S}_{b}^{HF}$ and $\mathcal{S}_{b}^{QP}$, the quartic phonon interactions still have a contribution due to the fact that the self-consistent Green's function is evaluated using the loop diagram. The effect of thermal expansion is not included in this section.

\begin{figure}[h]
\includegraphics[width=0.40\columnwidth]{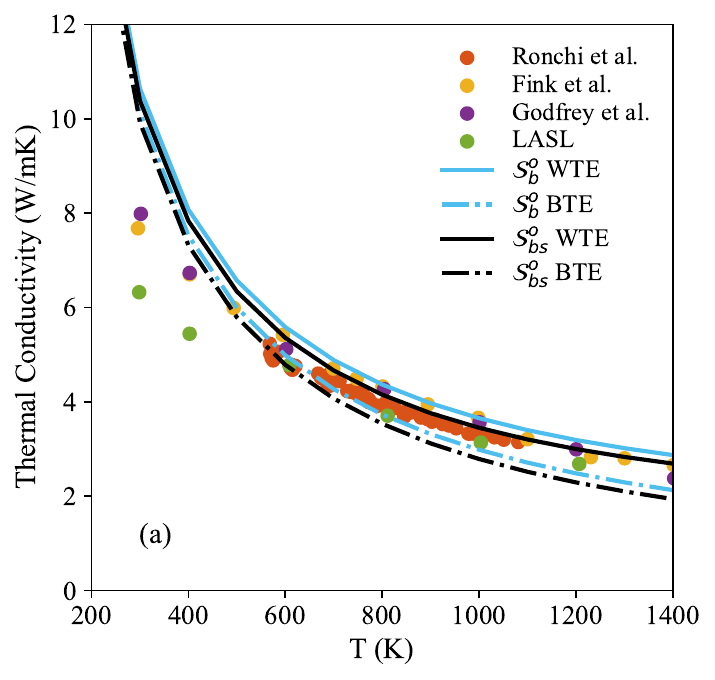}
\includegraphics[width=0.40\columnwidth]{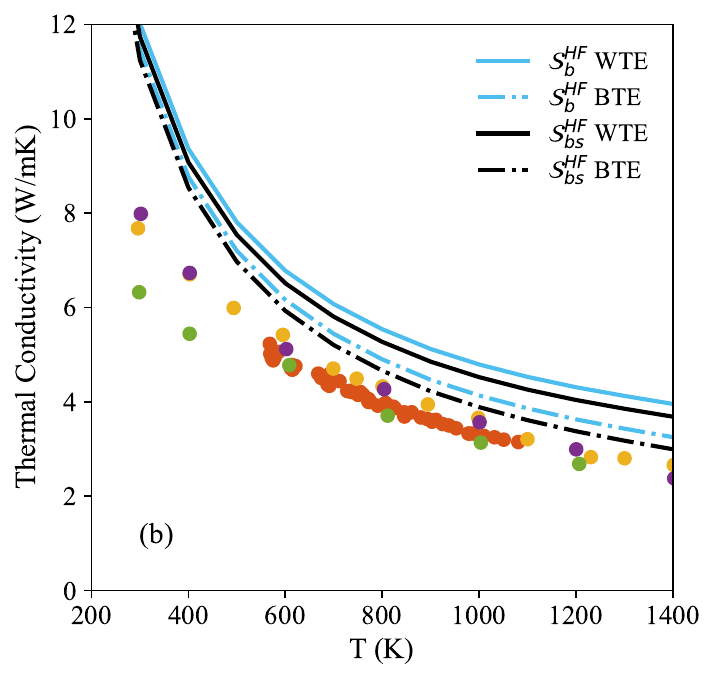}
\includegraphics[width=0.40\columnwidth]{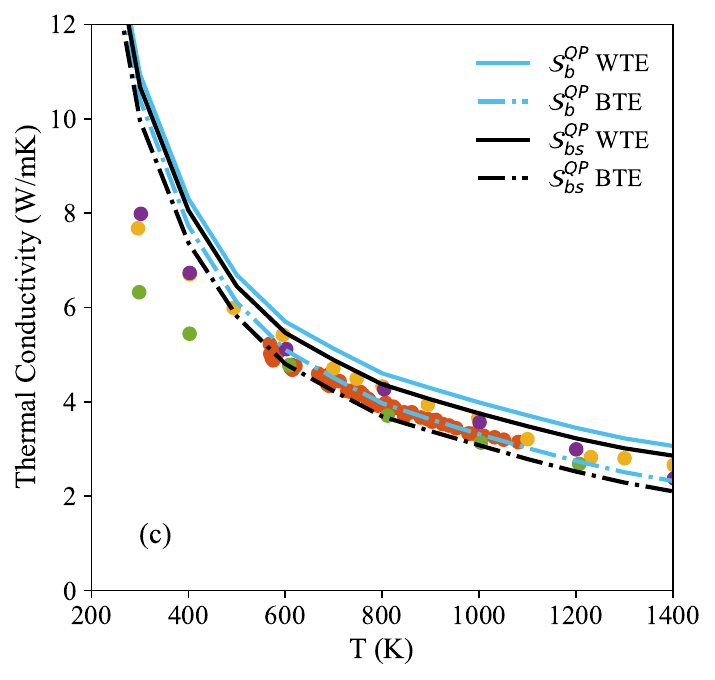}
\caption{\label{fig:sm_ther_cond_hfqp} Thermal conductivity, computed using GGA+$U$+SOC ($U=4$ eV) within (a) bare perturbation theory, (b) HF self-consistent perturbation theory, and (c) QP self-consistent perturbation theory, and comparing with experiments \cite{finkThermophysicalPropertiesUranium2000, batesVisibleInfraredAbsorption1965, godfreyThermalConductivityUranium1965, ronchiEffectBurnupThermal2004}. 
Thermal expansion is not included.}
\end{figure}
\clearpage

\section{\label{sec:smtherexpan}The effect of thermal expansion on thermal conductivity}

To address the effect of thermal expansion on thermal conductivity of UO$_2$, the irreducible derivatives are computed at four different cell volumes, where the lattice parameters $a=5.5461$, $5.5602$, $5.5717$, and $5.5942$\AA{} are used, corresponding to the experimental percentage change in the lattice parameter at $T=0$, $360$, $600$, and $1000$ K \cite{mominHighTemperatureXray1991}, respectively. These computed results are linearly interpolated or extrapolated to temperatures from $0$ to $1400$ K. 

\begin{figure}[h]
\includegraphics[width=0.71\columnwidth]{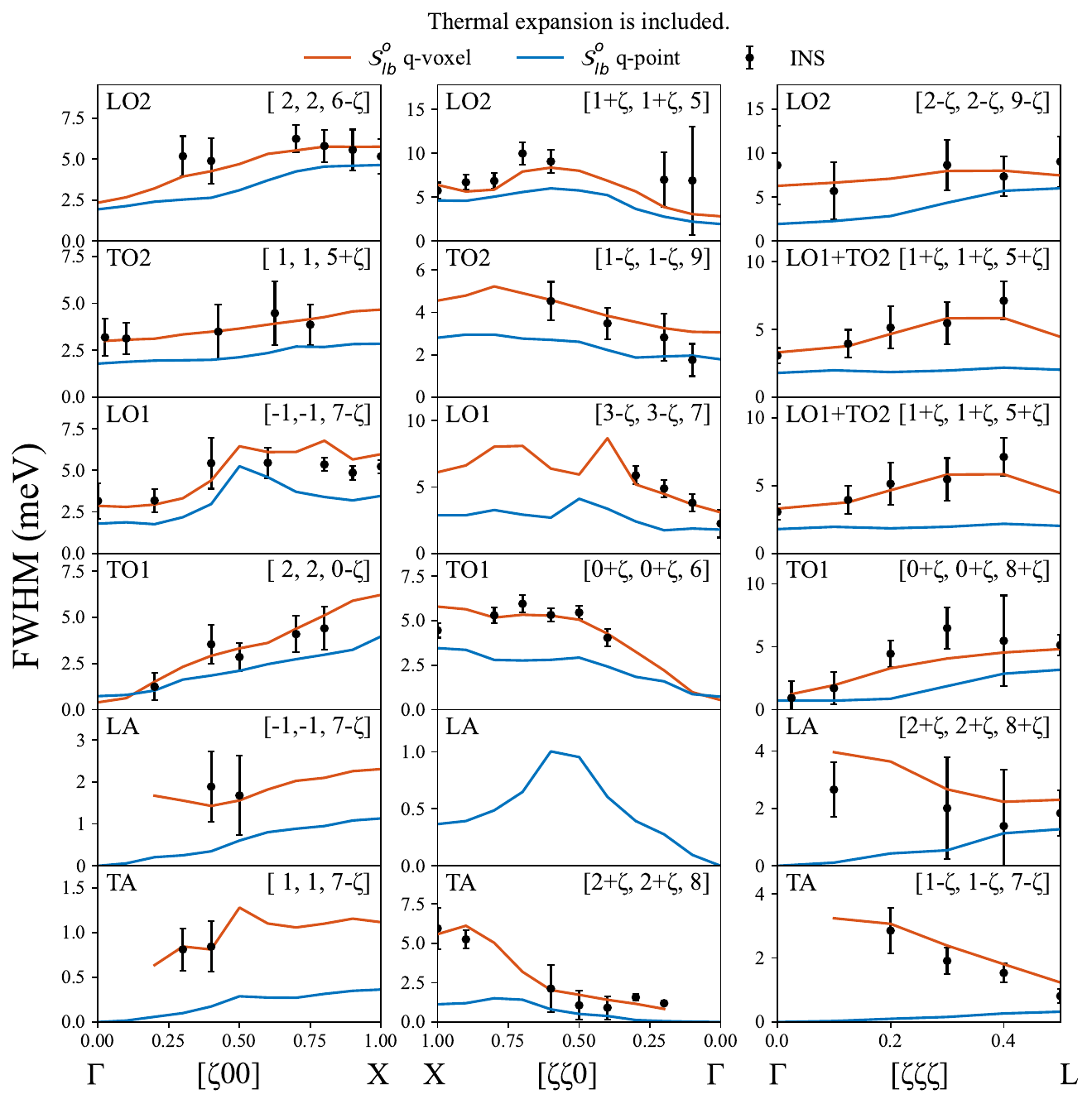}
\caption{\label{fig:sm_linewidth_pt} 
FWHMs of the $S(\mathbf{Q},\omega)$ peaks as a function of $q$ in various zones for UO$_2$ at
$T=600$ K. Thermal expansion is included when computing the irreducible derivatives. The $\mathcal{S}_{lb}^{o}$ $q$-point and $q$-voxel results are shown as blue and red curves, respectively. INS results are shown as black points. }
\end{figure}

\begin{figure}[h]
\includegraphics[width=0.71\columnwidth]{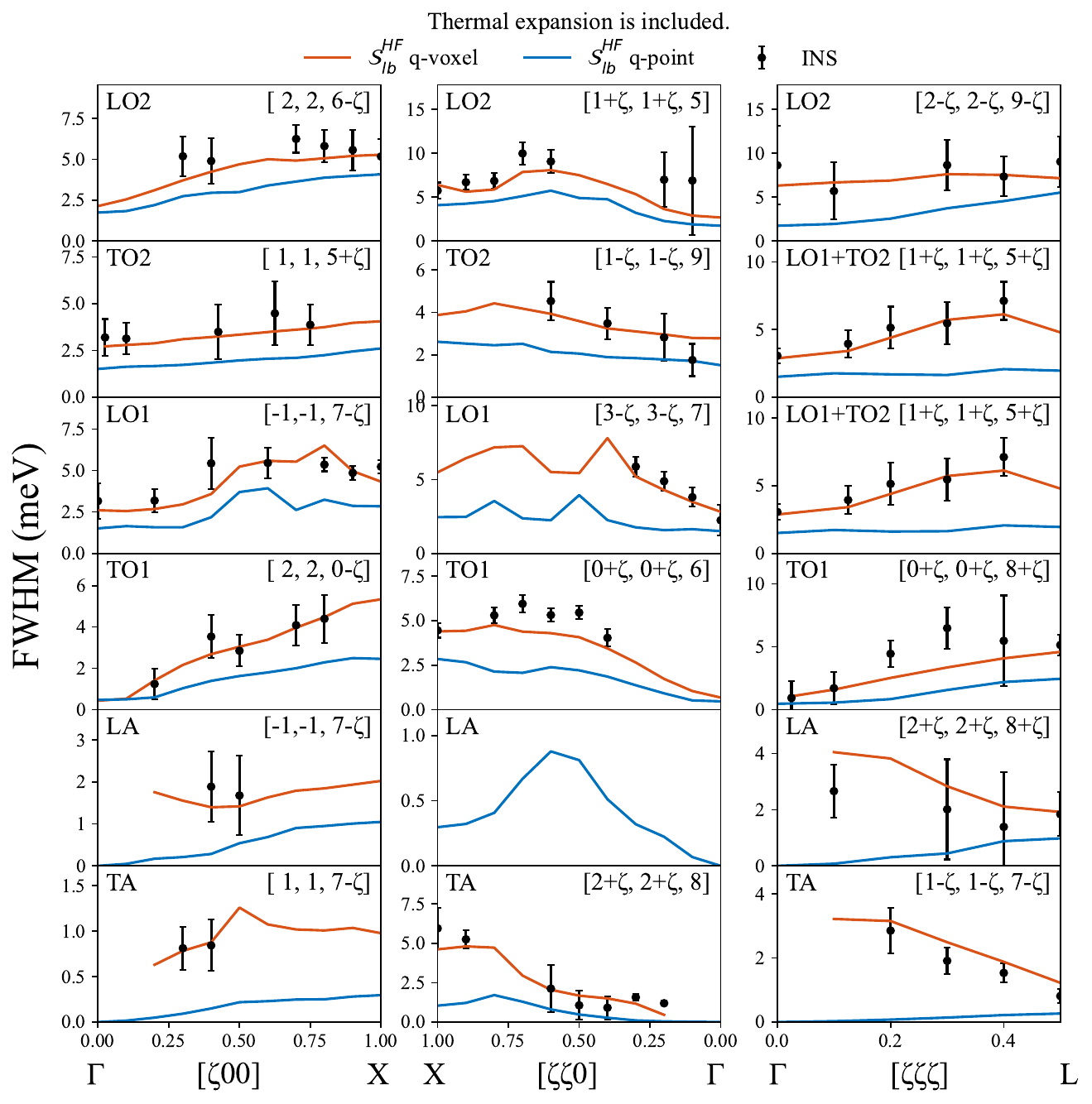}
\caption{\label{fig:sm_linewidth_hf} 
FWHMs of the $S(\mathbf{Q},\omega)$ peaks as a function of $q$ in various zones for UO$_2$ at
$T=600$ K. Thermal expansion is included when computing the irreducible derivatives. The $\mathcal{S}_{lb}^{HF}$ $q$-point and $q$-voxel results are shown as blue and red curves, respectively. INS results are shown as black points. }
\end{figure}

\begin{figure}[h]
\includegraphics[width=0.71\columnwidth]{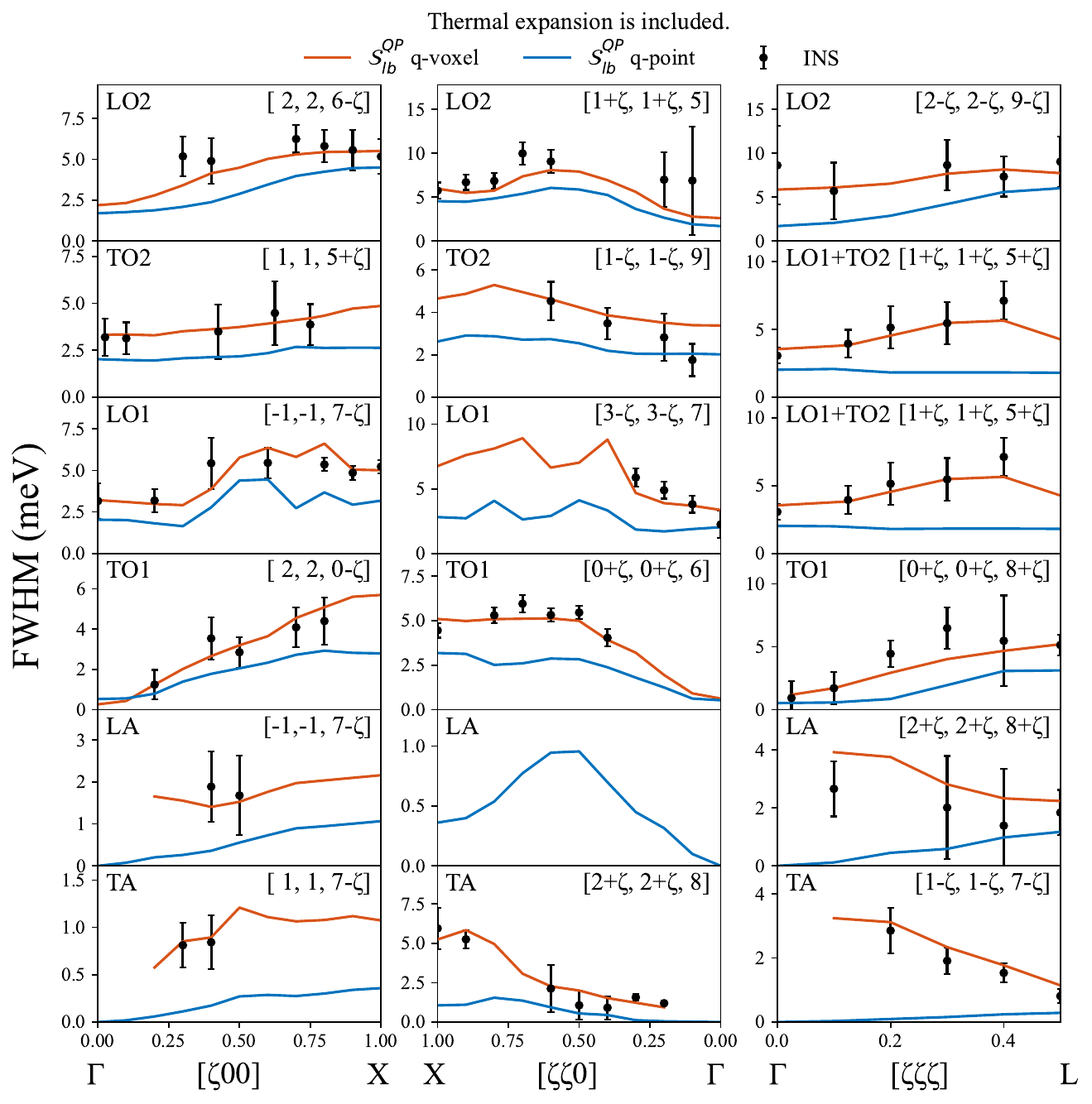}
\caption{\label{fig:sm_linewidth_qp} 
FWHMs of the $S(\mathbf{Q},\omega)$ peaks as a function of $q$ in various zones for UO$_2$ at
$T=600$ K. Thermal expansion is included when computing the irreducible derivatives. The $\mathcal{S}_{lb}^{QP}$ $q$-point and $q$-voxel results are shown as blue and red curves, respectively. INS results are shown as black points. }
\end{figure}

\begin{figure}[h]
\includegraphics[width=0.40\columnwidth]{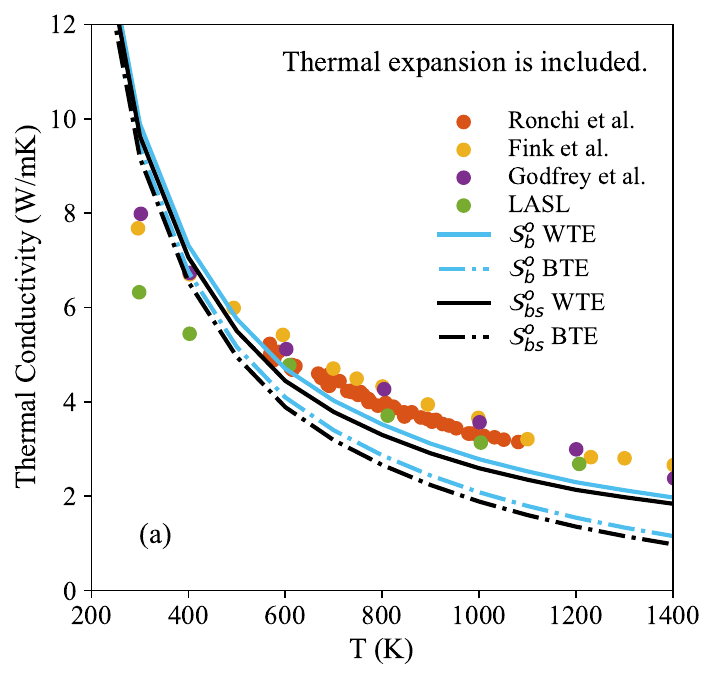}
\includegraphics[width=0.40\columnwidth]{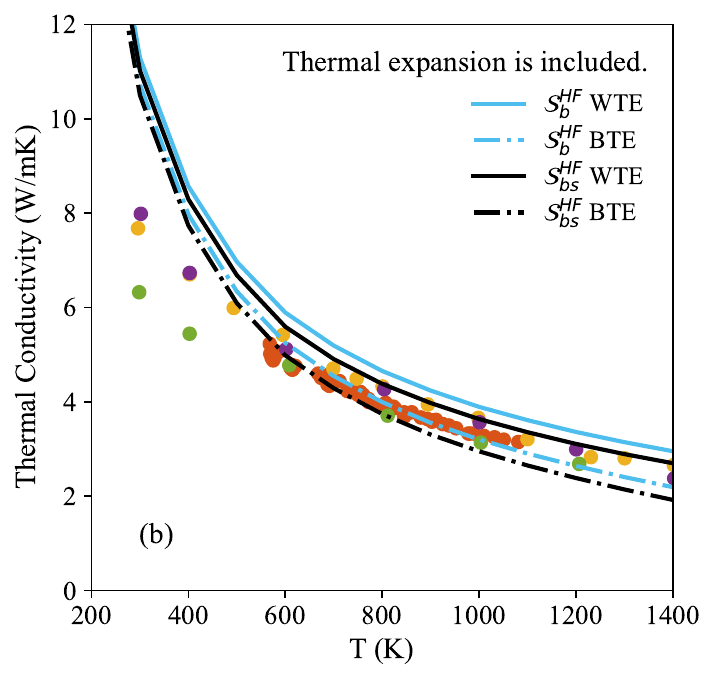}
\includegraphics[width=0.40\columnwidth]{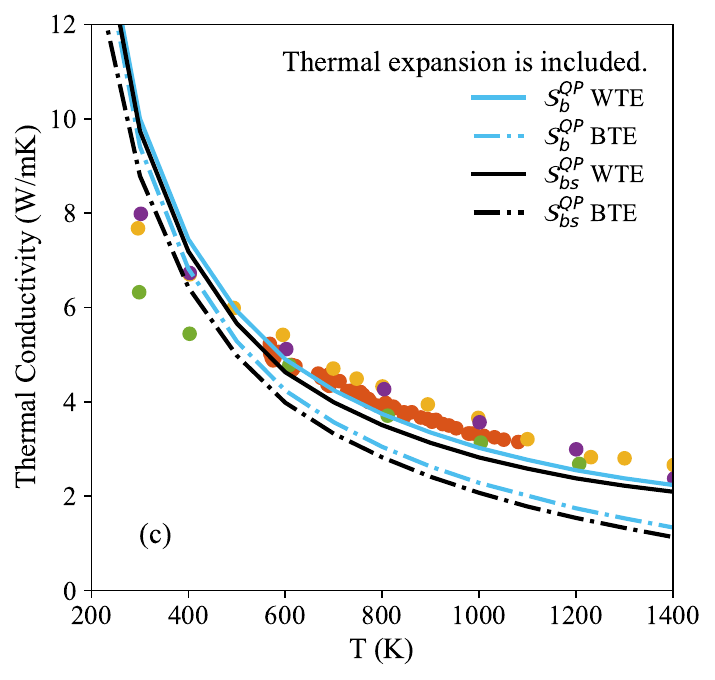}
\caption{\label{fig:sm_ther_cond_vc} Thermal conductivity, computed using GGA+$U$+SOC ($U=4$ eV) within (a) bare perturbation theory, (b) HF self-consistent perturbation theory, and (c) QP self-consistent perturbation theory, and comparing with experiments \cite{finkThermophysicalPropertiesUranium2000, batesVisibleInfraredAbsorption1965, godfreyThermalConductivityUranium1965, ronchiEffectBurnupThermal2004}. 
Thermal expansion is included when computing the irreducible derivatives. }
\end{figure}

We first present the phonon linewidth at $T = 600$K, and compare with INS experiment (see Fig~\ref{fig:sm_linewidth_pt} for $\mathcal{S}_{lb}^{o}$, Fig~\ref{fig:sm_linewidth_hf} for $\mathcal{S}_{lb}^{HF}$, and Fig~\ref{fig:sm_linewidth_qp} for $\mathcal{S}_{lb}^{QP}$ results). The differences between bare and self-consistent perturbation theory are modest. Finally, the thermal conductivity is calculated using bare perturbation theory and self-consistent perturbation theory (see Fig~\ref{fig:sm_ther_cond_vc}). While the thermal conductivity increase due to interband transition and the decrease due to thermal expansion are both non-trivial, they tend to cancel, thus combining both effects only slightly decrease the thermal conductivity in UO$_2$.

\clearpage

\section{\label{sec:smpbesol}Energy, structure, and phonon calculations using PBEsol}

In our previous work \cite{zhouCapturingGroundState2022a}, we presented the results of energy difference between different magnetic ordering, in addition to the lattice parameter, oxygen cage distortion, and phonon dispersion in \tkafm state, using both  PBE+$U$ and LDA+$U$. Recent studies \cite{torres_comparative_2020,torresThermalConductivityDiffusion2019} reported that PBEsol+$U$ predicts a more accurate lattice parameter than PBE+$U$ or LDA+$U$. Therefore, we performed PBEsol+$U$+SOC calculations for aforementioned energetic, structural, and phonon properties, and made a comparison with PBE+$U$+SOC and LDA+$U$+SOC.

\begin{figure}[h]
\includegraphics[width=0.37\columnwidth]{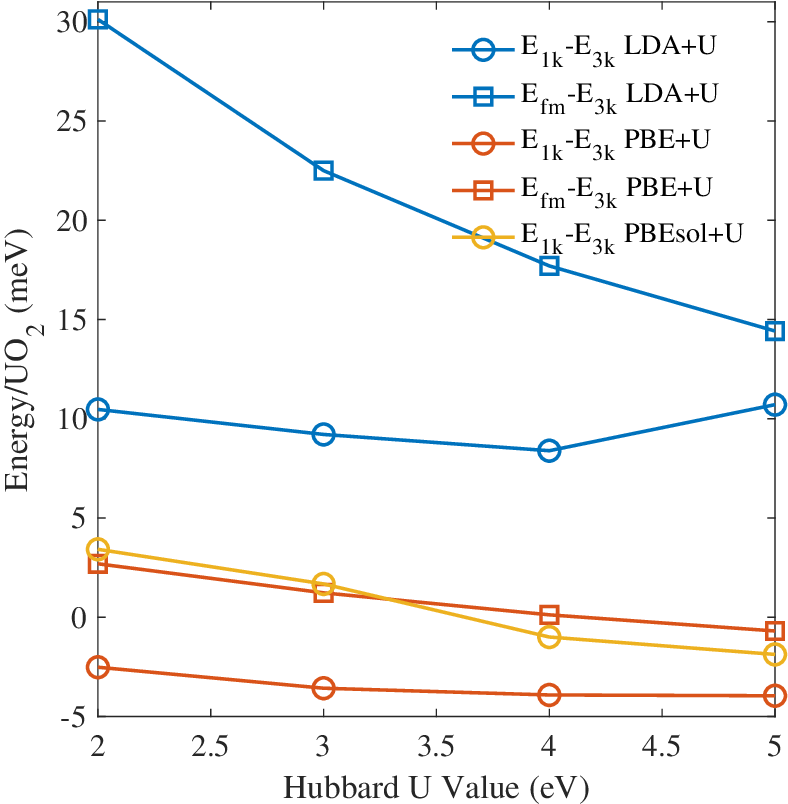}
\caption{\label{fig:smpbesolenergy} Calculated energy difference of UO$_2$ for
FM and 1\textbf{k} AFM relative to 3\textbf{k} AFM using LDA+$U$, PBE+$U$, and PBEsol+$U$ with SOC, as a function of the Hubbard $U$.}
\end{figure}

We first compare the energies of 1\textbf{k} AFM and \tkafm states using PBEsol+$U$+SOC calculation. The occupation matrix $\mathbb{S}_0$, computed by PBE \cite{zhouCapturingGroundState2022a}, is used to initialize the PBEsol calculation. 
Unsurprisingly, during PBEsol calculations, the resulting occupation matrix only slightly changes from the initial value. The energy differences between 1\textbf{k} AFM and \tkafm ordering are computed using $U=2- 5$ eV (see Fig. ~\ref{fig:smpbesolenergy}). In PBEsol+$U$, the \tkafm state has the lowest energy when $U<3.6$ eV, and the 1\textbf{k} AFM state has the lowest energy when $U>3.6$ eV. At $U=4$ eV, the 1\textbf{k} AFM and the \tkafm state are essentially degenerate, as the energy difference is only 1 meV.

\begin{figure}[h]
\includegraphics[width=0.37\columnwidth]{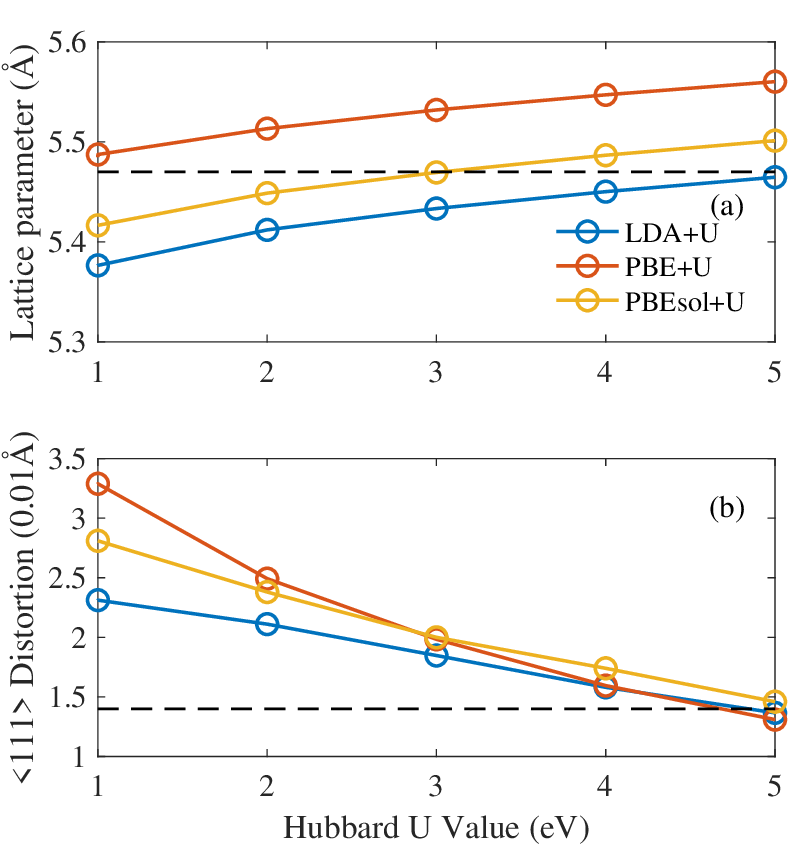}
\caption{\label{fig:smpbesolstructure} Calculated (a) lattice parameter and (b) $<$111$>$ oxygen cage distortion in
the 3\textbf{k} AFM state, using LDA+$U$+SOC, PBE+$U$+SOC, and PBEsol+$U$+SOC, as a function of the Hubbard $U$. The horizontal dashed lines represent the experimental values of the lattice parameter from Ref. \cite{idiri_behavior_2004} and the distortion from Ref.  \cite{santini_multipolar_2009}.}
\end{figure}

\begin{figure}[h]
\includegraphics[width=0.44\columnwidth]{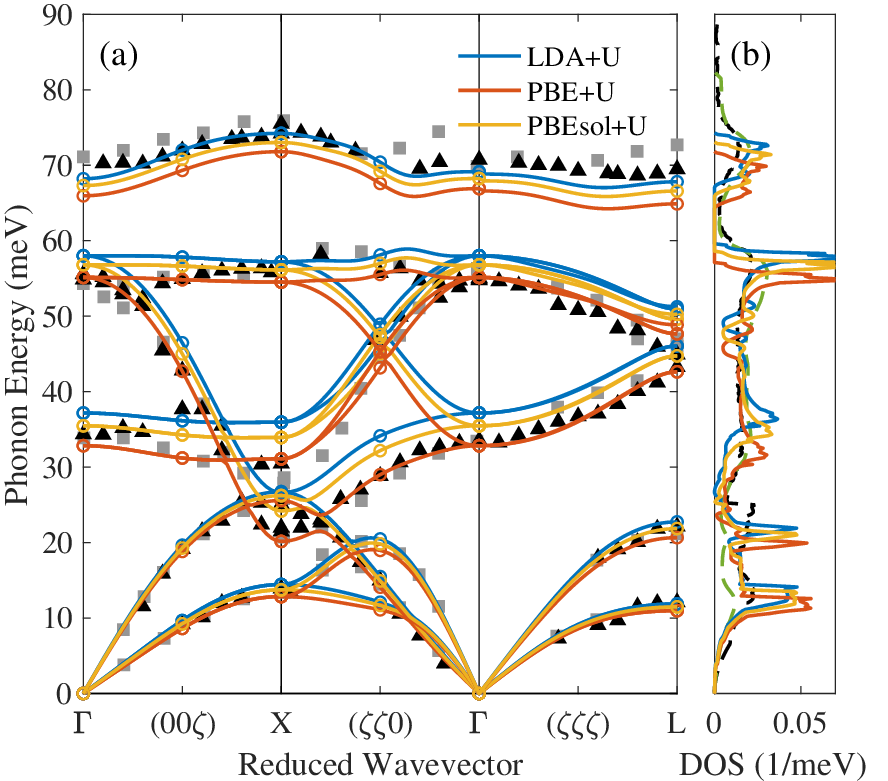}
\caption{\label{fig:smpbesolphonon} Phonons of  3\textbf{k} AFM UO$_2$. (a) The unfolded phonon dispersion curves and (b) density of states of 
LDA+$U$+SOC, PBE+$U$+SOC, and PBEsol+$U$+SOC ($U=4$ eV) (solid lines), compared with inelastic neutron
scattering data from Ref. \cite{pang_phonon_2013} at $T=300$ K (grey squares), Ref.  \cite{zhouCapturingGroundState2022a} at $T=600$ K (black triangles), Ref. \cite{zhouCapturingGroundState2022a} at $T=77$ K (dashed black curves), and Ref. \cite{bryan_impact_2019} at $T=10$ K (dashed green
curves). The hollow points were directly computed using DFT+$U$, while the 
corresponding lines are Fourier interpolations. }
\end{figure}

Additionally, the lattice parameter and oxygen cage distortion are computed in the \tkafm{}state (see Fig. ~\ref{fig:smpbesolstructure}). Generally, the predictions of the lattice parameter and oxygen cage distortion from PBEsol are very close to LDA and PBE. For the lattice parameter, PBEsol's prediction is between PBE and LDA, and $U=3$ eV gives excellent agreement with experiment. The oxygen cage distortion of PBEsol is slightly larger than both LDA and PBE for $U=4-5$ eV, and  $U=5$ eV gives favorable agreement with experiment. 
Finally, the phonon dispersion is computed in the \tkafm{} state using $U=4$ eV and $\hat{\mathbf{S}}_{BZ}=2\hat{\mathbf{S}}_{C}$ (see Fig. ~\ref{fig:smpbesolphonon}). The phonons computed by PBEsol are also between PBE and LDA phonons, mainly due to the different cell volume. For acoustic branches, LDA, PBE, and PBEsol have close predictions, however, for optical branches, PBE still has the overall best agreements except the highest optical branch. Therefore, we consider PBE to have the overall best description of the phonons for UO$_2$, and all results in the main text are computed using PBE.  

\clearpage

\section{\label{sec:u0phonon}Phonon thermal conductivity calculations using GGA ($U=0$)}

\begin{figure}[h]
\includegraphics[width=0.37\columnwidth]{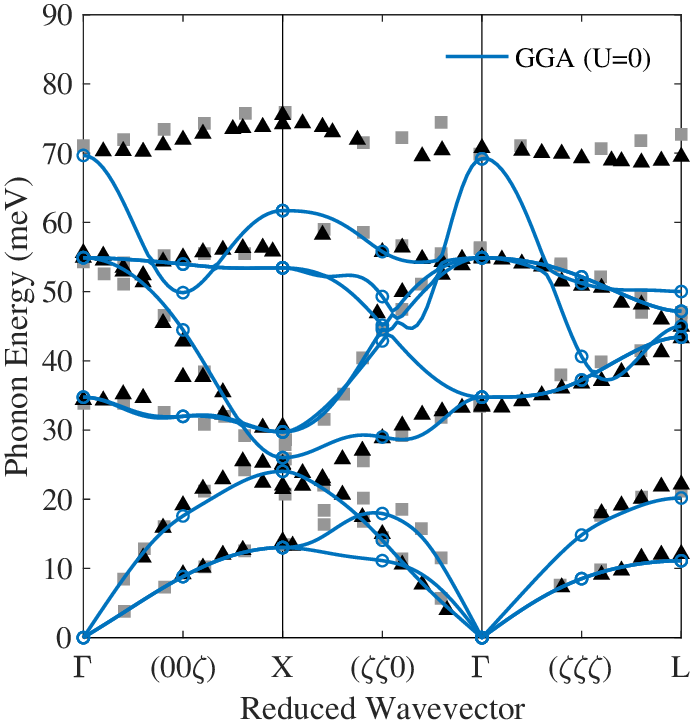}
\caption{\label{fig:u0phonon} Phonon dispersion of FM UO$_2$, calculated by GGA (i.e., $U=0$ without SOC), compared with inelastic neutron
scattering data from Ref. \cite{pang_phonon_2013} at $T=300$ K (grey squares) and Ref. \cite{zhouCapturingGroundState2022a} at $T=600$ K (black triangles). The hollow points were directly computed using DFT, while the 
corresponding lines are Fourier interpolations.}
\end{figure}

It is interesting to explore the thermal conductivity using GGA (i.e., $U=0$) in the FM state, despite the 
fact that it is incorrectly predicted to be a metal (though we still apply LO-TO splitting). In the Supplemental Materials of our previous work \cite{zhouCapturingGroundState2022a}, the phonon dispersion computed by GGA using  $\hat{\mathbf{S}}_{BZ}=2\hat{\mathbf{1}}$ is reported. Here the supercell is increased to $\hat{\mathbf{S}}_{BZ}=4\hat{\mathbf{1}}$ (see Fig. ~\ref{fig:u0phonon}). Notably, comparing with experiments, the highest acoustic phonon branch is slightly underestimated, and the highest optical branch is largely underestimated and has large oscillations. These anomalies clearly show that $U=0$ is not adequate to accurately capture the phonons of UO$_2$.

\begin{figure}[h]
\includegraphics[width=0.39\columnwidth]{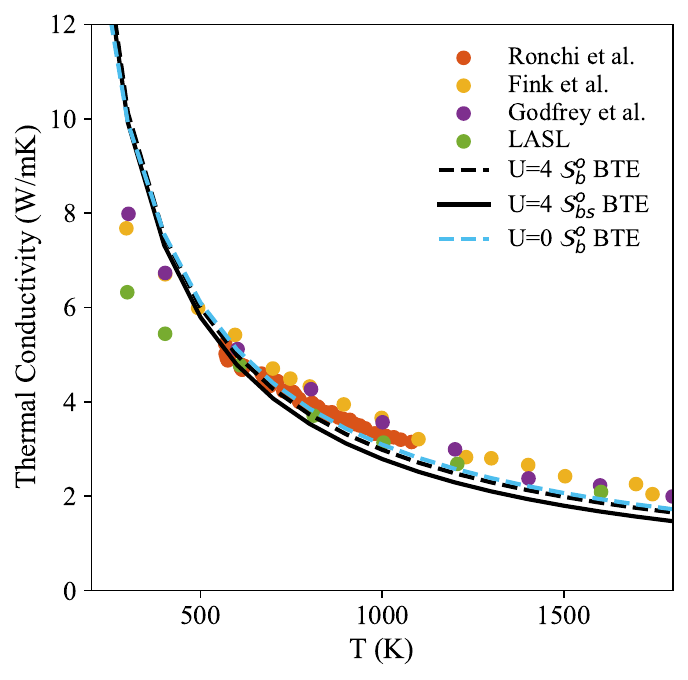}
\caption{\label{fig:u0ther_cond} Thermal conductivity of UO$_2$, computed
using GGA+$U$+SOC ($U=4$ eV) and GGA ($U=0$ without SOC), and comparing with experiments \cite{finkThermophysicalPropertiesUranium2000, batesVisibleInfraredAbsorption1965, godfreyThermalConductivityUranium1965, ronchiEffectBurnupThermal2004}. 
}
\end{figure}

\begin{figure}[h]
\includegraphics[width=0.45\columnwidth]{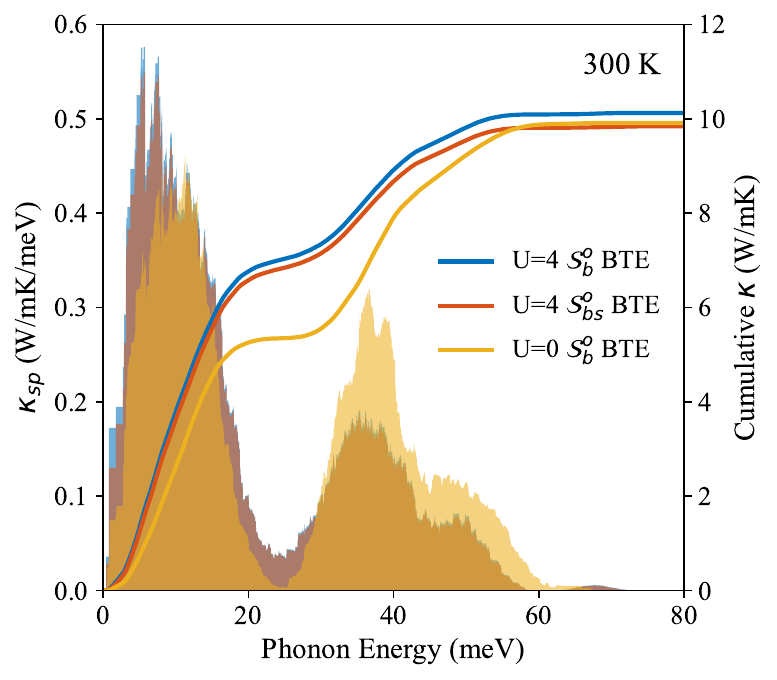}
\caption{\label{fig:u0mode_cond} Spectral and cumulative thermal conductivity
of UO$_2$ as functions of phonon energy at $T=300$K, computed using
GGA+$U$+SOC ($U=4$ eV) and GGA ($U=0$ without SOC).
}
\end{figure}

The third order IDs are computed using supercell $\hat{\mathbf{S}}_{BZ}=\hat{\mathbf{S}}_{O}$, and the thermal conductivity is computed using the RTA with  $\mathcal{S}_b^o$. Surprisingly, the thermal conductivity computed by GGA is very close to GGA+$U$ (see Fig. ~\ref{fig:u0ther_cond}). However, a significant difference between GGA and GGA+$U$ is observed in both the spectral and cumulative thermal conductivity at $T=300$ K (see Fig. ~\ref{fig:u0mode_cond}). For both $\mathcal{S}_b^o$ and $\mathcal{S}_{bs}^o$ with GGA+$U$, the optical phonons have a substantial contribution, quantified by phonon energies greater than 24 meV, accounting for \textasciitilde30\% of the total thermal conductivity, while the corresponding contribution is \textasciitilde50\% in GGA. Though the prediction of thermal conductivity using GGA at first seems accurate, it is achieved due to cancelling errors, rather than accurate descriptions of phonons and phonon interactions.

\clearpage

\section{\label{sec:smhightemp}Thermal conductivity contribution of electrons and photons}

\begin{figure}[h]
\includegraphics[width=0.4\columnwidth]{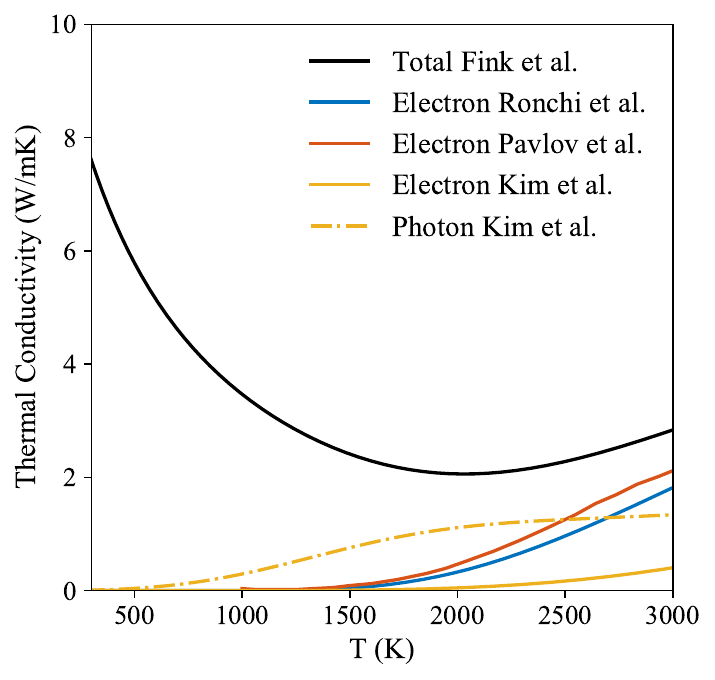}
\caption{\label{fig:smelectronphoton} Modeled electron and photon thermal conductivity in UO$_2$ by Ref. \cite{ronchiThermalConductivityUranium1999a,  kimLatticeThermalConductivity2014, pavlovMeasurementInterpretationThermophysical2017}, comparing to fitted total thermal conductivity by Ref. \cite{finkThermophysicalPropertiesUranium2000}}
\end{figure}

In experiments above room temperature, thermal conductivity of UO$_2$ decreases with increasing temperature until around $T=2000$K, then increases until the melting temperature \cite{finkThermophysicalPropertiesUranium2000}. As phonon thermal conductivity due to phonon-phonon interactions normally only decreases with increasing temperature, the increasing thermal conductivity at high temperature must be produced by other mechanisms. The potentially observable contributions from photon and electron (i.e., polaron, a thermally activated quasiparticle) are suggested \cite{international1965thermodynamic}. Semi-classical models have been applied to the thermal conductivity of UO$_2$, accounting for contributions of polarons \cite{ronchiThermalConductivityUranium1999a,  kimLatticeThermalConductivity2014, pavlovMeasurementInterpretationThermophysical2017} and photons \cite{kimLatticeThermalConductivity2014}, and their results are collected (see Fig. ~\ref{fig:smelectronphoton}). The total thermal conductivity is presented by Ref. \cite{finkThermophysicalPropertiesUranium2000}, fitting to experiments. Above $T=2000$ K, Ref. \cite{ronchiThermalConductivityUranium1999a,   pavlovMeasurementInterpretationThermophysical2017} claim the polaron contribution is significant, meanwhile \cite{kimLatticeThermalConductivity2014} claims the photon contribution is critical. A common understanding has not yet emerged, and modeling beyond the semi-classical theory level is still lacking. 

\clearpage

\section{\label{sec:smids}Computed irreducible derivatives}
This section contains all computed irreducible derivatives, and all information required to specify the irreducible derivatives.

\subsection{The labels of $q$ points and their standard displacement basis}


\begin{longtable*}{cccccccc}
\caption{The labels of $q$ points.} 
\\
\hline\hline\\[-0.8em]
\endhead
$\Gamma  = \left( 0 0 0\right) $ & 
\\[0.8em] 
$\Lambda = \left( \frac{1}{4} 0 0\right) $ & 
$\bar{\Lambda} = \left( \frac{3}{4} 0 0\right) $ & 
$\Lambda_{1} = \left( 0 \frac{1}{4} 0\right) $ & 
$\bar{\Lambda}_{1} = \left( 0 \frac{3}{4} 0\right) $ & 
$\Lambda_{2} = \left( 0 0 \frac{1}{4}\right) $ & 
$\bar{\Lambda}_{2} = \left( 0 0 \frac{3}{4}\right) $\\[0.8em] 
$\Lambda_{3} = \left( \frac{1}{4} \frac{1}{4} \frac{1}{4}\right) $ & 
$\bar{\Lambda}_{3} = \left( \frac{3}{4} \frac{3}{4} \frac{3}{4}\right) $ & 
\\[0.8em] 
$L  = \left( \frac{1}{2} 0 0\right) $ & 
$L_{1}  = \left( 0 \frac{1}{2} 0\right) $ & 
$L_{2}  = \left( 0 0 \frac{1}{2}\right) $ & 
$L_{3}  = \left( \frac{1}{2} \frac{1}{2} \frac{1}{2}\right) $ & 
\\[0.8em] 
$\Delta = \left( \frac{1}{4} \frac{1}{4} 0\right) $ & 
$\bar{\Delta} = \left( \frac{3}{4} \frac{3}{4} 0\right) $ & 
$\Delta_{1} = \left( \frac{1}{4} 0 \frac{1}{4}\right) $ & 
$\bar{\Delta}_{1} = \left( \frac{3}{4} 0 \frac{3}{4}\right) $ & 
$\Delta_{2} = \left( 0 \frac{1}{4} \frac{1}{4}\right) $ & 
$\bar{\Delta}_{2} = \left( 0 \frac{3}{4} \frac{3}{4}\right) $\\[0.8em] 
$B = \left( \frac{1}{2} \frac{1}{4} 0\right) $ & 
$\bar{B} = \left( \frac{1}{2} \frac{3}{4} 0\right) $ & 
$B_{1} = \left( \frac{1}{4} \frac{1}{2} 0\right) $ & 
$\bar{B}_{1} = \left( \frac{3}{4} \frac{1}{2} 0\right) $ & 
$B_{2} = \left( \frac{1}{2} 0 \frac{1}{4}\right) $ & 
$\bar{B}_{2} = \left( \frac{1}{2} 0 \frac{3}{4}\right) $\\[0.8em] 
$B_{3} = \left( \frac{3}{4} \frac{1}{4} \frac{1}{4}\right) $ & 
$\bar{B}_{3} = \left( \frac{1}{4} \frac{3}{4} \frac{3}{4}\right) $ & 
$B_{4} = \left( 0 \frac{1}{2} \frac{1}{4}\right) $ & 
$\bar{B}_{4} = \left( 0 \frac{1}{2} \frac{3}{4}\right) $ & 
$B_{5} = \left( \frac{1}{2} \frac{1}{2} \frac{1}{4}\right) $ & 
$\bar{B}_{5} = \left( \frac{1}{2} \frac{1}{2} \frac{3}{4}\right) $\\
\end{longtable*}
\normalsize
\clearpage
\begin{longtable*}{l|ccccccccc}
\caption{Symmetrized displacement amplitudes in terms of the standard displacement basis for $q$ points in UO$_2$} \\
\hline\hline\\[-0.8em]
& $u_{\textrm{U}}^x$ & $u_{\textrm{U}}^y$  & $u_{\textrm{U}}^z$ 
& $u_{\textrm{O}}^x$ & $u_{\textrm{O}}^y$  & $u_{\textrm{O}}^z$ 
& $u_{\textrm{O}}^x$ & $u_{\textrm{O}}^y$  & $u_{\textrm{O}}^z$ \\ 
\hline
\\[\dimexpr-\normalbaselineskip+0.4em]
\endfirsthead
\hline\hline\\[-0.8em]
& $u_{\textrm{U}}^x$ & $u_{\textrm{U}}^y$  & $u_{\textrm{U}}^z$ 
& $u_{\textrm{O}}^x$ & $u_{\textrm{O}}^y$  & $u_{\textrm{O}}^z$ 
& $u_{\textrm{O}}^x$ & $u_{\textrm{O}}^y$  & $u_{\textrm{O}}^z$ \\ 
\hline
\\[\dimexpr-\normalbaselineskip+0.4em]
\endhead
\hline\hline
\endfoot
${u_{\Gamma }^{T_{2g}^{}, {0}}}$ & $  0 $ & $  0 $ & $  0 $ & $  0 $ & $  0 $ & $  \frac{ \sqrt{2}}{2} $ & $  0 $ & $  0 $ & $ -\frac{ \sqrt{2}}{2} $ \\[0.4em]
${u_{\Gamma }^{T_{2g}^{}, {1}}}$ & $  0 $ & $  0 $ & $  0 $ & $  \frac{ \sqrt{2}}{2} $ & $  0 $ & $  0 $ & $ -\frac{ \sqrt{2}}{2} $ & $  0 $ & $  0 $ \\[0.4em]
${u_{\Gamma }^{T_{2g}^{}, {2}}}$ & $  0 $ & $  0 $ & $  0 $ & $  0 $ & $  \frac{ \sqrt{2}}{2} $ & $  0 $ & $  0 $ & $ -\frac{ \sqrt{2}}{2} $ & $  0 $ \\[0.4em]
${u_{\Gamma }^{T_{1u}^{}, {0}}}$ & $ -\frac{ \sqrt{6}}{3} $ & $  0 $ & $  0 $ & $  \frac{ \sqrt{6}}{6} $ & $  0 $ & $  0 $ & $  \frac{ \sqrt{6}}{6} $ & $  0 $ & $  0 $ \\[0.4em]
${u_{\Gamma }^{T_{1u}^{}, {1}}}$ & $  0 $ & $ -\frac{ \sqrt{6}}{3} $ & $  0 $ & $  0 $ & $  \frac{ \sqrt{6}}{6} $ & $  0 $ & $  0 $ & $  \frac{ \sqrt{6}}{6} $ & $  0 $ \\[0.4em]
${u_{\Gamma }^{T_{1u}^{}, {2}}}$ & $  0 $ & $  0 $ & $ -\frac{ \sqrt{6}}{3} $ & $  0 $ & $  0 $ & $  \frac{ \sqrt{6}}{6} $ & $  0 $ & $  0 $ & $  \frac{ \sqrt{6}}{6} $ \\[0.4em]
${u_{\Lambda}^{A_{1}^{}, {0}}}$ & $ -\frac{ \sqrt{3}}{3}i $ & $  \frac{ \sqrt{3}}{3}i $ & $  \frac{ \sqrt{3}}{3}i $ & $  0 $ & $  0 $ & $  0 $ & $  0 $ & $  0 $ & $  0 $ \\[0.4em]
${u_{\Lambda}^{\tensor*[^{1}]{\hspace{-0.2em}A}{}_{1}^{}, {0}}}$ & $  0 $ & $  0 $ & $  0 $ & $ -\frac{ \sqrt{3}}{6}-\frac{ \sqrt{3}}{6}i $ & $  \frac{ \sqrt{3}}{6}+\frac{ \sqrt{3}}{6}i $ & $  \frac{ \sqrt{3}}{6}+\frac{ \sqrt{3}}{6}i $ & $ -\frac{ \sqrt{3}}{6}-\frac{ \sqrt{3}}{6}i $ & $  \frac{ \sqrt{3}}{6}+\frac{ \sqrt{3}}{6}i $ & $  \frac{ \sqrt{3}}{6}+\frac{ \sqrt{3}}{6}i $ \\
\hline
\end{longtable*}

\subsection{The irreducible derivatives computed at $a=$ 5.5461 \AA, corresponding to $T=0$ K}

\begin{longtable*}{lll|ll|ll|ll}
\caption{Second order irreducible derivatives for
UO$_2$ computed at $a=$ 5.5461 \AA. The units are eV/\AA$^2$.} \\
\hline\hline\\[-1em]
& Derivative & Value & Derivative & Value & Derivative & Value & Derivative & Value \\[0.4em]
\hline
\\[\dimexpr-\normalbaselineskip+0.2em]
\endfirsthead
\hline\hline\\[-1em]
& Derivative & Value & Derivative & Value & Derivative & Value & Derivative & Value \\[0.4em]
\hline
\\[\dimexpr-\normalbaselineskip+0.2em]
\endhead
\hline\hline
\endfoot
& $ d^{T_{2g}^{}T_{2g}^{}}_{\Gamma \Gamma } $ & $  11.639047 $& $ d^{T_{1u}^{}T_{1u}^{}}_{\Gamma \Gamma } $ & $  10.913052 $& $ d^{A_{1g}^{}A_{1g}^{}}_{L L } $ & $  16.114994 $& $ d^{E_{g}^{}E_{g}^{}}_{L L } $ & $  8.693366 $\\[0.4em] 

& $ d^{A_{2u}^{}A_{2u}^{}}_{L L } $ & $  30.499107 $& $ d^{A_{2u}^{}\tensor*[^{1}]{\hspace{-0.2em}A}{}_{2u}^{}}_{L L } $ & $  6.627570 $& $ d^{\tensor*[^{1}]{\hspace{-0.2em}A}{}_{2u}^{}\tensor*[^{1}]{\hspace{-0.2em}A}{}_{2u}^{}}_{L L } $ & $  8.720557 $& $ d^{E_{u}^{}E_{u}^{}}_{L L } $ & $  7.929498 $\\[0.4em] 

& $ d^{E_{u}^{}\tensor*[^{1}]{\hspace{-0.2em}E}{}_{u}^{}}_{L L } $ & $ -2.621590 $& $ d^{\tensor*[^{1}]{\hspace{-0.2em}E}{}_{u}^{}\tensor*[^{1}]{\hspace{-0.2em}E}{}_{u}^{}}_{L L } $ & $  6.885507 $& $ d^{A_{1}^{}A_{1}^{}}_{\Delta\bar{\Delta}} $ & $  33.905151 $& $ d^{\tensor*[^{1}]{\hspace{-0.2em}A}{}_{1}^{}A_{1}^{}}_{\Delta\bar{\Delta}} $ & $  14.878378 $\\[0.4em] 

& $ d^{\tensor*[^{1}]{\hspace{-0.2em}A}{}_{1}^{}\tensor*[^{1}]{\hspace{-0.2em}A}{}_{1}^{}}_{\Delta\bar{\Delta}} $ & $  17.470587 $& $ d^{B_{2}^{}B_{2}^{}}_{\Delta\bar{\Delta}} $ & $  6.978432 $& $ d^{E_{}^{}E_{}^{}}_{\Delta\bar{\Delta}} $ & $  8.300181 $& $ d^{\tensor*[^{1}]{\hspace{-0.2em}E}{}_{}^{}E_{}^{}}_{\Delta\bar{\Delta}} $ & $ -3.576226 $\\[0.4em] 

& $ d^{\tensor*[^{2}]{\hspace{-0.2em}E}{}_{}^{}E_{}^{}}_{\Delta\bar{\Delta}} $ & $  0.337451 $& $ d^{\tensor*[^{1}]{\hspace{-0.2em}E}{}_{}^{}\tensor*[^{1}]{\hspace{-0.2em}E}{}_{}^{}}_{\Delta\bar{\Delta}} $ & $  3.552432 $& $ d^{\tensor*[^{2}]{\hspace{-0.2em}E}{}_{}^{}\tensor*[^{1}]{\hspace{-0.2em}E}{}_{}^{}}_{\Delta\bar{\Delta}} $ & $  0.887305 $& $ d^{\tensor*[^{2}]{\hspace{-0.2em}E}{}_{}^{}\tensor*[^{2}]{\hspace{-0.2em}E}{}_{}^{}}_{\Delta\bar{\Delta}} $ & $  11.411557 $\\
\hline
\end{longtable*}
\begin{longtable*}{lll|ll|ll|ll}
\caption{Third order irreducible derivatives for
UO$_2$ computed at $a=$ 5.5461 \AA. The units are eV/\AA$^3$.} \\
\hline\hline\\[-1em]
& Derivative & Value & Derivative & Value & Derivative & Value & Derivative & Value \\[0.4em]
\hline
\\[\dimexpr-\normalbaselineskip+0.2em]
\endfirsthead
\hline\hline\\[-1em]
& Derivative & Value & Derivative & Value & Derivative & Value & Derivative & Value \\[0.4em]
\hline
\\[\dimexpr-\normalbaselineskip+0.2em]
\endhead
\hline\hline
\endfoot
& $ d^{T_{2g}^{}T_{2g}^{}T_{2g}^{}}_{\Gamma \Gamma \Gamma } $ & $ -65.664187 $& $ d^{T_{1u}^{}T_{1u}^{}T_{2g}^{}}_{\Gamma \Gamma \Gamma } $ & $ -54.859204 $& $ d^{T_{2g}^{}A_{1}^{}A_{1}^{}}_{\Gamma \bar{A}A} $ & $  31.426600 $& $ d^{T_{2g}^{}A_{1}^{}\tensor*[^{1}]{\hspace{-0.2em}A}{}_{1}^{}}_{\Gamma \bar{A}A} $ & $ -8.202863 $\\[0.4em] 

& $ d^{T_{1u}^{}A_{1}^{}\tensor*[^{1}]{\hspace{-0.2em}A}{}_{1}^{}}_{\Gamma \bar{A}A} $ & $  2.079260i $& $ d^{T_{2g}^{}A_{1}^{}\tensor*[^{2}]{\hspace{-0.2em}A}{}_{1}^{}}_{\Gamma \bar{A}A} $ & $  6.222260 $& $ d^{T_{1u}^{}A_{1}^{}\tensor*[^{2}]{\hspace{-0.2em}A}{}_{1}^{}}_{\Gamma \bar{A}A} $ & $  34.714584i $& $ d^{T_{2g}^{}A_{1}^{}A_{2}^{}}_{\Gamma \bar{A}A} $ & $  3.760041i $\\[0.4em] 

& $ d^{T_{2g}^{}A_{1}^{}B_{1}^{}}_{\Gamma \bar{A}A} $ & $ -3.463657i $& $ d^{T_{1u}^{}A_{1}^{}B_{1}^{}}_{\Gamma \bar{A}A} $ & $ -20.947786 $& $ d^{T_{2g}^{}A_{1}^{}\tensor*[^{1}]{\hspace{-0.2em}B}{}_{1}^{}}_{\Gamma \bar{A}A} $ & $  21.622804i $& $ d^{T_{1u}^{}A_{1}^{}\tensor*[^{1}]{\hspace{-0.2em}B}{}_{1}^{}}_{\Gamma \bar{A}A} $ & $ -6.655611 $\\[0.4em] 

& $ d^{T_{2g}^{}A_{1}^{}\tensor*[^{2}]{\hspace{-0.2em}B}{}_{1}^{}}_{\Gamma \bar{A}A} $ & $  16.447136i $& $ d^{T_{1u}^{}A_{1}^{}\tensor*[^{2}]{\hspace{-0.2em}B}{}_{1}^{}}_{\Gamma \bar{A}A} $ & $ -5.080140 $& $ d^{T_{1u}^{}A_{1}^{}B_{2}^{}}_{\Gamma \bar{A}A} $ & $ -10.954692i $& $ d^{T_{1u}^{}A_{1}^{}\tensor*[^{1}]{\hspace{-0.2em}B}{}_{2}^{}}_{\Gamma \bar{A}A} $ & $  4.202561i $\\[0.4em] 

& $ d^{T_{2g}^{}\tensor*[^{1}]{\hspace{-0.2em}A}{}_{1}^{}\tensor*[^{1}]{\hspace{-0.2em}A}{}_{1}^{}}_{\Gamma \bar{A}A} $ & $ -2.588961 $& $ d^{T_{2g}^{}\tensor*[^{1}]{\hspace{-0.2em}A}{}_{1}^{}\tensor*[^{2}]{\hspace{-0.2em}A}{}_{1}^{}}_{\Gamma \bar{A}A} $ & $ -4.364649 $& $ d^{T_{1u}^{}\tensor*[^{1}]{\hspace{-0.2em}A}{}_{1}^{}\tensor*[^{2}]{\hspace{-0.2em}A}{}_{1}^{}}_{\Gamma \bar{A}A} $ & $ -20.523154i $& $ d^{T_{2g}^{}\tensor*[^{1}]{\hspace{-0.2em}A}{}_{1}^{}A_{2}^{}}_{\Gamma \bar{A}A} $ & $  8.508160i $\\
\hline
\end{longtable*}
\begin{longtable*}{lll|ll|ll|ll}
\caption{Fourth order irreducible derivatives for
UO$_2$ computed at $a=$ 5.5461 \AA. The units are eV/\AA$^4$.} \\
\hline\hline\\[-1em]
& Derivative & Value & Derivative & Value & Derivative & Value & Derivative & Value \\[0.4em]
\hline
\\[\dimexpr-\normalbaselineskip+0.2em]
\endfirsthead
\hline\hline\\[-1em]
& Derivative & Value & Derivative & Value & Derivative & Value & Derivative & Value \\[0.4em]
\hline
\\[\dimexpr-\normalbaselineskip+0.2em]
\endhead
\hline\hline
\endfoot
& $ d^{T_{2g}^{}T_{2g}^{}T_{2g}^{}T_{2g}^{}}_{\Gamma \Gamma \Gamma \Gamma } $ & $  227.370114 $& $ ^{1}d^{T_{2g}^{}T_{2g}^{}T_{2g}^{}T_{2g}^{}}_{\Gamma \Gamma \Gamma \Gamma } $ & $ -125.546164 $& $ d^{T_{1u}^{}T_{1u}^{}T_{2g}^{}T_{2g}^{}}_{\Gamma \Gamma \Gamma \Gamma } $ & $  18.321959 $& $ ^{1}d^{T_{1u}^{}T_{1u}^{}T_{2g}^{}T_{2g}^{}}_{\Gamma \Gamma \Gamma \Gamma } $ & $  66.519027 $\\[0.4em] 

& $ ^{2}d^{T_{1u}^{}T_{1u}^{}T_{2g}^{}T_{2g}^{}}_{\Gamma \Gamma \Gamma \Gamma } $ & $  84.798437 $& $ d^{T_{1u}^{}T_{1u}^{}T_{1u}^{}T_{1u}^{}}_{\Gamma \Gamma \Gamma \Gamma } $ & $ -49.043355 $& $ ^{1}d^{T_{1u}^{}T_{1u}^{}T_{1u}^{}T_{1u}^{}}_{\Gamma \Gamma \Gamma \Gamma } $ & $  726.994669 $& $ d^{T_{2g}^{}T_{2g}^{}A_{1g}^{}A_{1g}^{}}_{\Gamma \Gamma X X } $ & $  14.402123 $\\[0.4em] 

& $ ^{1}d^{T_{2g}^{}T_{2g}^{}A_{1g}^{}A_{1g}^{}}_{\Gamma \Gamma X X } $ & $  120.078631 $& $ d^{T_{1u}^{}T_{1u}^{}A_{1g}^{}A_{1g}^{}}_{\Gamma \Gamma X X } $ & $ -9.151089 $& $ ^{1}d^{T_{1u}^{}T_{1u}^{}A_{1g}^{}A_{1g}^{}}_{\Gamma \Gamma X X } $ & $  87.278954 $& $ d^{T_{2g}^{}T_{2g}^{}A_{1g}^{}E_{g}^{}}_{\Gamma \Gamma X X } $ & $  2.329857 $\\[0.4em] 

& $ d^{T_{1u}^{}T_{1u}^{}A_{1g}^{}E_{g}^{}}_{\Gamma \Gamma X X } $ & $  112.631276 $& $ d^{T_{1u}^{}T_{2g}^{}A_{1g}^{}\tensor*[^{2}]{\hspace{-0.2em}A}{}_{2u}^{}}_{\Gamma \Gamma X X } $ & $ -69.953506 $& $ d^{T_{1u}^{}T_{2g}^{}A_{1g}^{}B_{1u}^{}}_{\Gamma \Gamma X X } $ & $ -27.598335 $& $ ^{1}d^{T_{1u}^{}T_{2g}^{}A_{1g}^{}B_{1u}^{}}_{\Gamma \Gamma X X } $ & $  32.405873 $\\[0.4em] 

& $ d^{T_{1u}^{}T_{2g}^{}A_{1g}^{}E_{u}^{}}_{\Gamma \Gamma X X } $ & $  45.342299 $& $ ^{1}d^{T_{1u}^{}T_{2g}^{}A_{1g}^{}E_{u}^{}}_{\Gamma \Gamma X X } $ & $  42.391718 $& $ d^{T_{1u}^{}T_{2g}^{}A_{1g}^{}\tensor*[^{1}]{\hspace{-0.2em}E}{}_{u}^{}}_{\Gamma \Gamma X X } $ & $ -63.493913 $& $ ^{1}d^{T_{1u}^{}T_{2g}^{}A_{1g}^{}\tensor*[^{1}]{\hspace{-0.2em}E}{}_{u}^{}}_{\Gamma \Gamma X X } $ & $ -64.947117 $\\
\hline
\end{longtable*}

\subsection{The irreducible derivatives computed at $a=$ 5.5602 \AA, corresponding to $T=360$ K}

\begin{longtable*}{lll|ll|ll|ll}
\caption{Second order irreducible derivatives for
UO$_2$ computed at $a=$ 5.5602 \AA. The units are eV/\AA$^2$.} \\
\hline\hline\\[-1em]
& Derivative & Value & Derivative & Value & Derivative & Value & Derivative & Value \\[0.4em]
\hline
\\[\dimexpr-\normalbaselineskip+0.2em]
\endfirsthead
\hline\hline\\[-1em]
& Derivative & Value & Derivative & Value & Derivative & Value & Derivative & Value \\[0.4em]
\hline
\\[\dimexpr-\normalbaselineskip+0.2em]
\endhead
\hline\hline
\endfoot
& $ d^{T_{2g}^{}T_{2g}^{}}_{\Gamma \Gamma } $ & $  11.203346 $& $ d^{T_{1u}^{}T_{1u}^{}}_{\Gamma \Gamma } $ & $  10.316931 $& $ d^{A_{1g}^{}A_{1g}^{}}_{L L } $ & $  15.803315 $& $ d^{E_{g}^{}E_{g}^{}}_{L L } $ & $  8.418678 $\\[0.4em] 

& $ d^{A_{2u}^{}A_{2u}^{}}_{L L } $ & $  29.862475 $& $ d^{A_{2u}^{}\tensor*[^{1}]{\hspace{-0.2em}A}{}_{2u}^{}}_{L L } $ & $  6.812328 $& $ d^{\tensor*[^{1}]{\hspace{-0.2em}A}{}_{2u}^{}\tensor*[^{1}]{\hspace{-0.2em}A}{}_{2u}^{}}_{L L } $ & $  8.506553 $& $ d^{E_{u}^{}E_{u}^{}}_{L L } $ & $  7.522995 $\\[0.4em] 

& $ d^{E_{u}^{}\tensor*[^{1}]{\hspace{-0.2em}E}{}_{u}^{}}_{L L } $ & $ -2.342394 $& $ d^{\tensor*[^{1}]{\hspace{-0.2em}E}{}_{u}^{}\tensor*[^{1}]{\hspace{-0.2em}E}{}_{u}^{}}_{L L } $ & $  6.676590 $& $ d^{A_{1}^{}A_{1}^{}}_{\Delta\bar{\Delta}} $ & $  33.320266 $& $ d^{\tensor*[^{1}]{\hspace{-0.2em}A}{}_{1}^{}A_{1}^{}}_{\Delta\bar{\Delta}} $ & $  14.635534 $\\[0.4em] 

& $ d^{\tensor*[^{1}]{\hspace{-0.2em}A}{}_{1}^{}\tensor*[^{1}]{\hspace{-0.2em}A}{}_{1}^{}}_{\Delta\bar{\Delta}} $ & $  17.191623 $& $ d^{B_{2}^{}B_{2}^{}}_{\Delta\bar{\Delta}} $ & $  6.741835 $& $ d^{E_{}^{}E_{}^{}}_{\Delta\bar{\Delta}} $ & $  7.917001 $& $ d^{\tensor*[^{1}]{\hspace{-0.2em}E}{}_{}^{}E_{}^{}}_{\Delta\bar{\Delta}} $ & $ -3.432796 $\\[0.4em] 

& $ d^{\tensor*[^{2}]{\hspace{-0.2em}E}{}_{}^{}E_{}^{}}_{\Delta\bar{\Delta}} $ & $  0.130189 $& $ d^{\tensor*[^{1}]{\hspace{-0.2em}E}{}_{}^{}\tensor*[^{1}]{\hspace{-0.2em}E}{}_{}^{}}_{\Delta\bar{\Delta}} $ & $  3.394426 $& $ d^{\tensor*[^{2}]{\hspace{-0.2em}E}{}_{}^{}\tensor*[^{1}]{\hspace{-0.2em}E}{}_{}^{}}_{\Delta\bar{\Delta}} $ & $  0.867236 $& $ d^{\tensor*[^{2}]{\hspace{-0.2em}E}{}_{}^{}\tensor*[^{2}]{\hspace{-0.2em}E}{}_{}^{}}_{\Delta\bar{\Delta}} $ & $  11.141199 $\\
\hline
\end{longtable*}
\begin{longtable*}{lll|ll|ll|ll}
\caption{Third order irreducible derivatives for
UO$_2$ computed at $a=$ 5.5602 \AA. The units are eV/\AA$^3$.} \\
\hline\hline\\[-1em]
& Derivative & Value & Derivative & Value & Derivative & Value & Derivative & Value \\[0.4em]
\hline
\\[\dimexpr-\normalbaselineskip+0.2em]
\endfirsthead
\hline\hline\\[-1em]
& Derivative & Value & Derivative & Value & Derivative & Value & Derivative & Value \\[0.4em]
\hline
\\[\dimexpr-\normalbaselineskip+0.2em]
\endhead
\hline\hline
\endfoot
& $ d^{T_{2g}^{}T_{2g}^{}T_{2g}^{}}_{\Gamma \Gamma \Gamma } $ & $ -41.546756 $& $ d^{T_{1u}^{}T_{1u}^{}T_{2g}^{}}_{\Gamma \Gamma \Gamma } $ & $ -50.936806 $& $ d^{T_{2g}^{}A_{1}^{}A_{1}^{}}_{\Gamma \bar{A}A} $ & $  31.106155 $& $ d^{T_{2g}^{}A_{1}^{}\tensor*[^{1}]{\hspace{-0.2em}A}{}_{1}^{}}_{\Gamma \bar{A}A} $ & $ -8.091566 $\\[0.4em] 

& $ d^{T_{1u}^{}A_{1}^{}\tensor*[^{1}]{\hspace{-0.2em}A}{}_{1}^{}}_{\Gamma \bar{A}A} $ & $  1.914188i $& $ d^{T_{2g}^{}A_{1}^{}\tensor*[^{2}]{\hspace{-0.2em}A}{}_{1}^{}}_{\Gamma \bar{A}A} $ & $  5.525065 $& $ d^{T_{1u}^{}A_{1}^{}\tensor*[^{2}]{\hspace{-0.2em}A}{}_{1}^{}}_{\Gamma \bar{A}A} $ & $  34.302036i $& $ d^{T_{2g}^{}A_{1}^{}A_{2}^{}}_{\Gamma \bar{A}A} $ & $  3.963269i $\\[0.4em] 

& $ d^{T_{2g}^{}A_{1}^{}B_{1}^{}}_{\Gamma \bar{A}A} $ & $ -2.846732i $& $ d^{T_{1u}^{}A_{1}^{}B_{1}^{}}_{\Gamma \bar{A}A} $ & $ -20.277643 $& $ d^{T_{2g}^{}A_{1}^{}\tensor*[^{1}]{\hspace{-0.2em}B}{}_{1}^{}}_{\Gamma \bar{A}A} $ & $  20.882804i $& $ d^{T_{1u}^{}A_{1}^{}\tensor*[^{1}]{\hspace{-0.2em}B}{}_{1}^{}}_{\Gamma \bar{A}A} $ & $ -7.387762 $\\[0.4em] 

& $ d^{T_{2g}^{}A_{1}^{}\tensor*[^{2}]{\hspace{-0.2em}B}{}_{1}^{}}_{\Gamma \bar{A}A} $ & $  15.518296i $& $ d^{T_{1u}^{}A_{1}^{}\tensor*[^{2}]{\hspace{-0.2em}B}{}_{1}^{}}_{\Gamma \bar{A}A} $ & $ -3.488960 $& $ d^{T_{1u}^{}A_{1}^{}B_{2}^{}}_{\Gamma \bar{A}A} $ & $ -11.050003i $& $ d^{T_{1u}^{}A_{1}^{}\tensor*[^{1}]{\hspace{-0.2em}B}{}_{2}^{}}_{\Gamma \bar{A}A} $ & $  4.429284i $\\[0.4em] 

& $ d^{T_{2g}^{}\tensor*[^{1}]{\hspace{-0.2em}A}{}_{1}^{}\tensor*[^{1}]{\hspace{-0.2em}A}{}_{1}^{}}_{\Gamma \bar{A}A} $ & $ -2.226293 $& $ d^{T_{2g}^{}\tensor*[^{1}]{\hspace{-0.2em}A}{}_{1}^{}\tensor*[^{2}]{\hspace{-0.2em}A}{}_{1}^{}}_{\Gamma \bar{A}A} $ & $ -4.029561 $& $ d^{T_{1u}^{}\tensor*[^{1}]{\hspace{-0.2em}A}{}_{1}^{}\tensor*[^{2}]{\hspace{-0.2em}A}{}_{1}^{}}_{\Gamma \bar{A}A} $ & $ -19.684185i $& $ d^{T_{2g}^{}\tensor*[^{1}]{\hspace{-0.2em}A}{}_{1}^{}A_{2}^{}}_{\Gamma \bar{A}A} $ & $  8.348201i $\\
\hline
\end{longtable*}
\begin{longtable*}{lll|ll|ll|ll}
\caption{Fourth order irreducible derivatives for
UO$_2$ computed at $a=$ 5.5602 \AA. The units are eV/\AA$^4$.} \\
\hline\hline\\[-1em]
& Derivative & Value & Derivative & Value & Derivative & Value & Derivative & Value \\[0.4em]
\hline
\\[\dimexpr-\normalbaselineskip+0.2em]
\endfirsthead
\hline\hline\\[-1em]
& Derivative & Value & Derivative & Value & Derivative & Value & Derivative & Value \\[0.4em]
\hline
\\[\dimexpr-\normalbaselineskip+0.2em]
\endhead
\hline\hline
\endfoot
& $ d^{T_{2g}^{}T_{2g}^{}T_{2g}^{}T_{2g}^{}}_{\Gamma \Gamma \Gamma \Gamma } $ & $  204.066428 $& $ ^{1}d^{T_{2g}^{}T_{2g}^{}T_{2g}^{}T_{2g}^{}}_{\Gamma \Gamma \Gamma \Gamma } $ & $ -163.708334 $& $ d^{T_{1u}^{}T_{1u}^{}T_{2g}^{}T_{2g}^{}}_{\Gamma \Gamma \Gamma \Gamma } $ & $  24.039865 $& $ ^{1}d^{T_{1u}^{}T_{1u}^{}T_{2g}^{}T_{2g}^{}}_{\Gamma \Gamma \Gamma \Gamma } $ & $  18.248430 $\\[0.4em] 

& $ ^{2}d^{T_{1u}^{}T_{1u}^{}T_{2g}^{}T_{2g}^{}}_{\Gamma \Gamma \Gamma \Gamma } $ & $  103.903736 $& $ d^{T_{1u}^{}T_{1u}^{}T_{1u}^{}T_{1u}^{}}_{\Gamma \Gamma \Gamma \Gamma } $ & $  127.276176 $& $ ^{1}d^{T_{1u}^{}T_{1u}^{}T_{1u}^{}T_{1u}^{}}_{\Gamma \Gamma \Gamma \Gamma } $ & $  601.839187 $& $ d^{T_{2g}^{}T_{2g}^{}A_{1g}^{}A_{1g}^{}}_{\Gamma \Gamma X X } $ & $ -53.460160 $\\[0.4em] 

& $ ^{1}d^{T_{2g}^{}T_{2g}^{}A_{1g}^{}A_{1g}^{}}_{\Gamma \Gamma X X } $ & $  85.586456 $& $ d^{T_{1u}^{}T_{1u}^{}A_{1g}^{}A_{1g}^{}}_{\Gamma \Gamma X X } $ & $  9.182876 $& $ ^{1}d^{T_{1u}^{}T_{1u}^{}A_{1g}^{}A_{1g}^{}}_{\Gamma \Gamma X X } $ & $  79.781704 $& $ d^{T_{2g}^{}T_{2g}^{}A_{1g}^{}E_{g}^{}}_{\Gamma \Gamma X X } $ & $  13.386701 $\\[0.4em] 

& $ d^{T_{1u}^{}T_{1u}^{}A_{1g}^{}E_{g}^{}}_{\Gamma \Gamma X X } $ & $  114.103614 $& $ d^{T_{1u}^{}T_{2g}^{}A_{1g}^{}\tensor*[^{2}]{\hspace{-0.2em}A}{}_{2u}^{}}_{\Gamma \Gamma X X } $ & $ -67.875605 $& $ d^{T_{1u}^{}T_{2g}^{}A_{1g}^{}B_{1u}^{}}_{\Gamma \Gamma X X } $ & $ -18.553171 $& $ ^{1}d^{T_{1u}^{}T_{2g}^{}A_{1g}^{}B_{1u}^{}}_{\Gamma \Gamma X X } $ & $  37.127425 $\\[0.4em] 

& $ d^{T_{1u}^{}T_{2g}^{}A_{1g}^{}E_{u}^{}}_{\Gamma \Gamma X X } $ & $  25.943192 $& $ ^{1}d^{T_{1u}^{}T_{2g}^{}A_{1g}^{}E_{u}^{}}_{\Gamma \Gamma X X } $ & $  24.275144 $& $ d^{T_{1u}^{}T_{2g}^{}A_{1g}^{}\tensor*[^{1}]{\hspace{-0.2em}E}{}_{u}^{}}_{\Gamma \Gamma X X } $ & $ -62.775122 $& $ ^{1}d^{T_{1u}^{}T_{2g}^{}A_{1g}^{}\tensor*[^{1}]{\hspace{-0.2em}E}{}_{u}^{}}_{\Gamma \Gamma X X } $ & $ -63.730592 $\\
\hline
\end{longtable*}

\subsection{The irreducible derivatives computed at $a=$ 5.5717 \AA, corresponding to $T=600$ K}

\begin{longtable*}{lll|ll|ll|ll}
\caption{Second order irreducible derivatives for
UO$_2$ computed at $a=$ 5.5717 \AA. The units are eV/\AA$^2$.} \\
\hline\hline\\[-1em]
& Derivative & Value & Derivative & Value & Derivative & Value & Derivative & Value \\[0.4em]
\hline
\\[\dimexpr-\normalbaselineskip+0.2em]
\endfirsthead
\hline\hline\\[-1em]
& Derivative & Value & Derivative & Value & Derivative & Value & Derivative & Value \\[0.4em]
\hline
\\[\dimexpr-\normalbaselineskip+0.2em]
\endhead
\hline\hline
\endfoot
& $ d^{T_{2g}^{}T_{2g}^{}}_{\Gamma \Gamma } $ & $  11.446056 $& $ d^{T_{1u}^{}T_{1u}^{}}_{\Gamma \Gamma } $ & $  9.921400 $& $ d^{A_{1g}^{}A_{1g}^{}}_{L L } $ & $  15.539400 $& $ d^{E_{g}^{}E_{g}^{}}_{L L } $ & $  8.206198 $\\[0.4em] 

& $ d^{A_{2u}^{}A_{2u}^{}}_{L L } $ & $  29.353995 $& $ d^{A_{2u}^{}\tensor*[^{1}]{\hspace{-0.2em}A}{}_{2u}^{}}_{L L } $ & $  6.926765 $& $ d^{\tensor*[^{1}]{\hspace{-0.2em}A}{}_{2u}^{}\tensor*[^{1}]{\hspace{-0.2em}A}{}_{2u}^{}}_{L L } $ & $  8.354098 $& $ d^{E_{u}^{}E_{u}^{}}_{L L } $ & $  7.226230 $\\[0.4em] 

& $ d^{E_{u}^{}\tensor*[^{1}]{\hspace{-0.2em}E}{}_{u}^{}}_{L L } $ & $ -2.124085 $& $ d^{\tensor*[^{1}]{\hspace{-0.2em}E}{}_{u}^{}\tensor*[^{1}]{\hspace{-0.2em}E}{}_{u}^{}}_{L L } $ & $  6.503046 $& $ d^{A_{1}^{}A_{1}^{}}_{\Delta\bar{\Delta}} $ & $  32.820606 $& $ d^{\tensor*[^{1}]{\hspace{-0.2em}A}{}_{1}^{}A_{1}^{}}_{\Delta\bar{\Delta}} $ & $  14.385186 $\\[0.4em] 

& $ d^{\tensor*[^{1}]{\hspace{-0.2em}A}{}_{1}^{}\tensor*[^{1}]{\hspace{-0.2em}A}{}_{1}^{}}_{\Delta\bar{\Delta}} $ & $  16.921490 $& $ d^{B_{2}^{}B_{2}^{}}_{\Delta\bar{\Delta}} $ & $  6.553042 $& $ d^{E_{}^{}E_{}^{}}_{\Delta\bar{\Delta}} $ & $  7.631204 $& $ d^{\tensor*[^{1}]{\hspace{-0.2em}E}{}_{}^{}E_{}^{}}_{\Delta\bar{\Delta}} $ & $ -3.288536 $\\[0.4em] 

& $ d^{\tensor*[^{2}]{\hspace{-0.2em}E}{}_{}^{}E_{}^{}}_{\Delta\bar{\Delta}} $ & $ -0.035593 $& $ d^{\tensor*[^{1}]{\hspace{-0.2em}E}{}_{}^{}\tensor*[^{1}]{\hspace{-0.2em}E}{}_{}^{}}_{\Delta\bar{\Delta}} $ & $  3.225325 $& $ d^{\tensor*[^{2}]{\hspace{-0.2em}E}{}_{}^{}\tensor*[^{1}]{\hspace{-0.2em}E}{}_{}^{}}_{\Delta\bar{\Delta}} $ & $  0.853115 $& $ d^{\tensor*[^{2}]{\hspace{-0.2em}E}{}_{}^{}\tensor*[^{2}]{\hspace{-0.2em}E}{}_{}^{}}_{\Delta\bar{\Delta}} $ & $  10.926715 $\\
\hline
\end{longtable*}
\begin{longtable*}{lll|ll|ll|ll}
\caption{Third order irreducible derivatives for
UO$_2$ computed at $a=$ 5.5717 \AA. The units are eV/\AA$^3$.} \\
\hline\hline\\[-1em]
& Derivative & Value & Derivative & Value & Derivative & Value & Derivative & Value \\[0.4em]
\hline
\\[\dimexpr-\normalbaselineskip+0.2em]
\endfirsthead
\hline\hline\\[-1em]
& Derivative & Value & Derivative & Value & Derivative & Value & Derivative & Value \\[0.4em]
\hline
\\[\dimexpr-\normalbaselineskip+0.2em]
\endhead
\hline\hline
\endfoot
& $ d^{T_{2g}^{}T_{2g}^{}T_{2g}^{}}_{\Gamma \Gamma \Gamma } $ & $ -54.076213 $& $ d^{T_{1u}^{}T_{1u}^{}T_{2g}^{}}_{\Gamma \Gamma \Gamma } $ & $ -51.143035 $& $ d^{T_{2g}^{}A_{1}^{}A_{1}^{}}_{\Gamma \bar{A}A} $ & $  30.349903 $& $ d^{T_{2g}^{}A_{1}^{}\tensor*[^{1}]{\hspace{-0.2em}A}{}_{1}^{}}_{\Gamma \bar{A}A} $ & $ -7.313766 $\\[0.4em] 

& $ d^{T_{1u}^{}A_{1}^{}\tensor*[^{1}]{\hspace{-0.2em}A}{}_{1}^{}}_{\Gamma \bar{A}A} $ & $  2.238495i $& $ d^{T_{2g}^{}A_{1}^{}\tensor*[^{2}]{\hspace{-0.2em}A}{}_{1}^{}}_{\Gamma \bar{A}A} $ & $  6.069685 $& $ d^{T_{1u}^{}A_{1}^{}\tensor*[^{2}]{\hspace{-0.2em}A}{}_{1}^{}}_{\Gamma \bar{A}A} $ & $  33.284826i $& $ d^{T_{2g}^{}A_{1}^{}A_{2}^{}}_{\Gamma \bar{A}A} $ & $  4.507174i $\\[0.4em] 

& $ d^{T_{2g}^{}A_{1}^{}B_{1}^{}}_{\Gamma \bar{A}A} $ & $ -3.437231i $& $ d^{T_{1u}^{}A_{1}^{}B_{1}^{}}_{\Gamma \bar{A}A} $ & $ -19.959802 $& $ d^{T_{2g}^{}A_{1}^{}\tensor*[^{1}]{\hspace{-0.2em}B}{}_{1}^{}}_{\Gamma \bar{A}A} $ & $  20.636224i $& $ d^{T_{1u}^{}A_{1}^{}\tensor*[^{1}]{\hspace{-0.2em}B}{}_{1}^{}}_{\Gamma \bar{A}A} $ & $ -6.772564 $\\[0.4em] 

& $ d^{T_{2g}^{}A_{1}^{}\tensor*[^{2}]{\hspace{-0.2em}B}{}_{1}^{}}_{\Gamma \bar{A}A} $ & $  15.493271i $& $ d^{T_{1u}^{}A_{1}^{}\tensor*[^{2}]{\hspace{-0.2em}B}{}_{1}^{}}_{\Gamma \bar{A}A} $ & $ -4.182787 $& $ d^{T_{1u}^{}A_{1}^{}B_{2}^{}}_{\Gamma \bar{A}A} $ & $ -10.659395i $& $ d^{T_{1u}^{}A_{1}^{}\tensor*[^{1}]{\hspace{-0.2em}B}{}_{2}^{}}_{\Gamma \bar{A}A} $ & $  4.376051i $\\[0.4em] 

& $ d^{T_{2g}^{}\tensor*[^{1}]{\hspace{-0.2em}A}{}_{1}^{}\tensor*[^{1}]{\hspace{-0.2em}A}{}_{1}^{}}_{\Gamma \bar{A}A} $ & $ -2.443805 $& $ d^{T_{2g}^{}\tensor*[^{1}]{\hspace{-0.2em}A}{}_{1}^{}\tensor*[^{2}]{\hspace{-0.2em}A}{}_{1}^{}}_{\Gamma \bar{A}A} $ & $ -4.215682 $& $ d^{T_{1u}^{}\tensor*[^{1}]{\hspace{-0.2em}A}{}_{1}^{}\tensor*[^{2}]{\hspace{-0.2em}A}{}_{1}^{}}_{\Gamma \bar{A}A} $ & $ -19.495775i $& $ d^{T_{2g}^{}\tensor*[^{1}]{\hspace{-0.2em}A}{}_{1}^{}A_{2}^{}}_{\Gamma \bar{A}A} $ & $  7.645277i $\\
\hline
\end{longtable*}
\begin{longtable*}{lll|ll|ll|ll}
\caption{Fourth order irreducible derivatives for
UO$_2$ computed at $a=$ 5.5717 \AA. The units are eV/\AA$^4$.} \\
\hline\hline\\[-1em]
& Derivative & Value & Derivative & Value & Derivative & Value & Derivative & Value \\[0.4em]
\hline
\\[\dimexpr-\normalbaselineskip+0.2em]
\endfirsthead
\hline\hline\\[-1em]
& Derivative & Value & Derivative & Value & Derivative & Value & Derivative & Value \\[0.4em]
\hline
\\[\dimexpr-\normalbaselineskip+0.2em]
\endhead
\hline\hline
\endfoot
& $ d^{T_{2g}^{}T_{2g}^{}T_{2g}^{}T_{2g}^{}}_{\Gamma \Gamma \Gamma \Gamma } $ & $  228.600647 $& $ ^{1}d^{T_{2g}^{}T_{2g}^{}T_{2g}^{}T_{2g}^{}}_{\Gamma \Gamma \Gamma \Gamma } $ & $ -14.782732 $& $ d^{T_{1u}^{}T_{1u}^{}T_{2g}^{}T_{2g}^{}}_{\Gamma \Gamma \Gamma \Gamma } $ & $  51.712189 $& $ ^{1}d^{T_{1u}^{}T_{1u}^{}T_{2g}^{}T_{2g}^{}}_{\Gamma \Gamma \Gamma \Gamma } $ & $  13.436510 $\\[0.4em] 

& $ ^{2}d^{T_{1u}^{}T_{1u}^{}T_{2g}^{}T_{2g}^{}}_{\Gamma \Gamma \Gamma \Gamma } $ & $  106.876258 $& $ d^{T_{1u}^{}T_{1u}^{}T_{1u}^{}T_{1u}^{}}_{\Gamma \Gamma \Gamma \Gamma } $ & $  149.463978 $& $ ^{1}d^{T_{1u}^{}T_{1u}^{}T_{1u}^{}T_{1u}^{}}_{\Gamma \Gamma \Gamma \Gamma } $ & $  555.615894 $& $ d^{T_{2g}^{}T_{2g}^{}A_{1g}^{}A_{1g}^{}}_{\Gamma \Gamma X X } $ & $  2.820577 $\\[0.4em] 

& $ ^{1}d^{T_{2g}^{}T_{2g}^{}A_{1g}^{}A_{1g}^{}}_{\Gamma \Gamma X X } $ & $  104.688354 $& $ d^{T_{1u}^{}T_{1u}^{}A_{1g}^{}A_{1g}^{}}_{\Gamma \Gamma X X } $ & $ -12.639582 $& $ ^{1}d^{T_{1u}^{}T_{1u}^{}A_{1g}^{}A_{1g}^{}}_{\Gamma \Gamma X X } $ & $  64.636932 $& $ d^{T_{2g}^{}T_{2g}^{}A_{1g}^{}E_{g}^{}}_{\Gamma \Gamma X X } $ & $  11.113146 $\\[0.4em] 

& $ d^{T_{1u}^{}T_{1u}^{}A_{1g}^{}E_{g}^{}}_{\Gamma \Gamma X X } $ & $  67.124909 $& $ d^{T_{1u}^{}T_{2g}^{}A_{1g}^{}\tensor*[^{2}]{\hspace{-0.2em}A}{}_{2u}^{}}_{\Gamma \Gamma X X } $ & $ -63.601668 $& $ d^{T_{1u}^{}T_{2g}^{}A_{1g}^{}B_{1u}^{}}_{\Gamma \Gamma X X } $ & $  1.095535 $& $ ^{1}d^{T_{1u}^{}T_{2g}^{}A_{1g}^{}B_{1u}^{}}_{\Gamma \Gamma X X } $ & $  31.342885 $\\[0.4em] 

& $ d^{T_{1u}^{}T_{2g}^{}A_{1g}^{}E_{u}^{}}_{\Gamma \Gamma X X } $ & $  34.335773 $& $ ^{1}d^{T_{1u}^{}T_{2g}^{}A_{1g}^{}E_{u}^{}}_{\Gamma \Gamma X X } $ & $  20.524324 $& $ d^{T_{1u}^{}T_{2g}^{}A_{1g}^{}\tensor*[^{1}]{\hspace{-0.2em}E}{}_{u}^{}}_{\Gamma \Gamma X X } $ & $ -64.334986 $& $ ^{1}d^{T_{1u}^{}T_{2g}^{}A_{1g}^{}\tensor*[^{1}]{\hspace{-0.2em}E}{}_{u}^{}}_{\Gamma \Gamma X X } $ & $ -75.530490 $\\
\hline
\end{longtable*}

\subsection{The irreducible derivatives computed at $a=$ 5.5942 \AA, corresponding to $T=1000$ K}

\begin{longtable*}{lll|ll|ll|ll}
\caption{Second order irreducible derivatives for
UO$_2$ computed at $a=$ 5.5942 \AA. The units are eV/\AA$^2$.} \\
\hline\hline\\[-1em]
& Derivative & Value & Derivative & Value & Derivative & Value & Derivative & Value \\[0.4em]
\hline
\\[\dimexpr-\normalbaselineskip+0.2em]
\endfirsthead
\hline\hline\\[-1em]
& Derivative & Value & Derivative & Value & Derivative & Value & Derivative & Value \\[0.4em]
\hline
\\[\dimexpr-\normalbaselineskip+0.2em]
\endhead
\hline\hline
\endfoot
& $ d^{T_{2g}^{}T_{2g}^{}}_{\Gamma \Gamma } $ & $  10.731368 $& $ d^{T_{1u}^{}T_{1u}^{}}_{\Gamma \Gamma } $ & $  9.139516 $& $ d^{A_{1g}^{}A_{1g}^{}}_{L L } $ & $  15.077629 $& $ d^{E_{g}^{}E_{g}^{}}_{L L } $ & $  7.841429 $\\[0.4em] 

& $ d^{A_{2u}^{}A_{2u}^{}}_{L L } $ & $  28.403984 $& $ d^{A_{2u}^{}\tensor*[^{1}]{\hspace{-0.2em}A}{}_{2u}^{}}_{L L } $ & $  7.159112 $& $ d^{\tensor*[^{1}]{\hspace{-0.2em}A}{}_{2u}^{}\tensor*[^{1}]{\hspace{-0.2em}A}{}_{2u}^{}}_{L L } $ & $  8.045605 $& $ d^{E_{u}^{}E_{u}^{}}_{L L } $ & $  6.656453 $\\[0.4em] 

& $ d^{E_{u}^{}\tensor*[^{1}]{\hspace{-0.2em}E}{}_{u}^{}}_{L L } $ & $ -1.707333 $& $ d^{\tensor*[^{1}]{\hspace{-0.2em}E}{}_{u}^{}\tensor*[^{1}]{\hspace{-0.2em}E}{}_{u}^{}}_{L L } $ & $  6.187658 $& $ d^{A_{1}^{}A_{1}^{}}_{\Delta\bar{\Delta}} $ & $  31.970357 $& $ d^{\tensor*[^{1}]{\hspace{-0.2em}A}{}_{1}^{}A_{1}^{}}_{\Delta\bar{\Delta}} $ & $  14.055809 $\\[0.4em] 

& $ d^{\tensor*[^{1}]{\hspace{-0.2em}A}{}_{1}^{}\tensor*[^{1}]{\hspace{-0.2em}A}{}_{1}^{}}_{\Delta\bar{\Delta}} $ & $  16.521171 $& $ d^{B_{2}^{}B_{2}^{}}_{\Delta\bar{\Delta}} $ & $  6.250437 $& $ d^{E_{}^{}E_{}^{}}_{\Delta\bar{\Delta}} $ & $  7.038060 $& $ d^{\tensor*[^{1}]{\hspace{-0.2em}E}{}_{}^{}E_{}^{}}_{\Delta\bar{\Delta}} $ & $ -3.012463 $\\[0.4em] 

& $ d^{\tensor*[^{2}]{\hspace{-0.2em}E}{}_{}^{}E_{}^{}}_{\Delta\bar{\Delta}} $ & $ -0.352854 $& $ d^{\tensor*[^{1}]{\hspace{-0.2em}E}{}_{}^{}\tensor*[^{1}]{\hspace{-0.2em}E}{}_{}^{}}_{\Delta\bar{\Delta}} $ & $  2.984175 $& $ d^{\tensor*[^{2}]{\hspace{-0.2em}E}{}_{}^{}\tensor*[^{1}]{\hspace{-0.2em}E}{}_{}^{}}_{\Delta\bar{\Delta}} $ & $  0.858252 $& $ d^{\tensor*[^{2}]{\hspace{-0.2em}E}{}_{}^{}\tensor*[^{2}]{\hspace{-0.2em}E}{}_{}^{}}_{\Delta\bar{\Delta}} $ & $  10.537277 $\\
\hline
\end{longtable*}
\begin{longtable*}{lll|ll|ll|ll}
\caption{Third order irreducible derivatives for
UO$_2$ computed at $a=$ 5.5942 \AA. The units are eV/\AA$^3$.} \\
\hline\hline\\[-1em]
& Derivative & Value & Derivative & Value & Derivative & Value & Derivative & Value \\[0.4em]
\hline
\\[\dimexpr-\normalbaselineskip+0.2em]
\endfirsthead
\hline\hline\\[-1em]
& Derivative & Value & Derivative & Value & Derivative & Value & Derivative & Value \\[0.4em]
\hline
\\[\dimexpr-\normalbaselineskip+0.2em]
\endhead
\hline\hline
\endfoot
& $ d^{T_{2g}^{}T_{2g}^{}T_{2g}^{}}_{\Gamma \Gamma \Gamma } $ & $ -63.996229 $& $ d^{T_{1u}^{}T_{1u}^{}T_{2g}^{}}_{\Gamma \Gamma \Gamma } $ & $ -47.006050 $& $ d^{T_{2g}^{}A_{1}^{}A_{1}^{}}_{\Gamma \bar{A}A} $ & $  29.346713 $& $ d^{T_{2g}^{}A_{1}^{}\tensor*[^{1}]{\hspace{-0.2em}A}{}_{1}^{}}_{\Gamma \bar{A}A} $ & $ -7.150878 $\\[0.4em] 

& $ d^{T_{1u}^{}A_{1}^{}\tensor*[^{1}]{\hspace{-0.2em}A}{}_{1}^{}}_{\Gamma \bar{A}A} $ & $  1.943981i $& $ d^{T_{2g}^{}A_{1}^{}\tensor*[^{2}]{\hspace{-0.2em}A}{}_{1}^{}}_{\Gamma \bar{A}A} $ & $  6.161541 $& $ d^{T_{1u}^{}A_{1}^{}\tensor*[^{2}]{\hspace{-0.2em}A}{}_{1}^{}}_{\Gamma \bar{A}A} $ & $  31.864691i $& $ d^{T_{2g}^{}A_{1}^{}A_{2}^{}}_{\Gamma \bar{A}A} $ & $  3.811150i $\\[0.4em] 

& $ d^{T_{2g}^{}A_{1}^{}B_{1}^{}}_{\Gamma \bar{A}A} $ & $ -3.032661i $& $ d^{T_{1u}^{}A_{1}^{}B_{1}^{}}_{\Gamma \bar{A}A} $ & $ -18.639198 $& $ d^{T_{2g}^{}A_{1}^{}\tensor*[^{1}]{\hspace{-0.2em}B}{}_{1}^{}}_{\Gamma \bar{A}A} $ & $  19.323621i $& $ d^{T_{1u}^{}A_{1}^{}\tensor*[^{1}]{\hspace{-0.2em}B}{}_{1}^{}}_{\Gamma \bar{A}A} $ & $ -6.047773 $\\[0.4em] 

& $ d^{T_{2g}^{}A_{1}^{}\tensor*[^{2}]{\hspace{-0.2em}B}{}_{1}^{}}_{\Gamma \bar{A}A} $ & $  14.985033i $& $ d^{T_{1u}^{}A_{1}^{}\tensor*[^{2}]{\hspace{-0.2em}B}{}_{1}^{}}_{\Gamma \bar{A}A} $ & $ -3.730084 $& $ d^{T_{1u}^{}A_{1}^{}B_{2}^{}}_{\Gamma \bar{A}A} $ & $ -10.876785i $& $ d^{T_{1u}^{}A_{1}^{}\tensor*[^{1}]{\hspace{-0.2em}B}{}_{2}^{}}_{\Gamma \bar{A}A} $ & $  3.940739i $\\[0.4em] 

& $ d^{T_{2g}^{}\tensor*[^{1}]{\hspace{-0.2em}A}{}_{1}^{}\tensor*[^{1}]{\hspace{-0.2em}A}{}_{1}^{}}_{\Gamma \bar{A}A} $ & $ -2.202430 $& $ d^{T_{2g}^{}\tensor*[^{1}]{\hspace{-0.2em}A}{}_{1}^{}\tensor*[^{2}]{\hspace{-0.2em}A}{}_{1}^{}}_{\Gamma \bar{A}A} $ & $ -4.388093 $& $ d^{T_{1u}^{}\tensor*[^{1}]{\hspace{-0.2em}A}{}_{1}^{}\tensor*[^{2}]{\hspace{-0.2em}A}{}_{1}^{}}_{\Gamma \bar{A}A} $ & $ -18.478635i $& $ d^{T_{2g}^{}\tensor*[^{1}]{\hspace{-0.2em}A}{}_{1}^{}A_{2}^{}}_{\Gamma \bar{A}A} $ & $  7.367873i $\\
\hline
\end{longtable*}
\begin{longtable*}{lll|ll|ll|ll}
\caption{Fourth order irreducible derivatives for
UO$_2$ computed at $a=$ 5.5942 \AA. The units are eV/\AA$^4$.} \\
\hline\hline\\[-1em]
& Derivative & Value & Derivative & Value & Derivative & Value & Derivative & Value \\[0.4em]
\hline
\\[\dimexpr-\normalbaselineskip+0.2em]
\endfirsthead
\hline\hline\\[-1em]
& Derivative & Value & Derivative & Value & Derivative & Value & Derivative & Value \\[0.4em]
\hline
\\[\dimexpr-\normalbaselineskip+0.2em]
\endhead
\hline\hline
\endfoot
& $ d^{T_{2g}^{}T_{2g}^{}T_{2g}^{}T_{2g}^{}}_{\Gamma \Gamma \Gamma \Gamma } $ & $  150.033730 $& $ ^{1}d^{T_{2g}^{}T_{2g}^{}T_{2g}^{}T_{2g}^{}}_{\Gamma \Gamma \Gamma \Gamma } $ & $  67.412321 $& $ d^{T_{1u}^{}T_{1u}^{}T_{2g}^{}T_{2g}^{}}_{\Gamma \Gamma \Gamma \Gamma } $ & $  29.572698 $& $ ^{1}d^{T_{1u}^{}T_{1u}^{}T_{2g}^{}T_{2g}^{}}_{\Gamma \Gamma \Gamma \Gamma } $ & $  14.974370 $\\[0.4em] 

& $ ^{2}d^{T_{1u}^{}T_{1u}^{}T_{2g}^{}T_{2g}^{}}_{\Gamma \Gamma \Gamma \Gamma } $ & $  57.254761 $& $ d^{T_{1u}^{}T_{1u}^{}T_{1u}^{}T_{1u}^{}}_{\Gamma \Gamma \Gamma \Gamma } $ & $  124.545021 $& $ ^{1}d^{T_{1u}^{}T_{1u}^{}T_{1u}^{}T_{1u}^{}}_{\Gamma \Gamma \Gamma \Gamma } $ & $  552.997707 $& $ d^{T_{2g}^{}T_{2g}^{}A_{1g}^{}A_{1g}^{}}_{\Gamma \Gamma X X } $ & $  3.209651 $\\[0.4em] 

& $ ^{1}d^{T_{2g}^{}T_{2g}^{}A_{1g}^{}A_{1g}^{}}_{\Gamma \Gamma X X } $ & $  64.165987 $& $ d^{T_{1u}^{}T_{1u}^{}A_{1g}^{}A_{1g}^{}}_{\Gamma \Gamma X X } $ & $ -23.315677 $& $ ^{1}d^{T_{1u}^{}T_{1u}^{}A_{1g}^{}A_{1g}^{}}_{\Gamma \Gamma X X } $ & $  55.829441 $& $ d^{T_{2g}^{}T_{2g}^{}A_{1g}^{}E_{g}^{}}_{\Gamma \Gamma X X } $ & $ -12.828741 $\\[0.4em] 

& $ d^{T_{1u}^{}T_{1u}^{}A_{1g}^{}E_{g}^{}}_{\Gamma \Gamma X X } $ & $  82.383466 $& $ d^{T_{1u}^{}T_{2g}^{}A_{1g}^{}\tensor*[^{2}]{\hspace{-0.2em}A}{}_{2u}^{}}_{\Gamma \Gamma X X } $ & $ -78.732685 $& $ d^{T_{1u}^{}T_{2g}^{}A_{1g}^{}B_{1u}^{}}_{\Gamma \Gamma X X } $ & $ -5.552620 $& $ ^{1}d^{T_{1u}^{}T_{2g}^{}A_{1g}^{}B_{1u}^{}}_{\Gamma \Gamma X X } $ & $  34.344391 $\\[0.4em] 

& $ d^{T_{1u}^{}T_{2g}^{}A_{1g}^{}E_{u}^{}}_{\Gamma \Gamma X X } $ & $  36.112898 $& $ ^{1}d^{T_{1u}^{}T_{2g}^{}A_{1g}^{}E_{u}^{}}_{\Gamma \Gamma X X } $ & $  26.813383 $& $ d^{T_{1u}^{}T_{2g}^{}A_{1g}^{}\tensor*[^{1}]{\hspace{-0.2em}E}{}_{u}^{}}_{\Gamma \Gamma X X } $ & $ -58.910826 $& $ ^{1}d^{T_{1u}^{}T_{2g}^{}A_{1g}^{}\tensor*[^{1}]{\hspace{-0.2em}E}{}_{u}^{}}_{\Gamma \Gamma X X } $ & $ -66.127896 $\\
\hline
\end{longtable*}

\bibliographystyle{apsrev4-1} 
\bibliography{2Supplementary.bbl} 